%% file: SydneyFinal.tex
\documentclass[11pt,onecolumn]{IEEEtran}
\usepackage[top=0.2in,bottom=0.5in,left=0.4in,right=0.5in]{geometry}
\usepackage{cite}
\usepackage{graphicx}
\pdfoutput=1
\usepackage{caption}
\usepackage{subcaption}
\usepackage{color}
\usepackage{inputenc}

\begin{document}
\title{Passenger Flow Prediction at Sydney International Airport : a data-driven queuing approach}
\author{Harold Nikoue, Aude Marzuoli, Dr. John-Paul Clarke, Dr. Eric Feron, Jim Peters }
\maketitle
\IEEEpeerreviewmaketitle
Airports constitute some of the most complex systems enabling mobility in today's society. The various components inside the airport system each have specific requirements and include numerous systems, processes and stakeholders. Stakeholders are comprised of government entities (security and customs), private bodies (airport owner, airlines) and customers (passengers and cargo).

An airport determines the traveler's first and last impression of a city. A positive airport experience is beneficial to sales and influences future travel choices (cite airport council international customer service). Airports have taken steps to increase their customer focus. An improved customer experience relies on technologies supporting better service. Two examples of such technologies are Radio-Frequency Identification (RFID), which would enable airports to track passengers and bags effectively \cite{devries2008state} \cite{wyld2005my}, and Bluetooth \cite{hansen2009location} to support passenger tracking. To facilitate the needs of airport customers and operators, such technologies need to be part of the activities of airport users at the airport. While at the airport, passengers engage in processing and discretionary activities. Processing activities are enforced to conform to the legal and regulatory requirements for air travel. They correspond to : check in, departure paperwork fill out, going through identity and security checkpoints, boarding and deboarding a plane. Passengers actually spend a small portion of their time at airports engaging in processing activities, including time spend waiting to be processed \cite{takakuwa2003modeling}. 

A common feature of service systems is that the demand for service varies throughout the day. 

Air terminal queues \cite{koopman1972air}

Staffing requirements are part of the design and management of the service system. In the long term planning horizon, managers set the system capacity. On the short term horizon, managers make agents scheduling decisions, indicating the number of agents working during specific hours, and breaking down the day in time intervals.

The scheduling decision is often made based on the solution of an integer linear program \cite{dantzig1955linear}, \cite{segal1974operator}, \cite{kolesar1975queuing}.
In real time, managers may make additional adjustements (flexing decisions) to move agents on and off the line of duty. This can be achieved if there are additional agents on site working on other tasks or if more agents can be called on short notice.

Robertson et al. \cite{robertson2002role} provided a detailed procedure to model passenger arrivals to estimate how many passengers arrived at the airport during each day and time of day. The raw passenger volume for each time interval was the final product and corresponded to the passenger arrival pattern. Further analysis provided access to passenger arrival patterns at different processing points (check-in, baggage security, security checkpoint...). The passenger arrival pattern for each checkpoint was computed using several inputs : passenger arrival behavior, flight schedules, aircraft capacity, load factors and transfer rates.

In a system where congestion can build up at peak hours. The number of servers is dynamically adjusted according to queue length. If the queue reaches an upper threshold, additional servers are opened. If some servers are idle, they get closed. By choosing appropriate thresholds, the queue length can be controlled in a certain range with high probability. This staffing policy is called congestion-based staffing \cite{zhang2009performance}.

\subsection{Literature Review}

A growing number of airports are providing Wi-Fi access to their passengers. Hence large volumes of signals from laptops, tablets and smart phones are picked up at the airport. By design, most Wireless devices aim at saving battery and therefore only periodically connect to the Wi-Fi network access. This leads to a set of data with discrete time location snapshots, and not a continuous set of location points of the device. In the 1990's, Lemer \cite{lemer1992measuring} stressed the need to develop performance measures at airports for different stakeholders, such as operators, airlines and passengers, who each have their set of measures. It suggested that queues, simulation and flow methods were adapted to modeling airports and hence measuring their performance. Using simulations and system dynamics, Manataki et al. \cite{manataki2009generic} modeled airport performance according to the staffing numbers required to process passengers and their waiting times. They later \cite{manataki2010assessing} surveyed existing analytical and simulation tools for airport analysis. They concluded that most analytical models focused on a particular area of the airport (e.g. check-in, baggage screening) but few tackled the entire airport terminal operations. Bluetooth has recently been used in the SPOPOS project to provide location-based services to passengers and airport operators \cite{hansen2009location}. From the airport perspective, it helped trigger alerts when queues were building up, to help passengers reach their plane on time.

\section{Model}
\subsection{Data sources}
Three different sources of information were used for the project:
  \begin{itemize}
\item The Flight Information Display System (FIDS) dataset contains information on gate, block, estimated and scheduled times for departing and arriving flights to Sydney International Airport. 
The records do not specify at what time the Estimated Time of Arrival (ETA) was recorded, nor how many times it was modified. Neither does it include the number of passengers on the flight, the time of arrival for an outbound flight or the time of departure for an inbound flight. \\
The simulation starts with the schedule of flights arrivals obtained from FIDS, and the passenger count estimated from the immigration files.

\item Passenger time stamps at immigration were recorded by the Australian Department of Immigration and Multicultural and Indigeneous Affairs (DIMIA). The DIMIA datasest consists in all border crossing activities for 2012. For any passenger, his or her nationality, the time stamp at immigration, his or her origin or destination airport and flight number are entered in the database.
The historic service rates at immigration can be derived from this information.\\
The flight number is used to compute the average number of passengers per flight by matching passengers to flights in FIDS. It allows us to generate a distribution of passenger occupancy per flight ID.
The dataset is also used to determine the service rate at each immigration desk at any hour during any day of the week. Note that each day of the week has a specific service rate distributin.\\
Since every DIMIA record contains the processing desk ID along with the time of the stamp, and other passenger information, the number of unique open desks can be estimated for a given time period. The service rate per desk per hour is the ratio of the number of passengers processed by the number of open desks.\\
The limitation of the dataset is the inconsistency of the manual recordings. Some entries are missing. At some places tail numbers are recorded in place of flight number, if that information is present. Many days in October and December are also missing from the data as shown on Figure \ref{fig:DIMIAYearly} 
\begin{figure}[!t]
\centering
\def\svgwidth{0.65\textwidth}
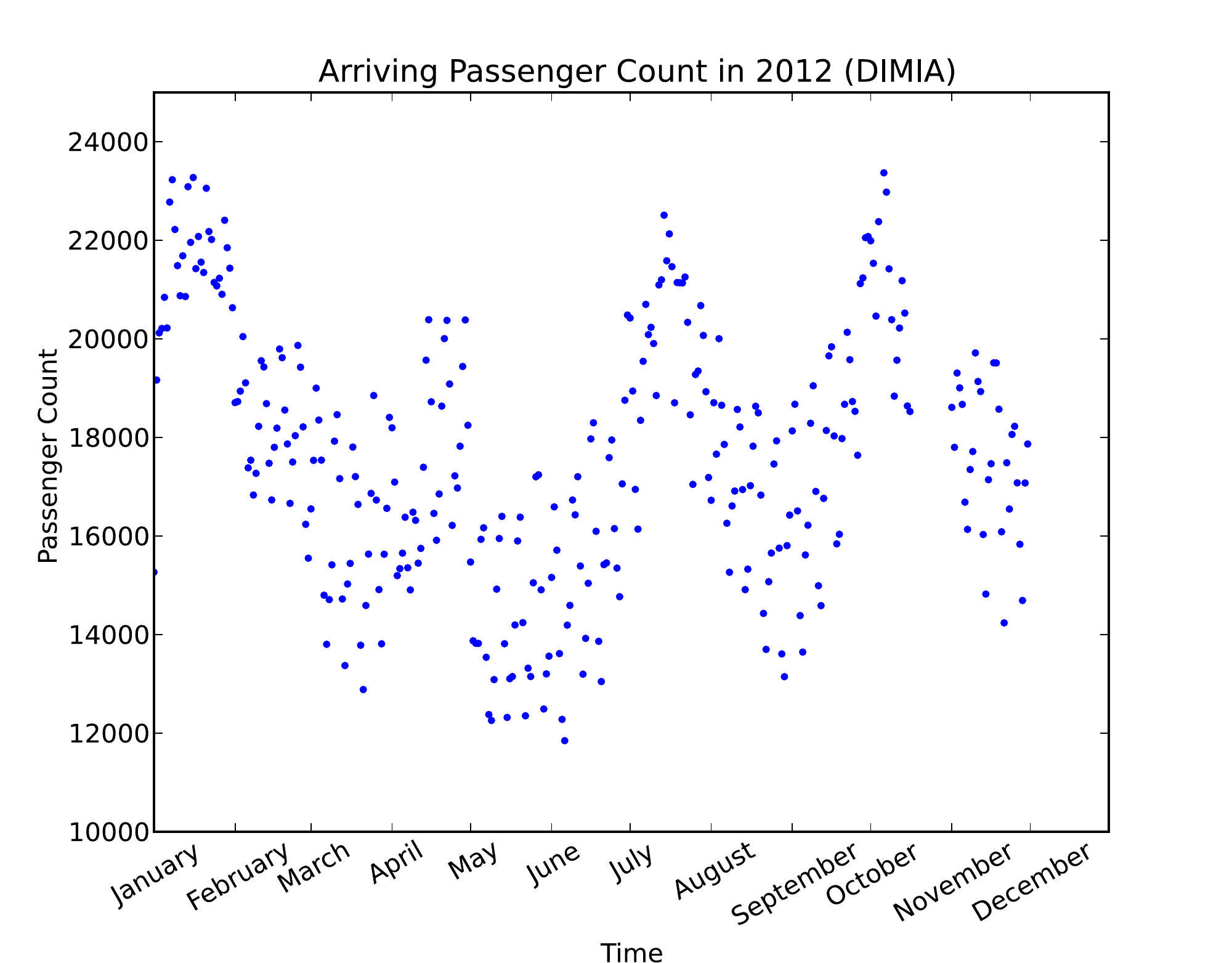
\caption{DIMIA count of arrivingpassengers stamps at Immigration in 2012}
\label{fig:DIMIAYearly}
\end{figure}
\begin{figure}[!t]
\centering
\def\svgwidth{0.65\textwidth}
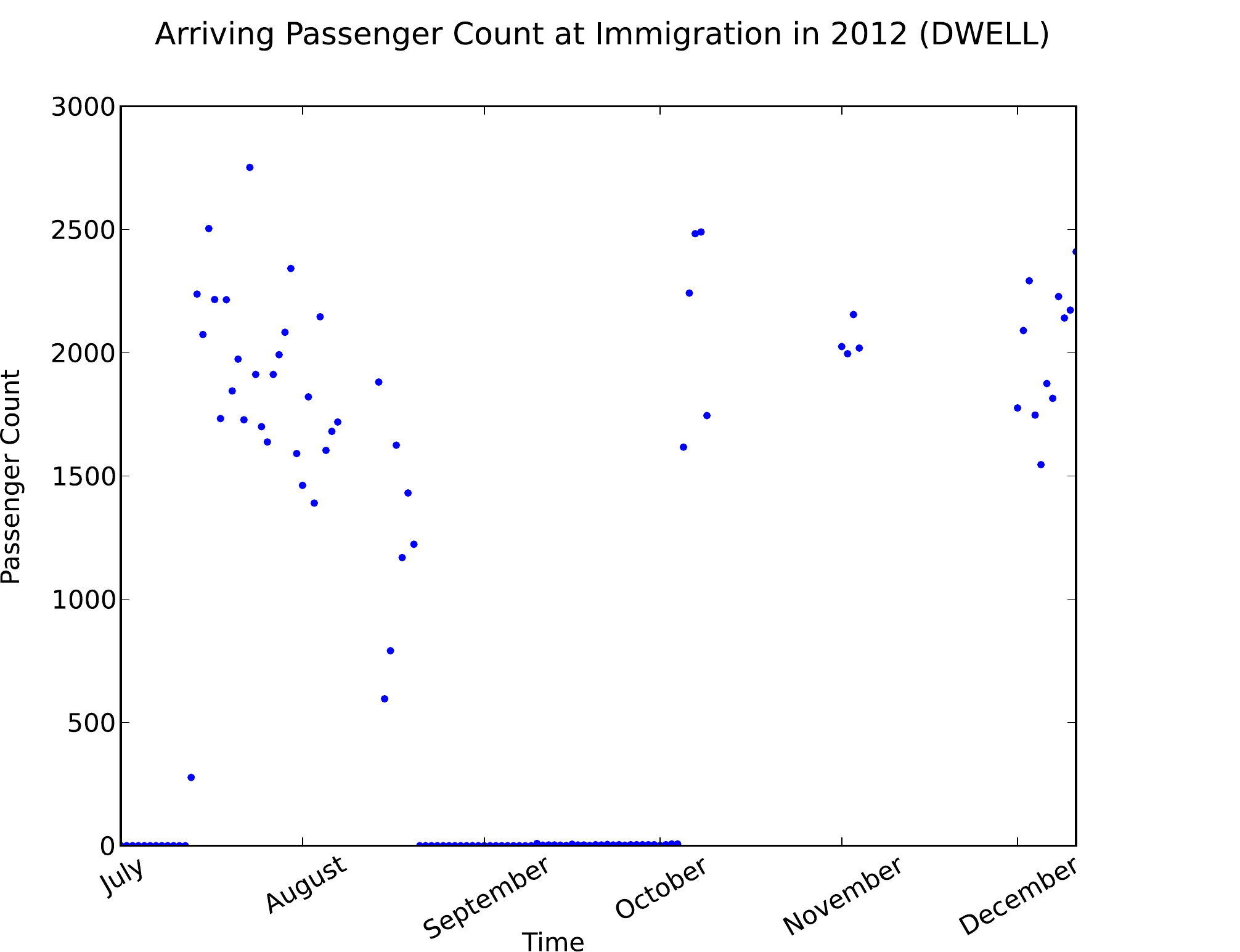
\caption{DWELL count of passengers recorded in Immigration zones in 2012}
\label{fig:DWELLYearly}
\end{figure}

\item  The airport is equipped with SITA iflow tool \cite{SITA:2012}, which returns anonymous Wi-Fi tracking information. 
The iflow tool consists in a network of more than 400 WiFi access points, 130 people-counters and 50 Bluetooth censors spread throughout the terminals \cite{SITA:2012}.

Wi-Fi tracking information includes (x,y) coordinates of the devices, the zone(s) assigned to the device by a triangulation algorithms and the time at which the device (e.g. computer, smart phone or tablet) is connected to the network. 
The results can lack precision due to the low accuracy of the triangulation and the low frequency of the signal updates.
 Many devices are observed only a couple of times at the airport at time intervals that can be as large as an hour. 
 A point is sometimes allocated to multiple neighbouring zones due to large uncertainties in measurements. Furthermore, some days have many entries and others very few. Within the same day, the quality of the data also varies by airport zones. Many zones did not have any recording of passengers  as illustrated on Table \ref{tab:DWELLhits} \\
\end{itemize}

%

Table \ref{tab:dataSources}  describes the content of the three data sets.
\begin{table}[Htp!]
	\caption{Data sources available on Sydney Airport.}
	\begin{tabular}{|l|p{5cm}|p{3cm}|p{3cm}|p{5cm}|}
	\hline
	Label &  Description & Date Range & Size &  Challenges\\
	\hline
	DIMIA &  
		Passenger time stamps at immigration for each border crossing & Jan. 2012 to May 2013 & 15,461,430 passengers(6,756,997 arrivals and 8,704,433 departures)  & Flight information or origin of the flight is not present\\
	\hline
	FIDS  &  Arriving and departing flight information including block, scheduled, estimated times and flight number &  Jan. 6th-Dec. 1st 2012 excluding May&  578,104(287,447 arrivals and 290,656 departures) &  No time of recording\\
	\hline
	DWELLL & Wi-Fi enabled devices tracking data for each  location(x,y) triangulated zone time stamp 	& July 1st-December 12th & 2,047,235 unique device IDs (827,474 arrivals and 1,236,372 departures overlapping)
	 & noisy information, inaccurate triangulations, unknown sampling\\
	\hline
	\end{tabular}
	\label{tab:dataSources}
\end{table}
The DWELL data lacks information for most of the days in August and September, as well as the second half of December as can be observed in figure \ref{fig:DWELLYearly}.
The walk times are fitted to a subset of the days that are contained in our records.

The DWELL data source is inconsistent between airport zones, see \ref{tab:DWELLhits}.
Due to the dearth of information for some zones of the airport, we decide to model the walk speed of passengers instead of their walkt times. A walk speed distribution is computed by dividing the walking times for all gates by the respective distances of these gates to immigration.
Figure \ref{fig:walkTimesGates50} illustrates the distributions of the walk times from three selected gates to immgigration. On figure \ref{fig:walkSpeed}, we can see that the shape of the distribution is well preserved for walk speeds.\\
\begin{center}
\begin{figure}[b]
\centering
\def\svgwidth{0.65\textwidth}
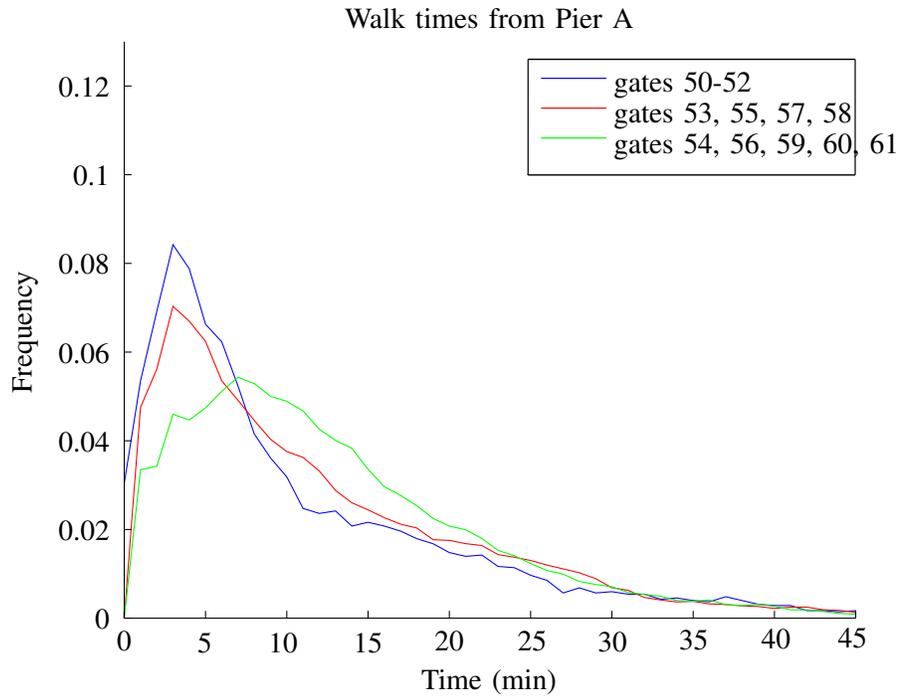
\caption{Walk times from gates in Pier A to immigration.}
\label{fig:walkTimesGates50}
\end{figure}

\begin{figure}[b]
\centering
\def\svgwidth{0.65\textwidth}
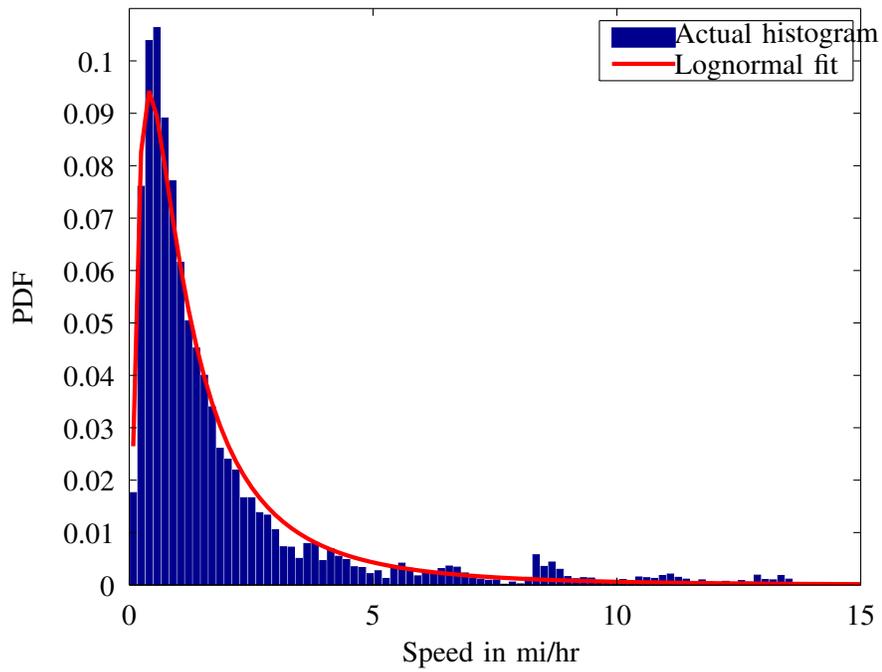
\caption{Walk Speeds  Distribution from all gates}
\label{fig:walkSpeed}
\end{figure}
\end{center}

\begin{center}
\begin{table}[H]
\caption{ Number of device ids recorded per zone in 2012}
\centering
\begin{tabular}{|l|l|l|l|}
\hline
Zone & Count & Zone & Count\\
\hline
AQIS PIER C	& 0	& Outbound-immigration	& 476,603\\
pierB-gate8	& 0	& dep-immi		& 287,381\\
pierB-gate9	& 0	& depart-dutyfree-all	& 294,722\\
pierB-gate10	& 0		& depart-dutyfree1	& 164,428\\
pierB-gate25	& 0		& depart-dutyfree2	& 109,532\\
pierB-gate30	& 0 		& depart-dutyfree3	& 135,641\\
pierB-gate31	& 0		& depart-dutyfree4	& 121,599\\
pierB-gate32	& 0		& depart-immigrationscreening	& 0 \\
pierB-gate33	& 0		& depart-foodcourt	& 0\\
pierB-gate34	& 0& depart-forum-all	& 0\\
pierB-gate35	& 0		& depart-landside	& 852,596 \\
pierB-gate36	& 0		& depart-landside-checkin1	& 0\\
pierB-gate37	& 0& depart-landside-checkin2	& 0\\
pierB-east	& 130,905	& depart-landside-checkin3	& 0\\
pierB-north	& 0		& depart-sec		& 280,048\\
pier B Inbound duty free & 0		& depart-staff area	& 0\\
pierB-North Arrivals & 0		& Departures-Check-in	& 883,611\\
pierB-East and South Arrivals	& 0		& Departures-North-Concourse	& 244,207\\
pierB-south	& 211,429		& Arrivals-Landside-all	& 0 \\
pierC-gate50	& 0		& Arrivals-Gates-08-and-09	& 0\\
pierC-gate51 	& 0		& Arrivals-Gates-24-and-25	& 0\\
pierC-gate53	& 0		& arrivals-immiB	& 98,223\\
pierC-gate54	& 0		& arrival-immiC		& 115,231\\
pierC-gate55 	& 0		& arrivals-PierB-North	& 18,460\\
pierC-gate56	& 0		& arrivals-PierB-south	& 17,452\\
pierC-gate57	& 0& arrivals-PierB-west		& 20,617\\
pierC-gate58	& 0& arrivals-PierC-all		& 158,126\\
pierC-gate59 	& 0& Forum 	& 532,942\\
pierC-gate60	& 0 &pier C Inbound duty free & 0\\
pierC-gate61	& 0&&\\
pierC-gate63	& 0&&\\
pierC-all	& 430,112&&\\
pier C Arrivals & 0&&\\
pierC-corridor	& 70,951&&\\
\hline
\end{tabular}
\label{tab:DWELLhits}
\end{table}
\end{center}
\clearpage
\newpage

In our simulation, we use the DIMIA information, by far the most complete and reliable dataset available, to generate a distribution of the number of passengers by flight to complement the FIDS information. The DWELL information is only used to get a relative measure of walk time that is independent from the number of passengers. This failure to model the dependency of walk time on congestion is one of the weaknesses of the model.
The combination of the data sets and their cross validation provides a clearer and more accurate picture of passenger flows in the airport.

\subsection{Theory}
Our model possesses stochastic and dynamic state variables that change after events that occur at discrete time intervals.
 Our state variables are the numbers of passengers located in selected airport zones: at gates, in the immigration queues, at the immigration service desks, at check in counters and at landside. All other airport zones are not part of the system studied. 

The state variables are both stochastic and dynamic, which constitute the last two requirements of a discrete-event simulation. The physical transition from one zone to the next, and the time spent in a  zone follow time-dependent probability distribution. 
Changes in passenger count occur in batches, after the arrival or departure of a flight.
For these reasons, an event-based discrete-event simulation was chosen to represent passenger movements behaviour\cite{Leemis2004Discrete}.

In an event-based simulation, time progresses directly to the next scheduled event.
An event for the simulation can be the arrival of a flight, the departure of a flight from a gate, an arrival at immigration or a departure from the immigration zone for the arrival process. After an event, the state variables are updated.\\
Future Event Lists (FELs) \cite{Birta2007Simulation} are used to schedule events. They consist in a list of event notices containing the start time and duration of a future event such as arrival or departure. 

In the simulation, each passenger arrives at the next service node following an exponential distribution of inter-arrival times. The service node includes a queue and a time-varying number of servers. The service is First-Come First-Served (FCFS), and the first passenger at the queue is always served first.
Upon arrival, if all active desks are busy, the passenger is scheduled to be processed by the first open desk. The passenger must wait, and will enter service after the first scheduled departure time.
If at least one active desk is free and no passenger is waiting to be processed, the passenger is processed and transferred to a departure list, where its departure time is computed.
If one desk is free and the list of waiting passengers is not empty, the first passenger in the queue is scheduled for departure and removed from the queue.. 
The departure time from a desk follows an empirical service rate distribution that varies with time of day and day of the week.  \\
Queue statistics including departure times, wait times, throughput and queue length  can be derived from the state variables, and are aggregated into 15 minutes time bins. 
Waiting times are  computed as the difference between the arrival and departure time of a passenger. A delay corresponds to the time difference between arrival at the general queue and arrival at a given server. The length of a queue is computed as the number of passengers in the queue at the end of a 15 minutes time interval.\\
Passengers are modelled individually from one queue to another.
Passengers travelling together are not treated as a group. There is no consideration of the fact that groups of passengers may have larger processing times at the different service nodes and larger walk times.
Similarly, all passengers are assigned the same priority at the service node, as a generalization of the FCFS assumption. No special consideration is being given to Australian nationals as compared to foreigners in the current simulation.

\subsection{Model}
\subsubsection{Arrivals}

As constructed, the model assumes a single path from a given gate to immigration. A passenger goes through each zone of the system with probability 1. Time spent in the duty free shops, food courts or restrooms is assumed to be accounted for in the walking time distributions. Although all airports vary in the size of these zones and their configurations, the overall layout should be common among most airports and easily adjustable to study passenger flows at other airports than Sydney.\\

Following Kendall's notation\cite{Ross2010Introduction}, all queues are modelled as M(t)/M(t)/c(t) First-Come-First-Serve (FCFS) queues. The interval process is Markovian(Poisson) and the service distribution time is exponential. 
The time between succesive arrivals are independent.  
The arrival rates follow a Poisson distribution, where the service rates vary depending on the locations, the arriving flights and time of day.  The service rates follow an empirical distribution for the different sections. Arrival and departures times are assumed to be identically independently distributed (i.i.d.).
No bound was assigned to the length of the queue.
The number of servers varies with the staffing level used as a control variable.\\

All the servers are assumed to be independent. The model described above can be modified to take into account to the existence of  different immigration lines, for instance depending on the citizenship of the passenger. This can be achieved by dedicating some of the servers to a given type of passengers. To extend the model to the case where servers dedicated to national passengers can also serve foreigners if they are empty, some of the queues would become priority queues and no longer FCFS.

\begin{figure}[htp!]
{\footnotesize
\centering
\def\svgwidth{0.7\columnwidth}
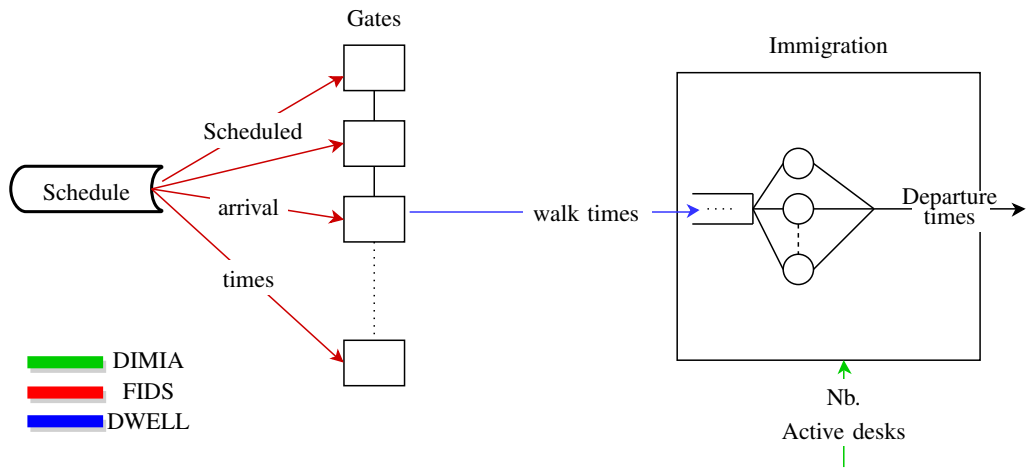
\caption{Arrival process data flow}
\label{fig:arriv_proc}
}
\end{figure}

For the arrival process, the simulation starts with flights arrivals at gates. Drawing from the DWELL information, the arrival times at immigration are computed.
Using service rates computed from the DIMIA information, departure times are finally computed. 
The process is illustrated on Figure \ref{fig:arriv_proc}.
\vspace{-2pt}
\vspace{-5pt}

The inter-arrival times between a given gate and the closest immigration zones are obtained by observing the walks of all passengers passing through these gates and arriving to the immigration zone.
The DWELL data is queriesd across all days to create two tables: one table for all passengers passing through a given gate  and another table for all passengers passing through immigration. The two tables are joined based on their device ID. The time difference between the last record at the gates and the first one at immigration gives us the walk time for one passenger.\\
 Few of the passengers could be traced from DWELL data alone. As can be observed in Table \ref{tab:passGateClassification}, the dataset is very sparse in some zones. This table compares the total number of passengers observed at immigration against the number of passengers who were found at immigration and any arrival gate.
\begin{table}[htp]
\caption{Number of passengers traced  by day}
\begin{tabular}{|l||c|c|c|c|c|c|}
\hline
& 4th Jul. '12 & 19th Jul. '12 & 15 Aug. '12 & 7 Dec. '12 & 8 Dec. '12 & 10 Dec. 12'\\
Total number of passengers 	& 81,857 	& 28,830 	& 88,195 	& 87671 & 88,195 	& 27,701\\
Number of passengers traced 	& 0 		& 0 		& 44 		& 773 	& 546 		& 747\\
\hline
\end{tabular}
\label{tab:passGateClassification}
\end{table}
\\
The service rate distribution is obtained from a subset of days with large delays at immigration. We use DIMIA information to compute the number of passengers at the immigration service nodes at different hours of the days for all days in DWELL. For each day of the week and time of the day, the days with the worst delays at the immigration service node are selected. 
The days in this set were used to compute the service rate per desk as a function of time of the day.
For each hour, the maximum service rate from that set was kept.
 The assumption was that during peak demands the servers operated at their highest throughput. The number of servers were  obtained by looking at specific days immigration data to recreate actual operations, then modified to mitigate delays\\

The simulation starts with the schedule of flights for a given day taken from FIDS. The block time and the walk time distribution are used to determine at what time the passengers reach immigration. The number of servers depends on the arrival time at immigration.
Several cases can arise. The case when there is at least one unoccupied server. In that situation then the next passenger is processed immediately. It can also happen that all servers are busy. The passenger is forced to wait for service in the queue.
The outputs of the simulation are the length of the queue at any time, the departure time from immigration, the time to be served, and the time spent in the queue.
\clearpage
\section{Analysis}
The propagation of delays inside the airport is examined in order to identify the different observable factors affecting it.
The analysis was performed by:
\begin{itemize}
\item Analyzing the impact of flight delays on passengers, based on flight delays  and passenger wait times at immigration.
\item Quantifying the effects of queue length on overall capacity, and the saturation of the queue beyond a certain occupancy.
\end{itemize}

\subsection{Delays Propagation}
To study the propagation of delays, we need means of measuring the effect of flight delays on passengers. This requires knowledge of information from flights, passengers and immigration. For this reason, the analysis was restricted to the 51 days with recorded data in all three databases.\\
The historic records of flights for 2012 are used to extract the daily arrivals of flights, which acts as a demand on our system. The demand distribution with respect to time is bimodal.
There is a large demand in the morning between 6am and noon, and a smaller peak in the afternoon between 3pm and 6pm. 
The average flight delays are also fully observable from that database.
An average delay of 26 minutes flight delay for an average of 812 flights per day is observed across  all days. Because the delays are only the difference between scheduled time of arrival at the gates, and actual arrival times at the gates, they encompass en-route and taxi delays.\\
The operations at immigrations are directly obtained from the immigration information. Low staffing levels were observed around noon across all days.
The low staffing period exarceberates the delays on the occasions where a delayed flight arrives early in the afternoon. It takes more time for the system to recover from such disruption. \\
After analyzing several days, we focus on three days in the dataset to illustrate the different trends in delays propagation: Sunday August 12th 2012, Saturday November 10th 2012 and Wednesday July 25th 2012.
For each of these days, we show the actual and the scheduled arrival flight times and the flight delays per hour of the day for the flights. We study the impacts of these delays on passengers by examining the throughput of the immigration services per hour of the day, together with the staffing levels.
\newpage
\subsubsection{August 12th}
\begin{figure}[htp!]
	\begin{subfigure}[b]{0.5\columnwidth}
		\centering
		\def\svgwidth{0.9\textwidth}	
		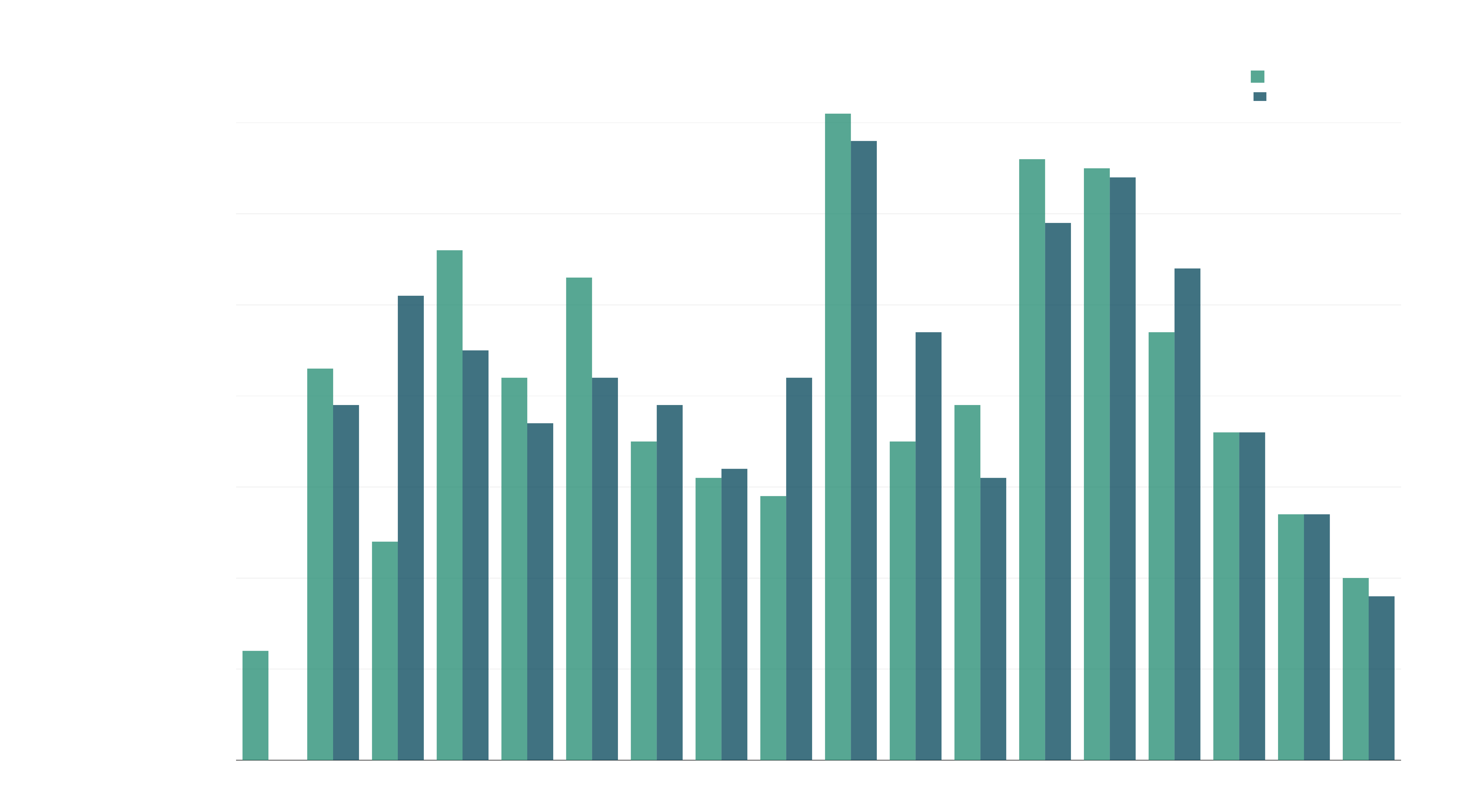
		\caption{Flight Arrival Times on August 12th (731 arrivals)}
		\label{fig:ActualVsSched12August}
	\end{subfigure}
	\begin{subfigure}[b]{0.5\columnwidth}
		\centering
		\def\svgwidth{0.9\textwidth}
		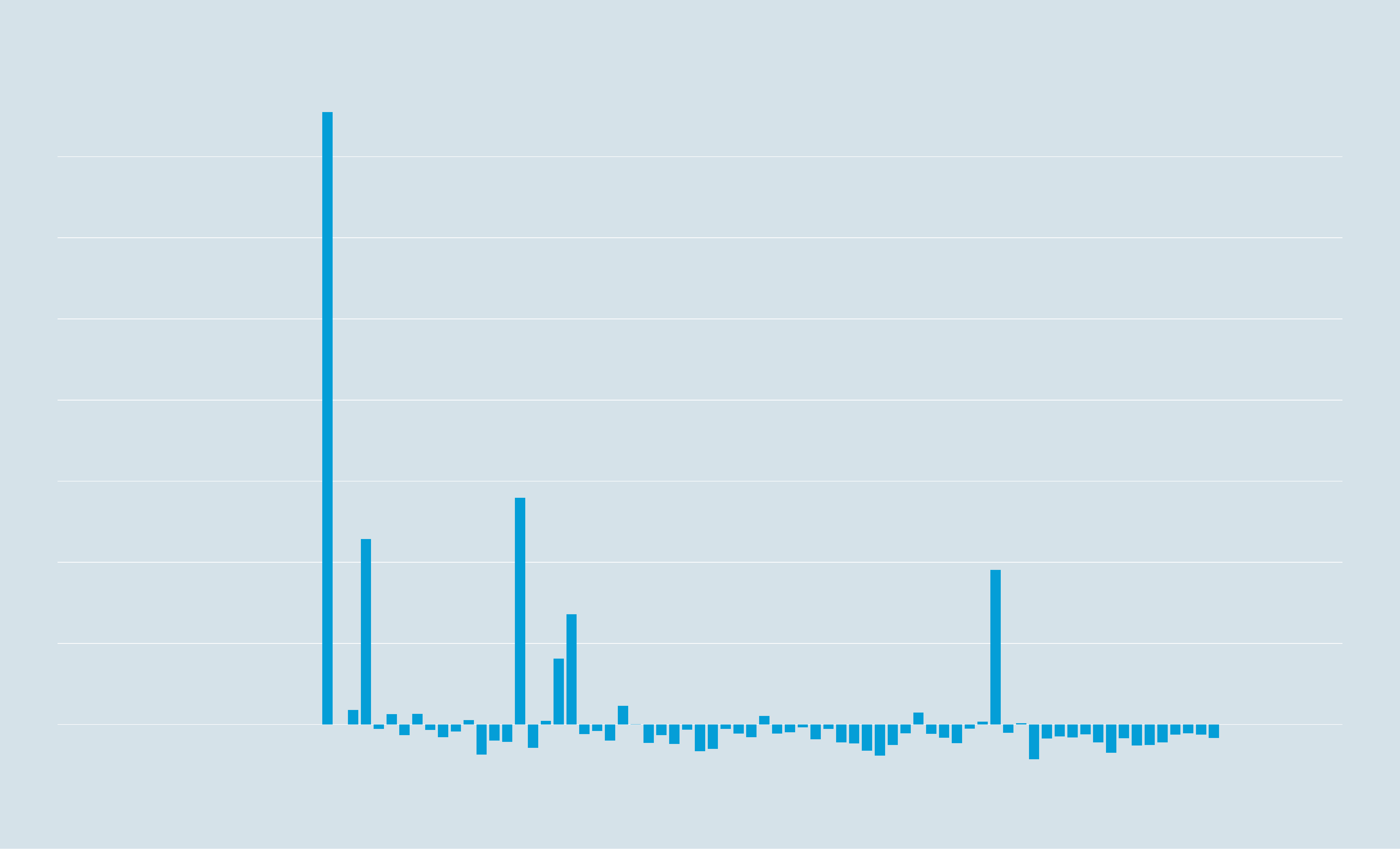
		\caption{Flight delays in minutes on August 12th}
		\label{fig:flightDelayAugust12th}
	\end{subfigure}\\
	\begin{subfigure}[b]{0.5\columnwidth}
		\centering
		\def\svgwidth{0.9\textwidth}	
			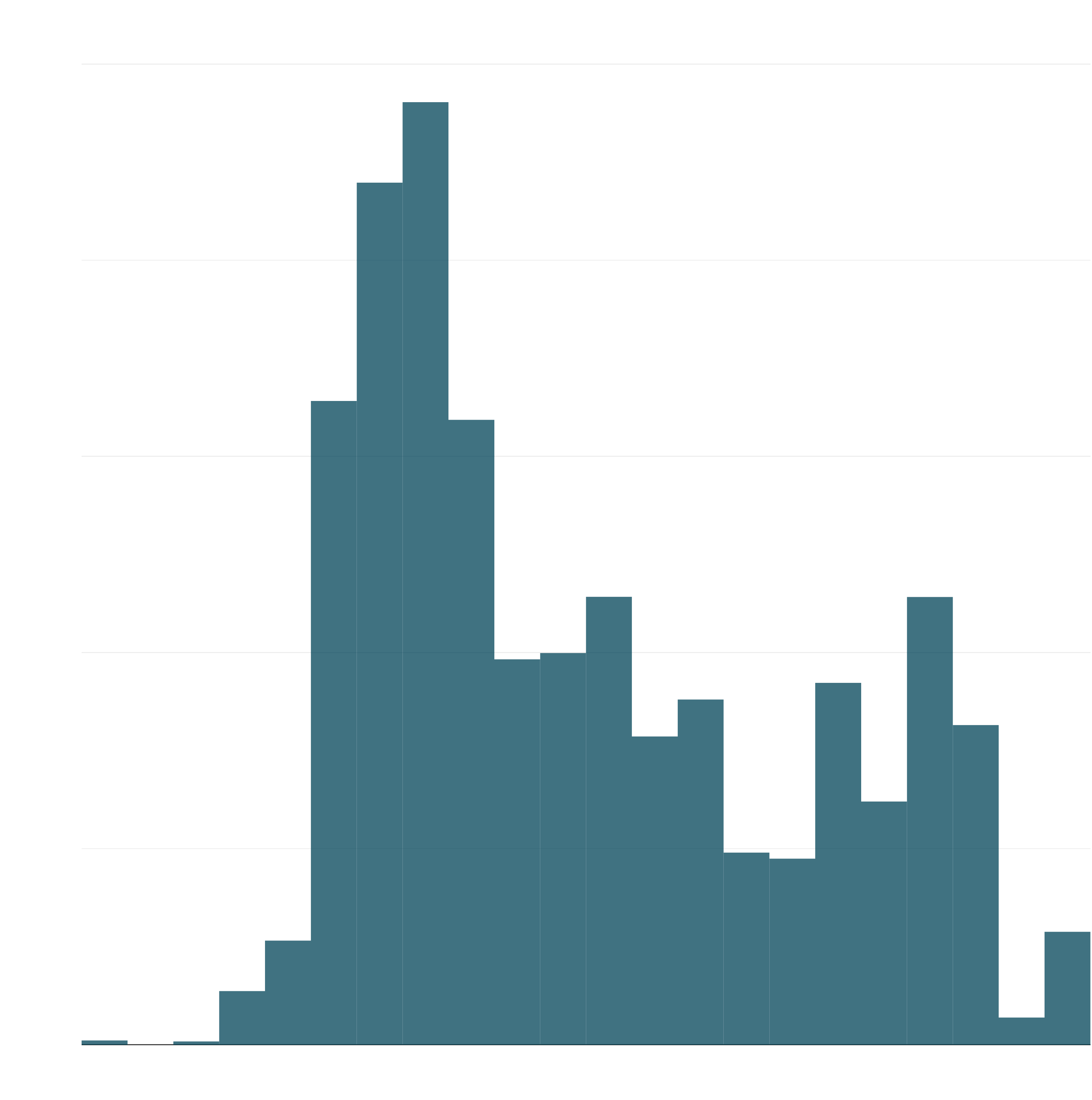
			\caption{Actual Departure Times from Immigration on August 12th}
			\label{fig:ActualDTimesImmi12August}
	\end{subfigure}
	\begin{subfigure}[b]{0.5\columnwidth}
		\centering
		\def\svgwidth{0.9\textwidth}	
			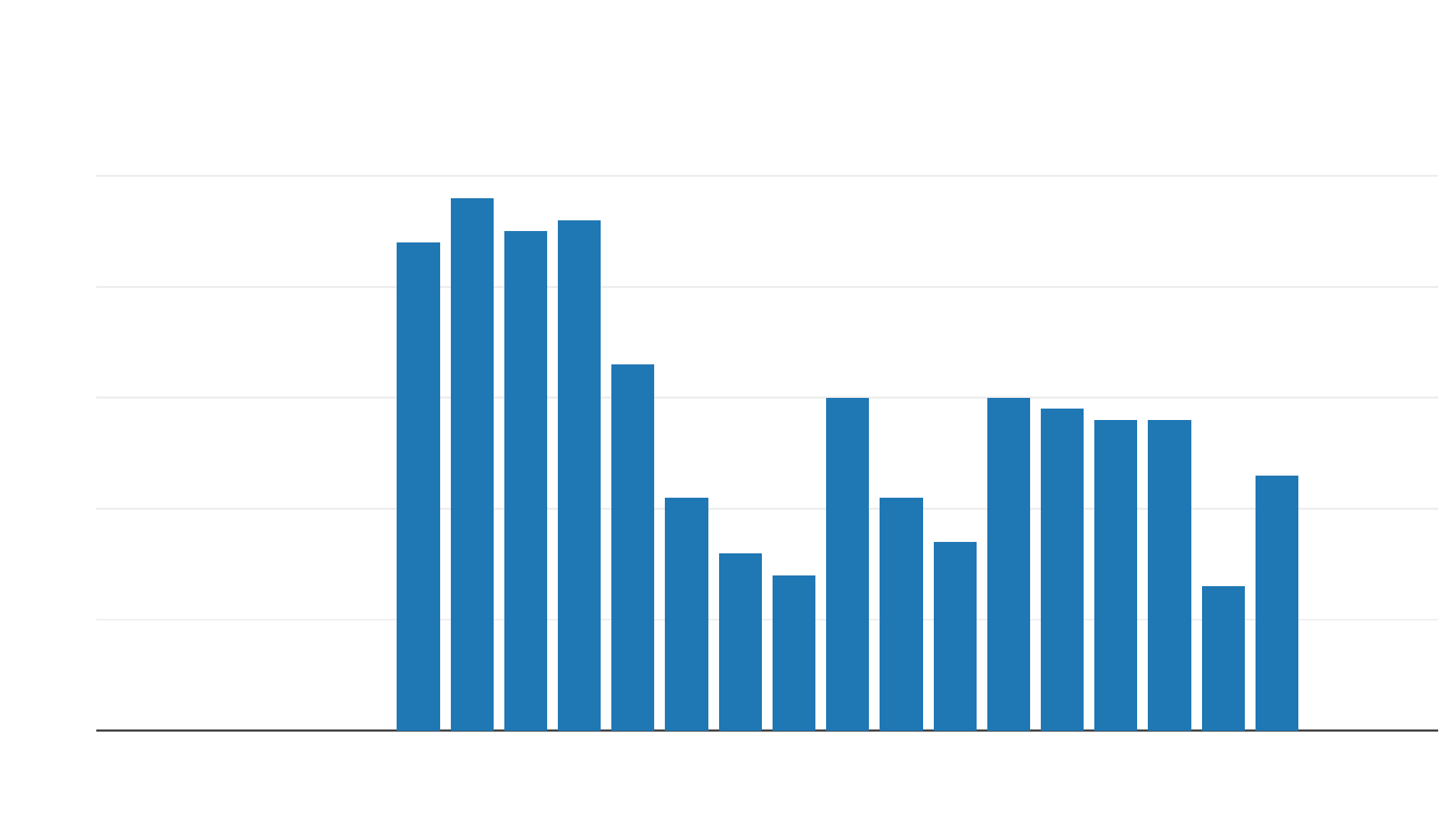
			\caption{Hourly Staffing levels on August 12th}
			\label{fig:StaffingLevel12August}
	\end{subfigure}
\label{fig:Analysis_Aug12}
\end{figure}

August 12th 2012 is a characteristic of most days in the dataset. The flights arrival times on Figure \ref{fig:ActualVsSched12August} show that most flights arrive within 15 minutes of their scheduled arrivals. 
The exception is at 5am see Figure \ref{fig:flightDelayAugust12th}, when three flights (AF8098, IB7705 and QF2) scheduled to arrive at 5:15PM are subject to a 9 hours delay.
As can be seen on Figure \ref{fig:StaffingLevel12August}, the staffing level has been set to accomodate the early stream of flights.
The second wave of departures from immigration occurs around 8pm, see Figure \ref{fig:ActualDTimesImmi12August}. That higher throughput indicates that more passengers are waiting to be served, and possibly that the immigration services are still processing passengers from flights that have arrived between 6 and 7 pm.\\
\clearpage
\subsection{November 10th}
\begin{figure}[htp!]
	\begin{subfigure}[b]{0.5\textwidth}
		\centering
		\def\svgwidth{0.9\textwidth}	
		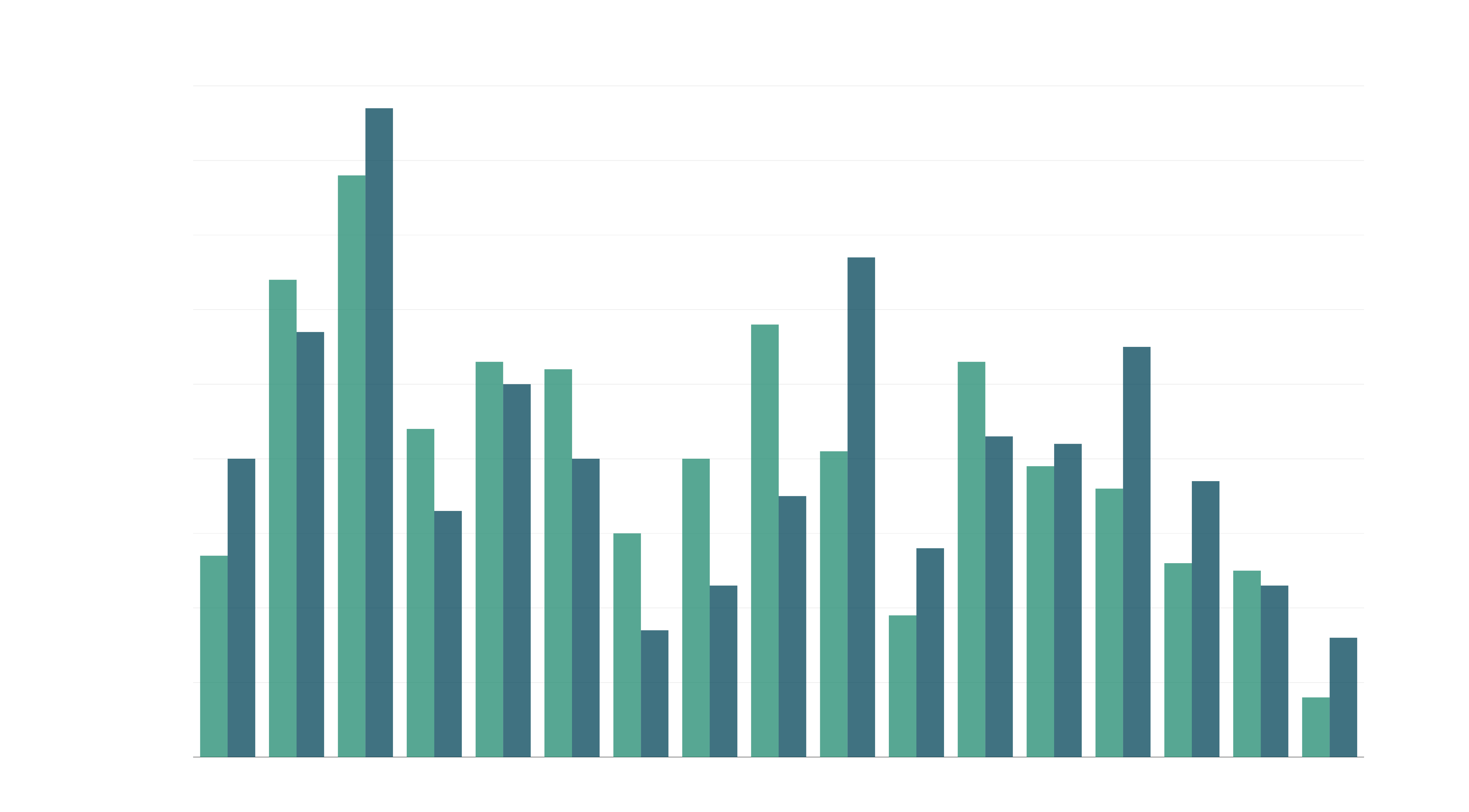
		\caption{Flight Arrival Times on Nov 10 (693 arrivals)}
		\label{fig:ActualVsSched10Nov}
	\end{subfigure}
	\begin{subfigure}[b]{0.5\textwidth}
		\centering
			\def\svgwidth{0.9\textwidth}
			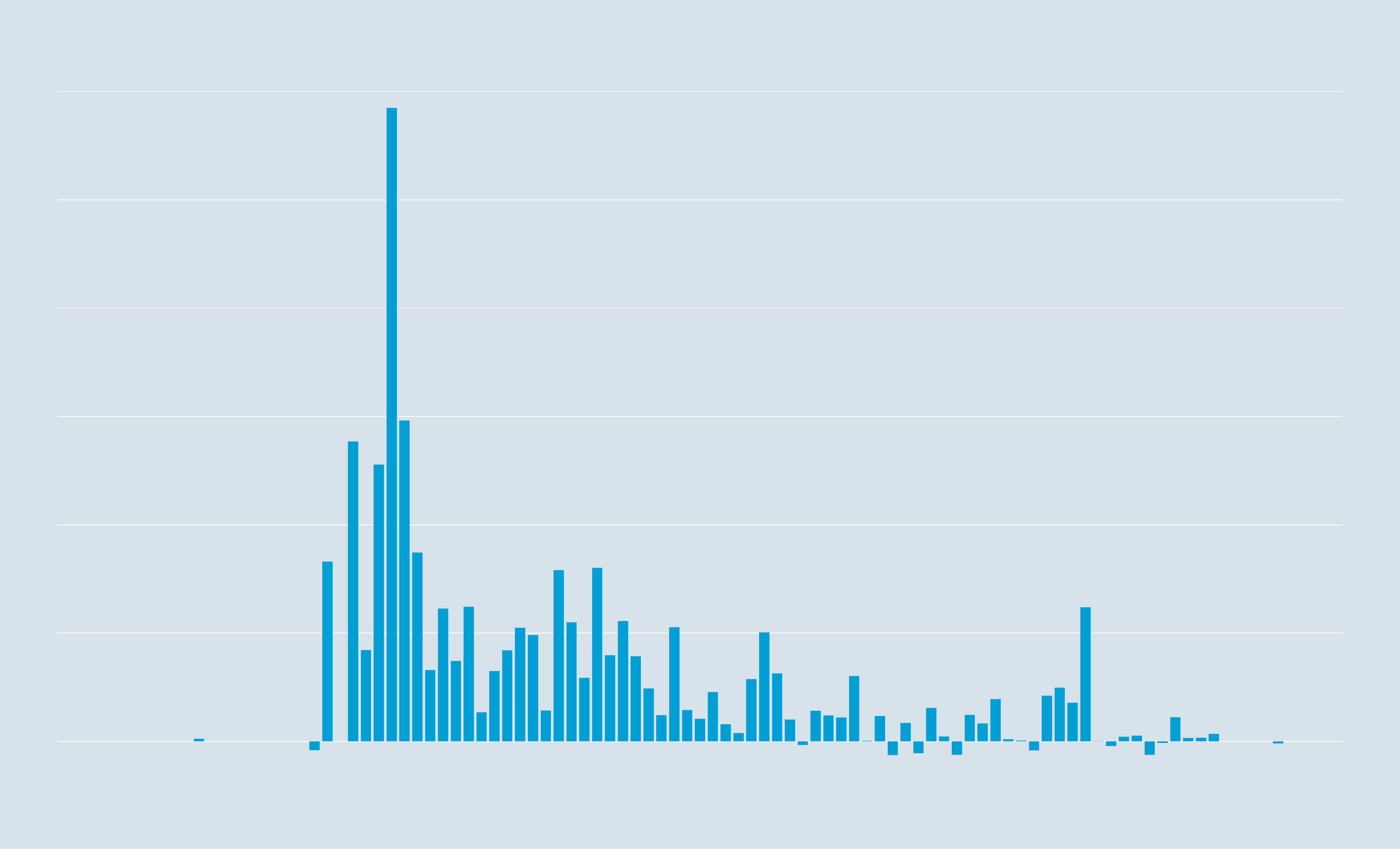
			\caption{Flight delays in minutes on November 10}
			\label{fig:flightDelay10Nov}
	\end{subfigure}\\
	\begin{subfigure}[b]{0.5\textwidth}
		\centering
		\def\svgwidth{0.9\textwidth}	
			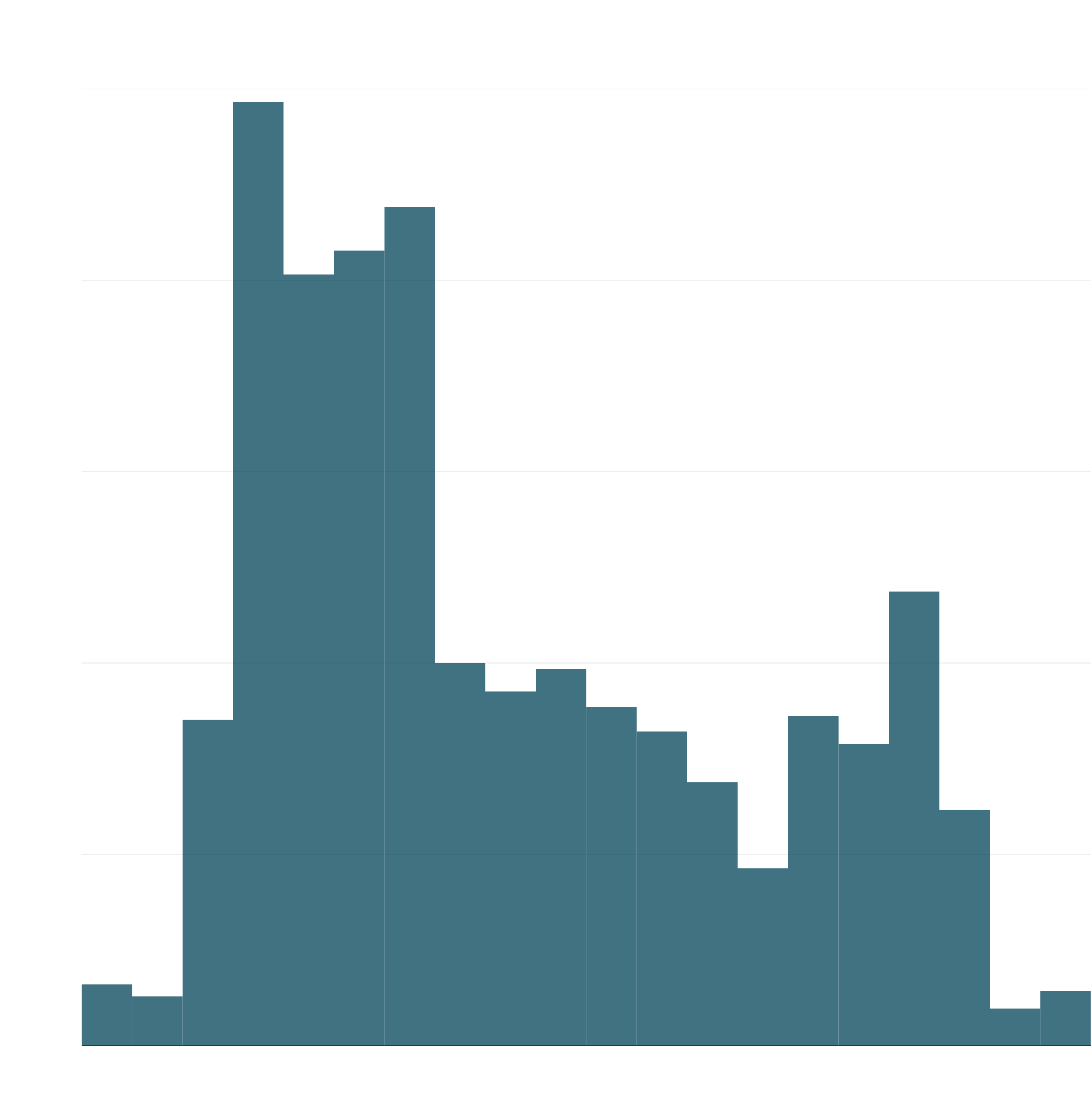
			\caption{Actual Departure Times from Immigration on November 10}
			\label{fig:ActualDTimesImmi10Nov}
	\end{subfigure}
	\begin{subfigure}[b]{0.5\textwidth}
		\centering
		\def\svgwidth{0.9\textwidth}	
			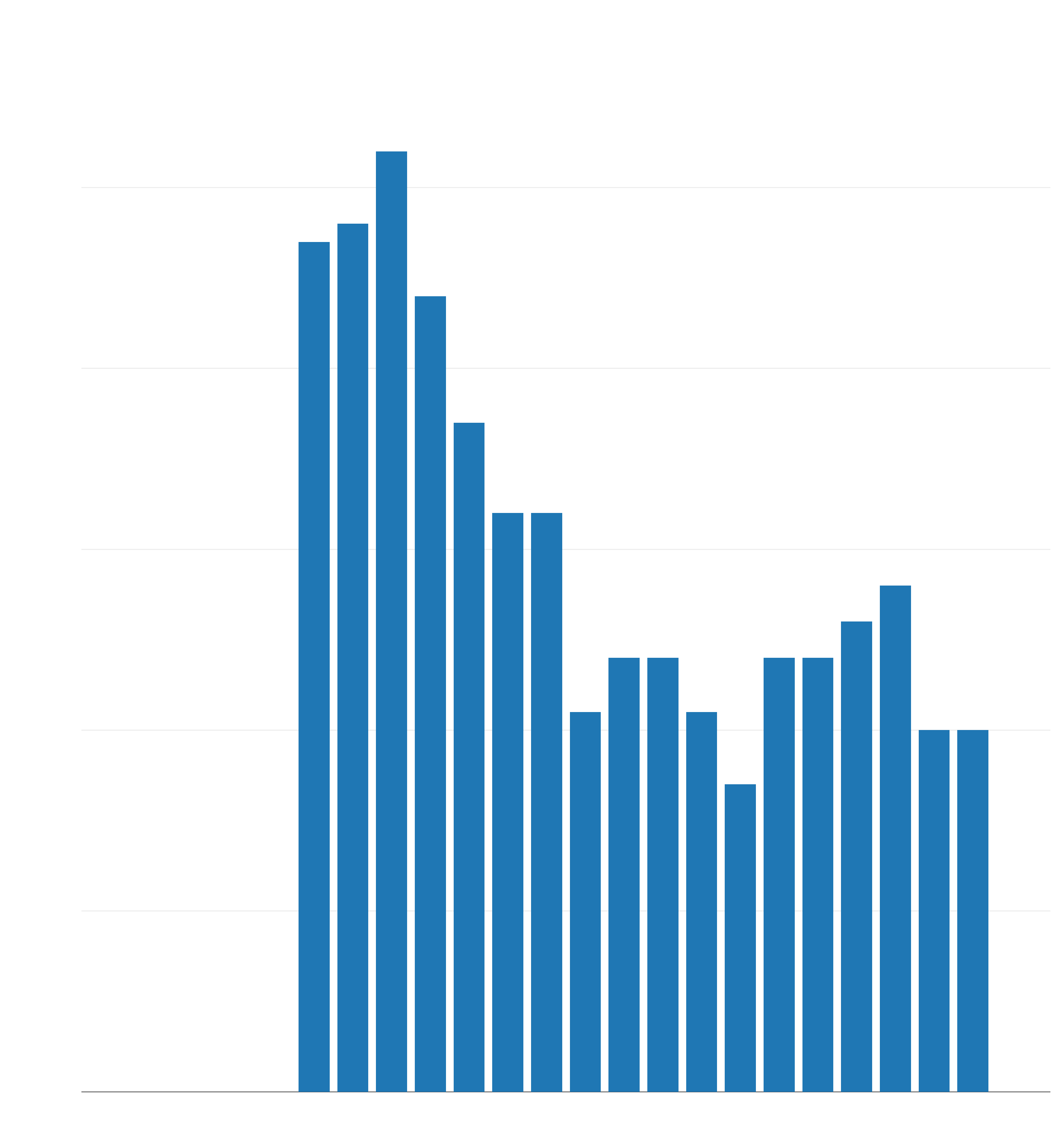
			\caption{Hourly Staffing levels on November 10}
			\label{fig:StaffingLevel10Nov}
	\end{subfigure}
\label{fig:Analysis_Nov10}
\end{figure}

The distribution of flight arrivals on November 10th on Figure \ref{fig:ActualVsSched10Nov} exhibits the bimodal trend mentioned above. With 693 arrivals compared to 731 on August 12th, the number of flights on November 10th is lower than the number of arrivals on August 12th. The flight delays are larger on August 12th. Since the throughput rates are lower, it indicates a high variability in passenger arrivals that is not accounted for by the actual staffing levels.\\ 
Figure \ref{fig:StaffingLevel10Nov} shows an augmentation in the number of open desks. Yet that reaction is not adequate to respond to the demand. The data clearly highlights a lack of predictability in demand affect passengers service, and results in over or understaffing atr different times of the day.
\clearpage
\subsection{ July 25th}
\begin{figure}[htp!]
	\begin{subfigure}[b]{0.5\textwidth}
		\centering
		\def\svgwidth{0.9\textwidth}	
		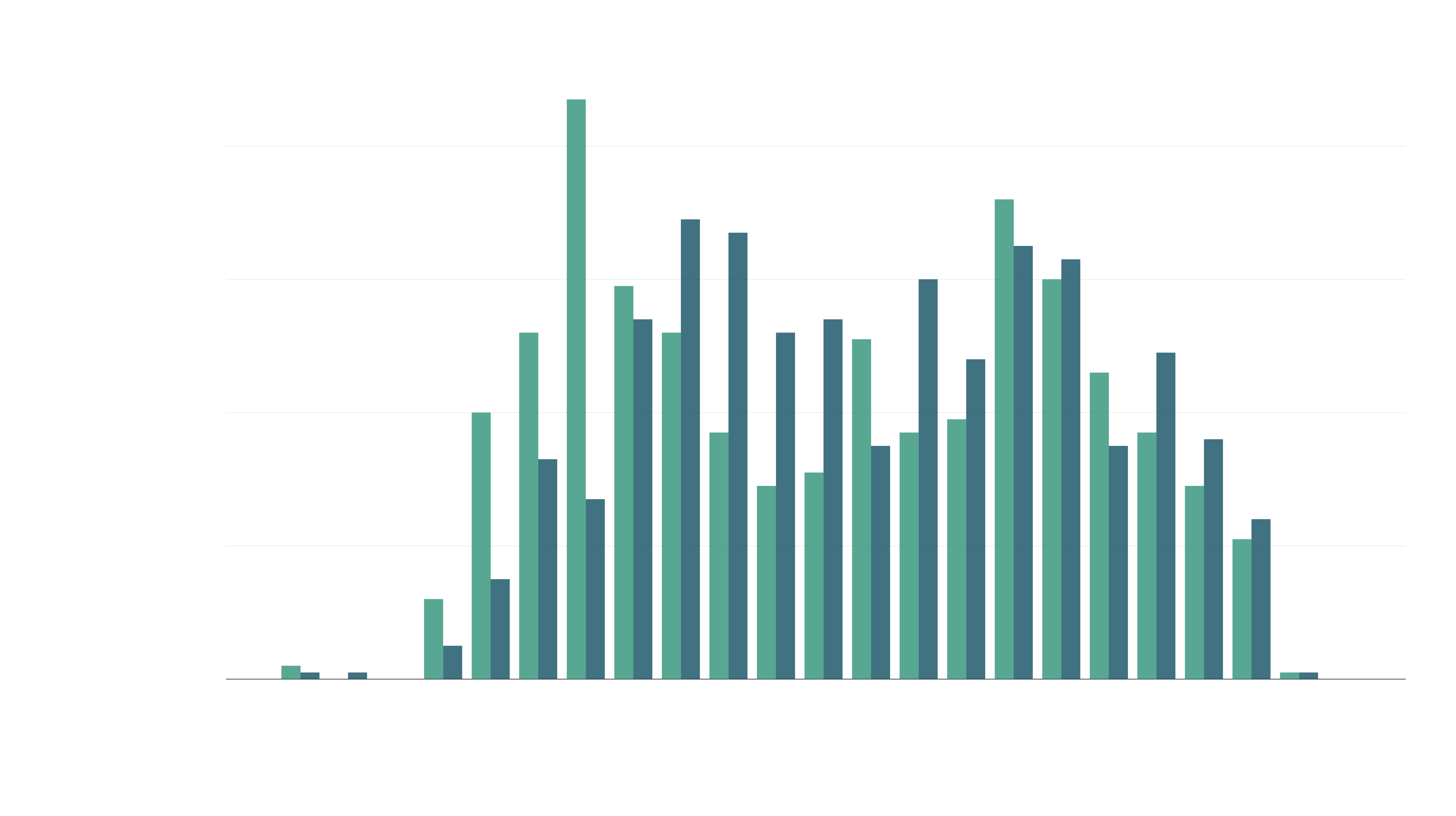
		\caption{Flight Arrival Times on July 25 (794 arrivals)}
		\label{fig:ActualVsSched25July}
	\end{subfigure}
	\begin{subfigure}[b]{0.5\textwidth}
		\footnotesize
		\centering
			\def\svgwidth{0.9\textwidth}
			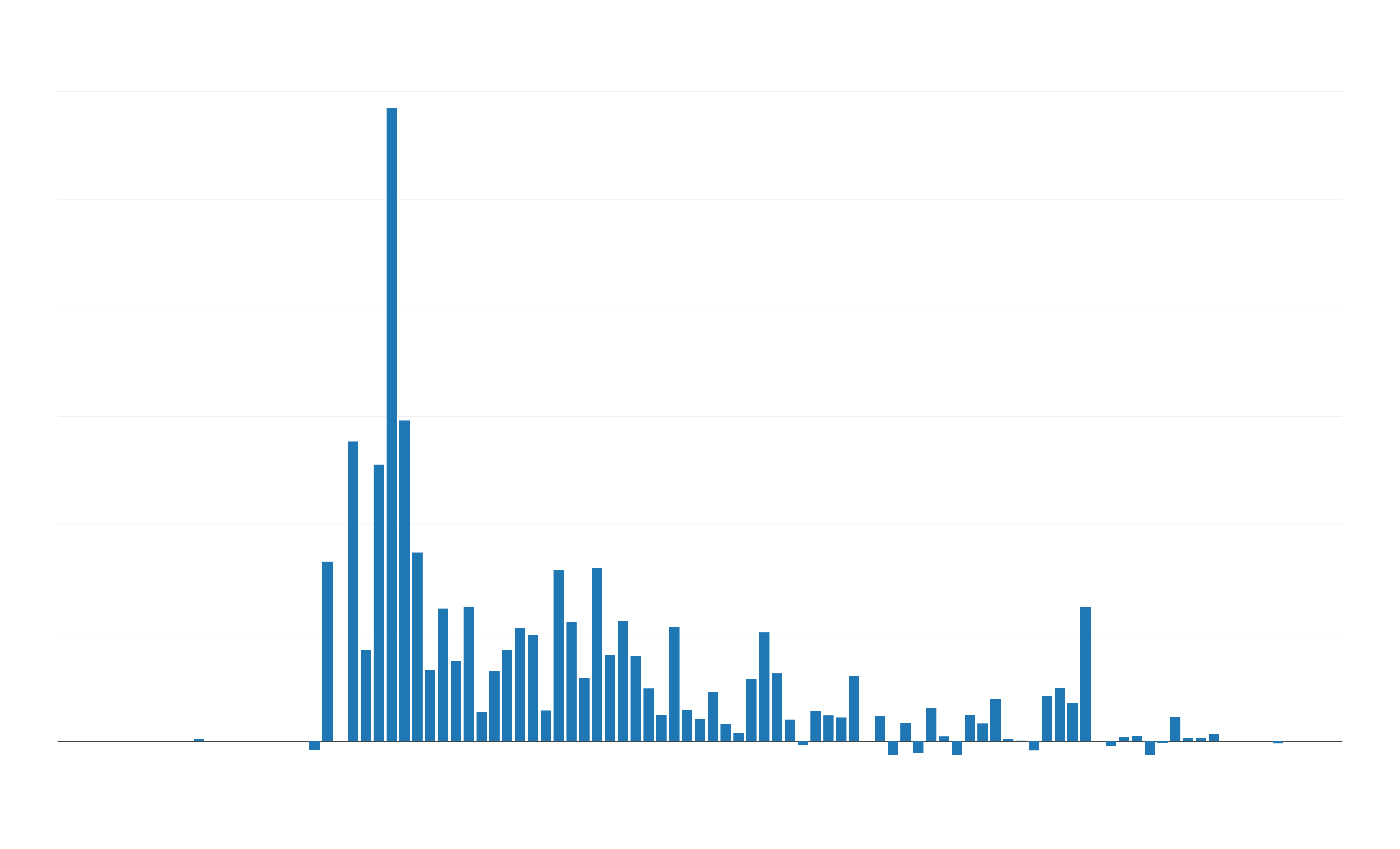
			\caption{Flight delays in minutes on July 25}
			\label{fig:flightDelaysJuly25th}
	\end{subfigure}
	\begin{subfigure}[b]{0.5\textwidth}
		\footnotesize
		\centering
		\def\svgwidth{0.9\textwidth}	
			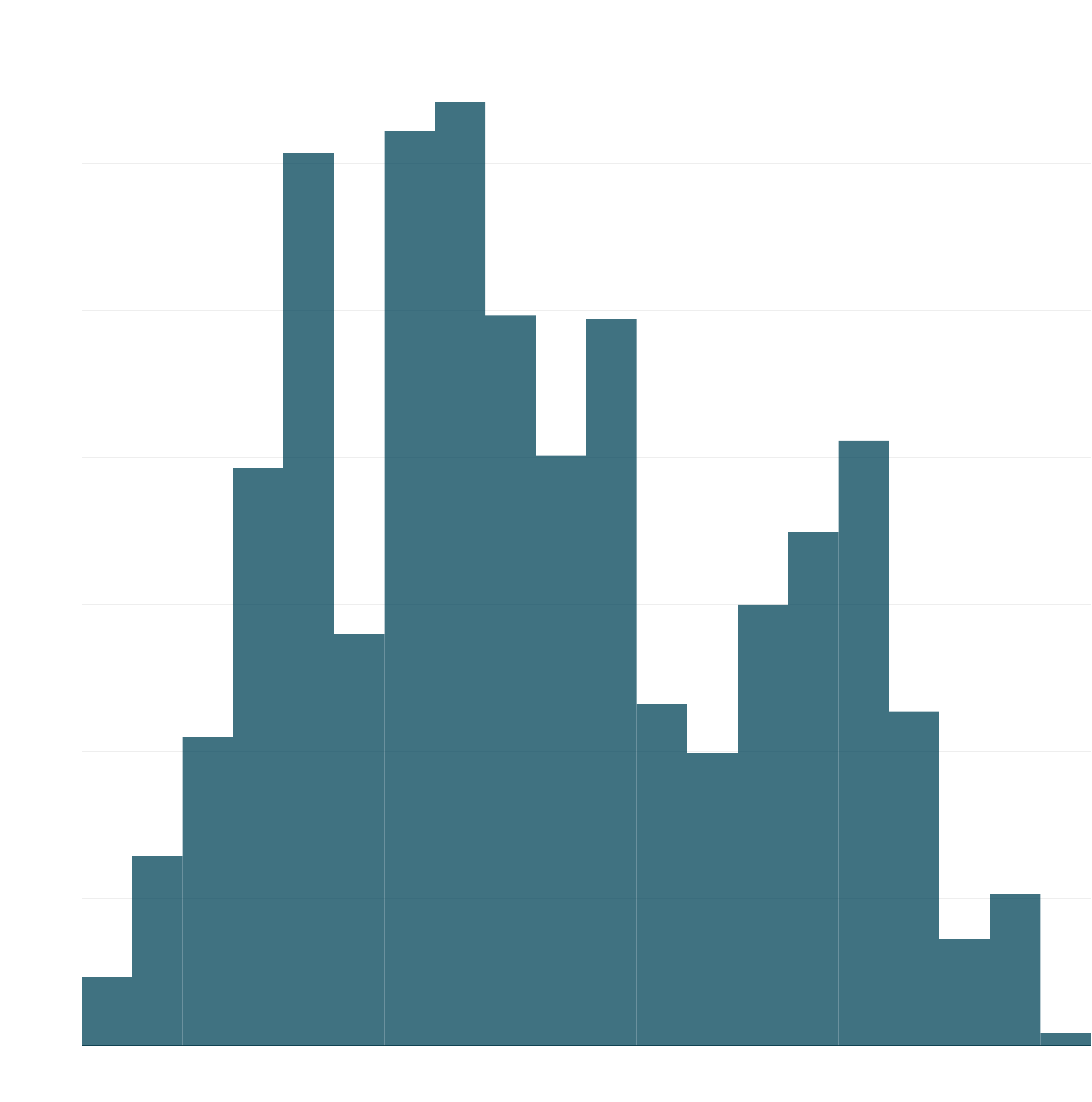
			\caption{Actual Departure Times from Immigration on July 25}
			\label{fig:ActualDTimesImmi25Jul}
	\end{subfigure}
	\begin{subfigure}[b]{0.5\textwidth}
		\footnotesize
		\centering
		\def\svgwidth{0.9\textwidth}	
			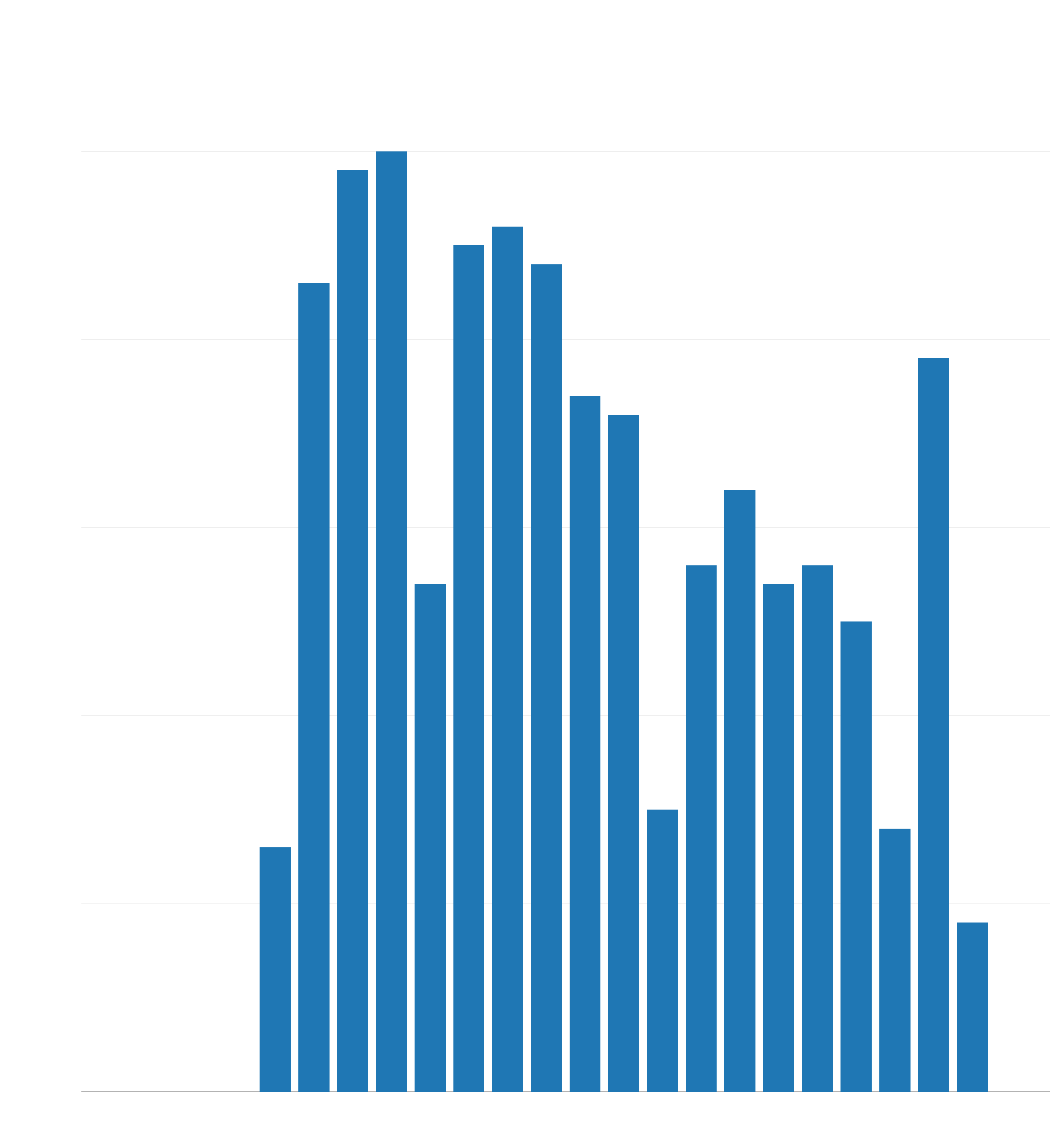
			\caption{Hourly Staffing levels on July 25}
			\label{fig:StaffingLevel25Jul}
	\end{subfigure}
\label{fig:Analysis_Jul25}
\end{figure}

July 25th was the day with the most flight delays in the dataset. The average delay per flight was over an hour.
Figure \ref{fig:flightDelaysJuly25th} shows the average flight delay for all airport arrivals for each 15 minutes time period. 
There are 9 periods with observed delays greater than 2 hours. \\

Most flight delays occured between 6 and 10 am, the high demand period of the airport, as illustrated on Figure \ref{fig:ActualVsSched25July}. 
It appears from figure \ref{fig:StaffingLevel25Jul}, that the number of open desks were increased in anticipation of the demand. However the throughput is lower than on November. This could be due to a saturation of the immigration services, faced with a larger demand. The phenomenon of saturation is presented in the next section.\\
 
Passengers service can suffer from large flight delays. With many delays, there is more uncertainty in the actual arrival times of the flights, and therefore a risk of overstaffing and understaffing key positions such as the immigration service desks. 
As will be shown in the next section, operating at  maximum demand also has negative effects. It forces the service system to operate near saturation. Therefore the service system is be unable to satisfy demand, and throughput is reduced. 
%

\clearpage
\subsection{Queue saturation}
\begin{figure}[htp!]
	\def\svgwidth{0.65\textwidth}
	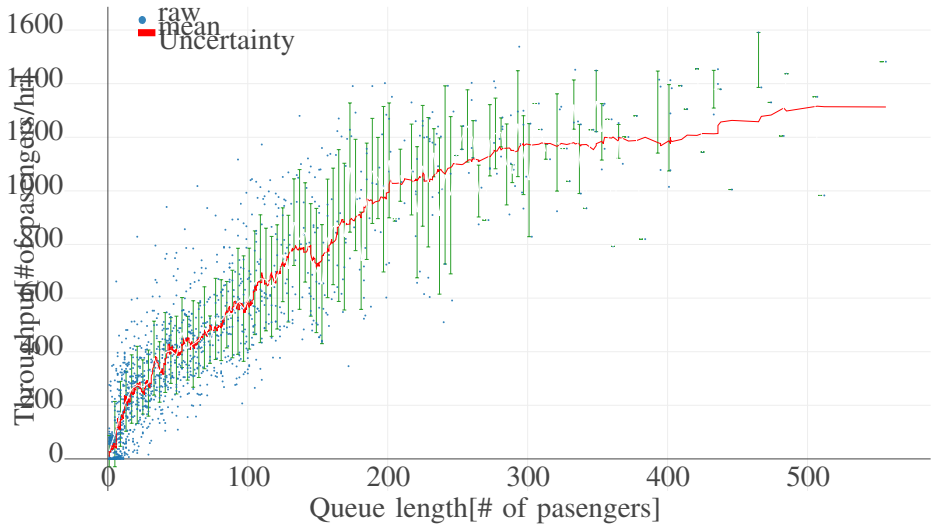
	\caption{Throughput vs. demand}
	\label{fig:throughputvsDemand}
\end{figure}

The passenger queue saturates after reaching a certain occupancy. The queue is said to saturate when the throughput stop increasing with an increase in demand. 
Passenger demand is the number of passengers in the immigration queue which is determined by the arrival rate.
This limit point can then be used in a threshold control policy, which goal would be to prevent the queue length from exceeding the saturation point.
It can be done strategically using predictions based on historic demand, and tactically by adapting to day to day operations.
The ability to predict passenger demand based on flight schedules is crucial for a tactical strategy.  \\
To obtain the throughput at immigration, the service rate of passengers at the immigration desks is used.
The demand is computed from the number of passengers arriving at immigration derived from DWELL.
After aggregating those two statistics for the whole year, we generate the throughput versus demand curve. 
Figure \ref{fig:throughputvsDemand} clearly illustrates the demand saturation occuring after the queue reaches 280 passengers.\\

\clearpage
\section{Calibration for the queueing model}
Passengers arrive at immigration with a rate $\lambda$ and depart with a rate $\mu$.
Arrival times and departure times are parametrized using the average walk speed from gates to immigration, and the observed throughput at immigration respectively.\\
\subsection{Modeling interarrival times from walk speed}
The average walk speed is defined  as the average speed of a passenger going from one gate to immigration.
The speed is computed based on all interarrival times between gates and immigration from August 2012 to October 2012. In the conversion from time to speed, the shortest distance from gate to the closer immgigration zone is used.
Because of the lack of granularity in the data, it is to be noted that the constructed distribution encompasses the walk to immigration, any sightseeing or wandering in between zones, and presumably some time spent deboarding the airplanes.\\ 
\subsubsection{Calibration process}
There is some uncertainty associated with the location of a device.
A device is often assigned to multiple zones, making it difficult to compute the exact time spent between the gates and immigration for any single device. As explained above, out of more than 80,000 devices observerd in a day less than 500 devices can be used to extract a path to immigration.
As a consequence of the lack of datapoints, the interarrival times for some gates cannot be computed. 
To obtain accurate arrival rates for all the gates, we choose to use walk speed instead of walk times to model for the airport. Arrival times are computed solely based on the distance to the immigration services. \\
\subsubsection{Walk Speeds Distributions}
\begin{figure}[bH!]
\centering
\def\svgwidth{0.65\textwidth}
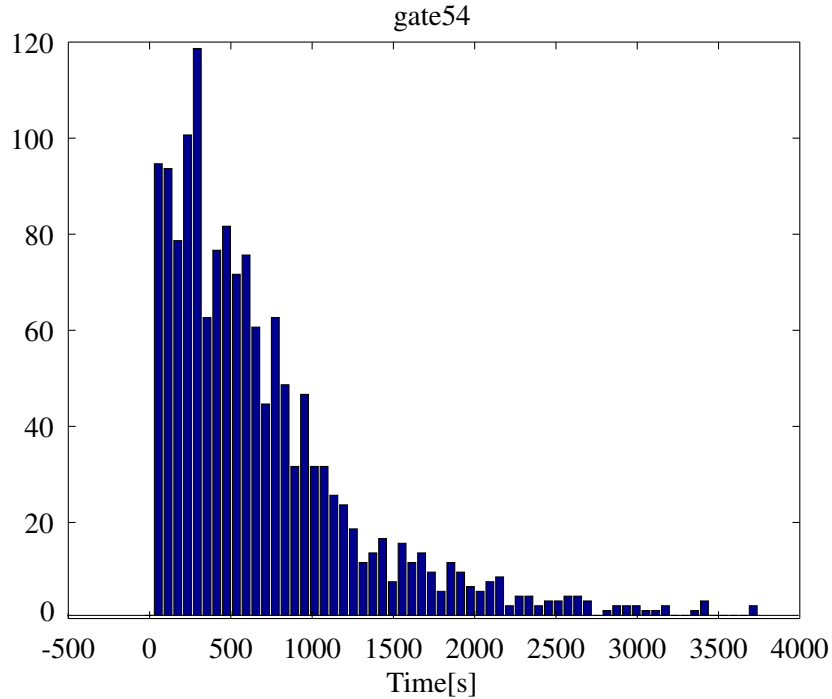
\caption{Walk Times for gate 54}
\label{fig:walkTimeGate54}
\end{figure}

\begin{figure}[bH!]
\centering
\def\svgwidth{0.65\textwidth}
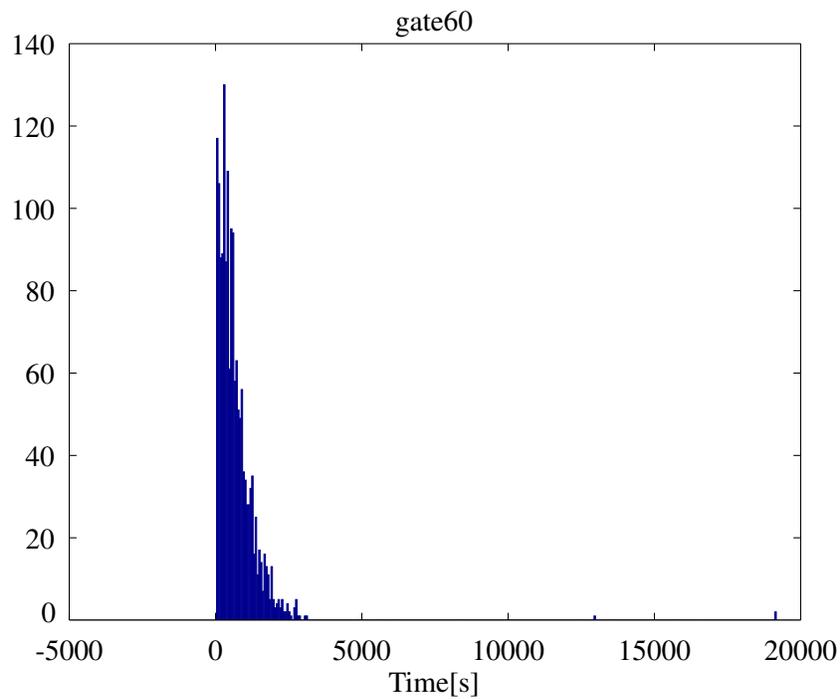
\caption{Walk Times for gate 60}
\label{fig:walkTimeGate60}
\end{figure}

When fitting for different distributions, we find that walk times follow an exponential distribution as can be seen on Figures  \ref{fig:walkTimeGate54} and \ref{fig:walkTimeGate60}.

\begin{figure}[bH!]
\centering
\def\svgwidth{0.65\textwidth}
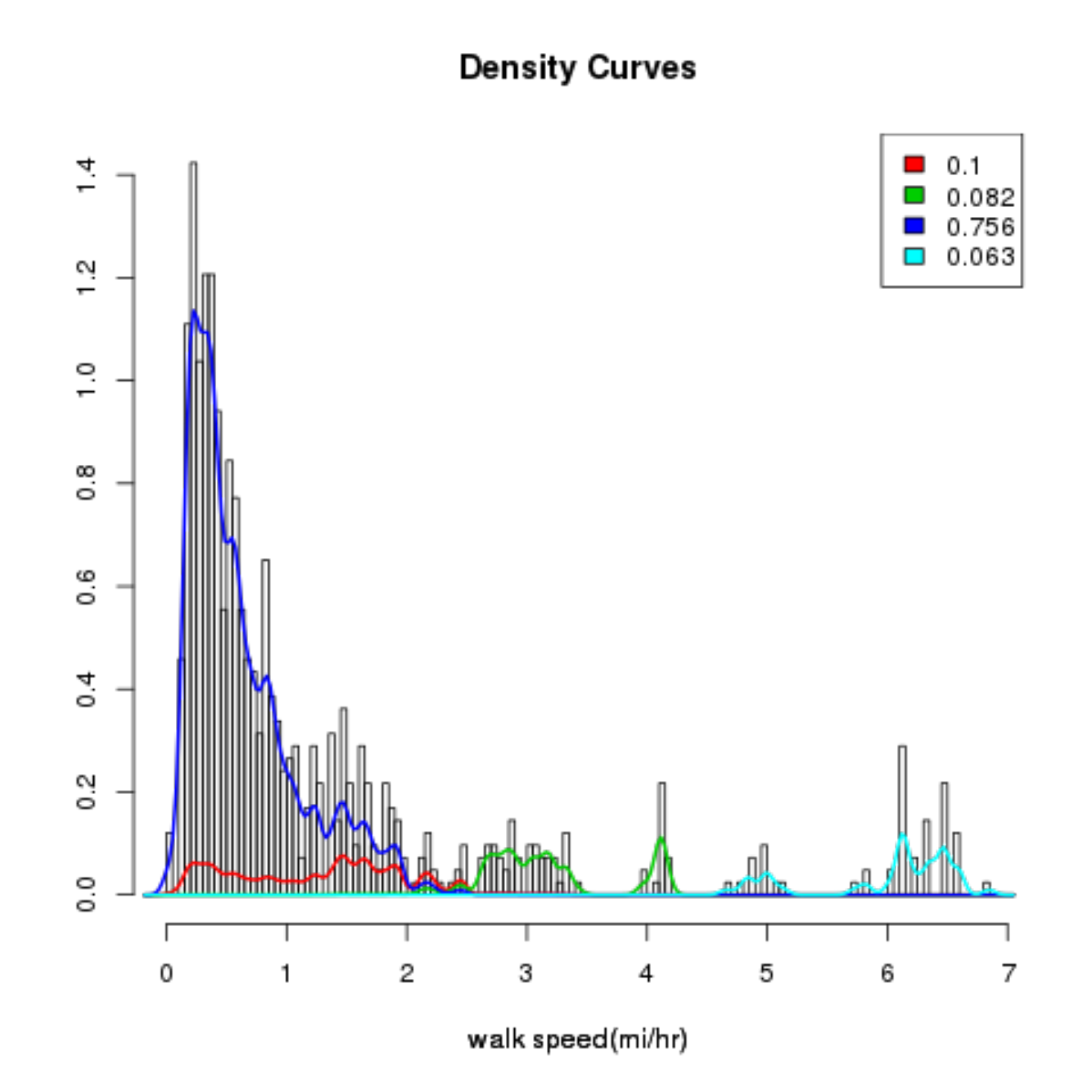
\caption{Walk Speed for gate 53}
\label{fig:walkSpeed53_npEM}
\end{figure}

 The walk speed distributions for individual gates contain multiple modes. 
For gate 53 on figure \ref{fig:walkSpeed53_npEM}, as many as 6 different modes can be observed. \\
We use a two-step process to capture the different modes and model the distribution as a mixture of logistic distributions:
\begin{enumerate}
\item Datapoints are clustered into differrent components using a nomparametric Expectation Maximization algorithm
\cite{Bishop:2006:PRM:1162264}.
 Using the posterior probabilities for each datapoint, we assign a point to a cluster if the posterior probability is higher than 0.05. 
This ensures that we account for the contribution of all components to a given walk speed. It implies that some of the clusters are overlapping, in agreement with what is observed in Figure \ref{fig:walkSpeed53_npEM}. 
\item A distribution is fit to each cluster, and the fit is evaluated using Aikake's Information Criterion(AIC).
	The logistic distribution was picked due to its finite support, and high Goodness of fit value as seen on Table \ref{tab:gate53_comp2FitResults}
\end{enumerate}
\begin{table}[H]
\caption{Results of the goodness of Fit test for gate 53}
\label{tab:gate53_comp2FitResults}
\centering
\begin{tabular}{l|c|c|c}
\hline
Component 	& Logistic & Lognormal & Gamma \\
1		& 1133.521 	& 807.2598	& 810.1791\\
2	   	& 161.0323	& 156.8922	& 158.1537\\
3		& 1133.521 	& 807.2598	& 810.1791\\
4		& 98.04888	& 103.859	& 101.8572\\
\hline
\end{tabular}
\end{table}

\begin{table}[H]
\caption{Clusters information for gate53}
\centering
\begin{tabular}{l|c|c|c|c}
\hline 
Mean & 1.26	& 3.16	& 0.638	& 6\\
Mixture Coefficient & 0.0999	& 0.7558	& 0.0627\\
\hline
\end{tabular}
\label{tab:gate53_clustersInfo}
\end{table}
\subsubsection{Walk Speeds Fit}
Based on the shape of the distribution, multiple possible fits are possible for the different mode. One is a standard lognormal that only takes into account the first mode. 
We also fit the data to a Gaussian and a Lognormal mixture model, to compare the performance. The results of the different fits are summarized in Table \ref{tab:gate53_comp2FitResults}.
\begin{figure}[bH!]
\centering
\def\svgwidth{0.65\textwidth}
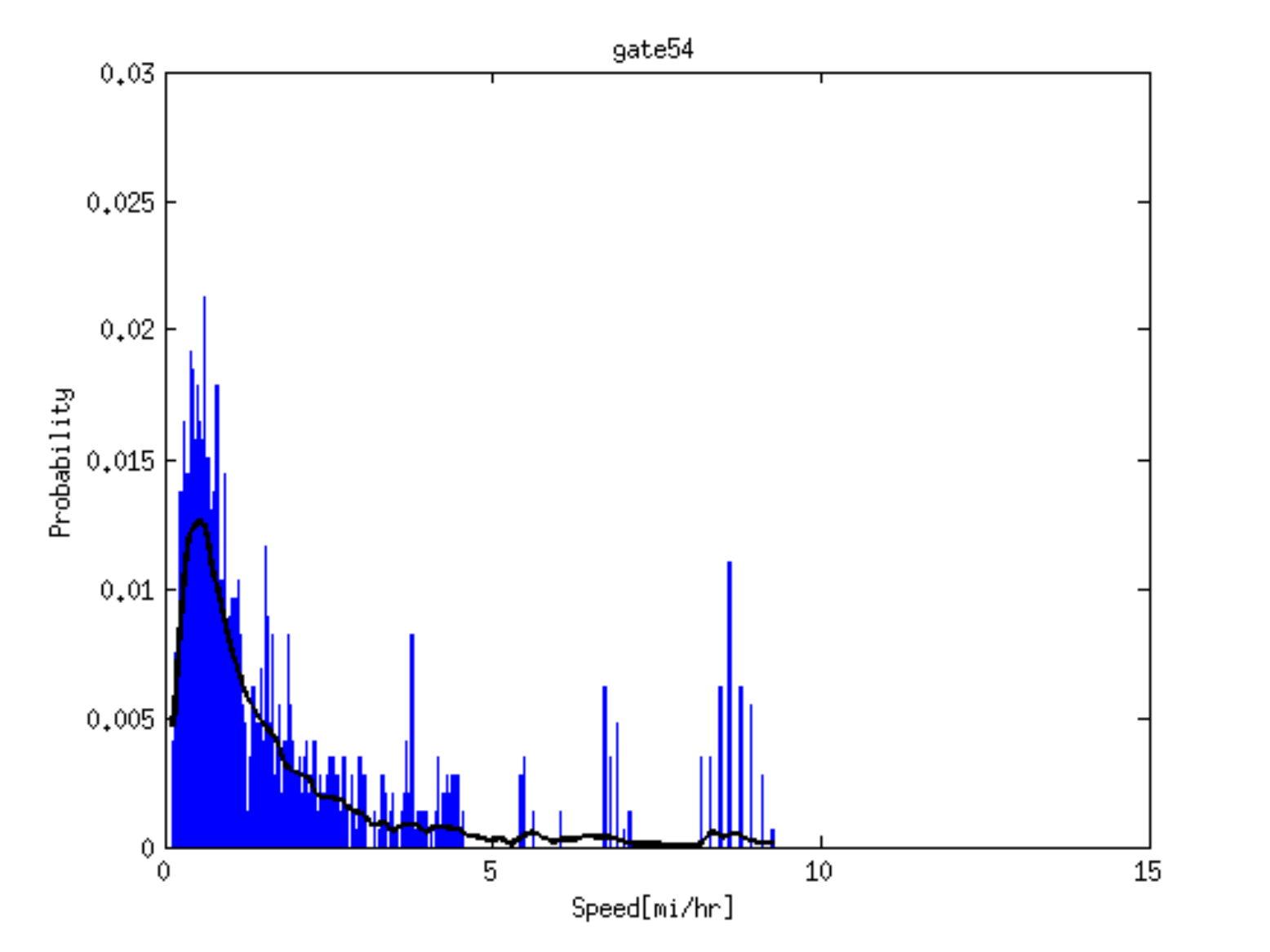
\caption{Walk Speed for gate 54}
\label{fig:walkSpeedGate54}
\end{figure}

\begin{figure}[bH!]
\centering
\def\svgwidth{0.65\textwidth}
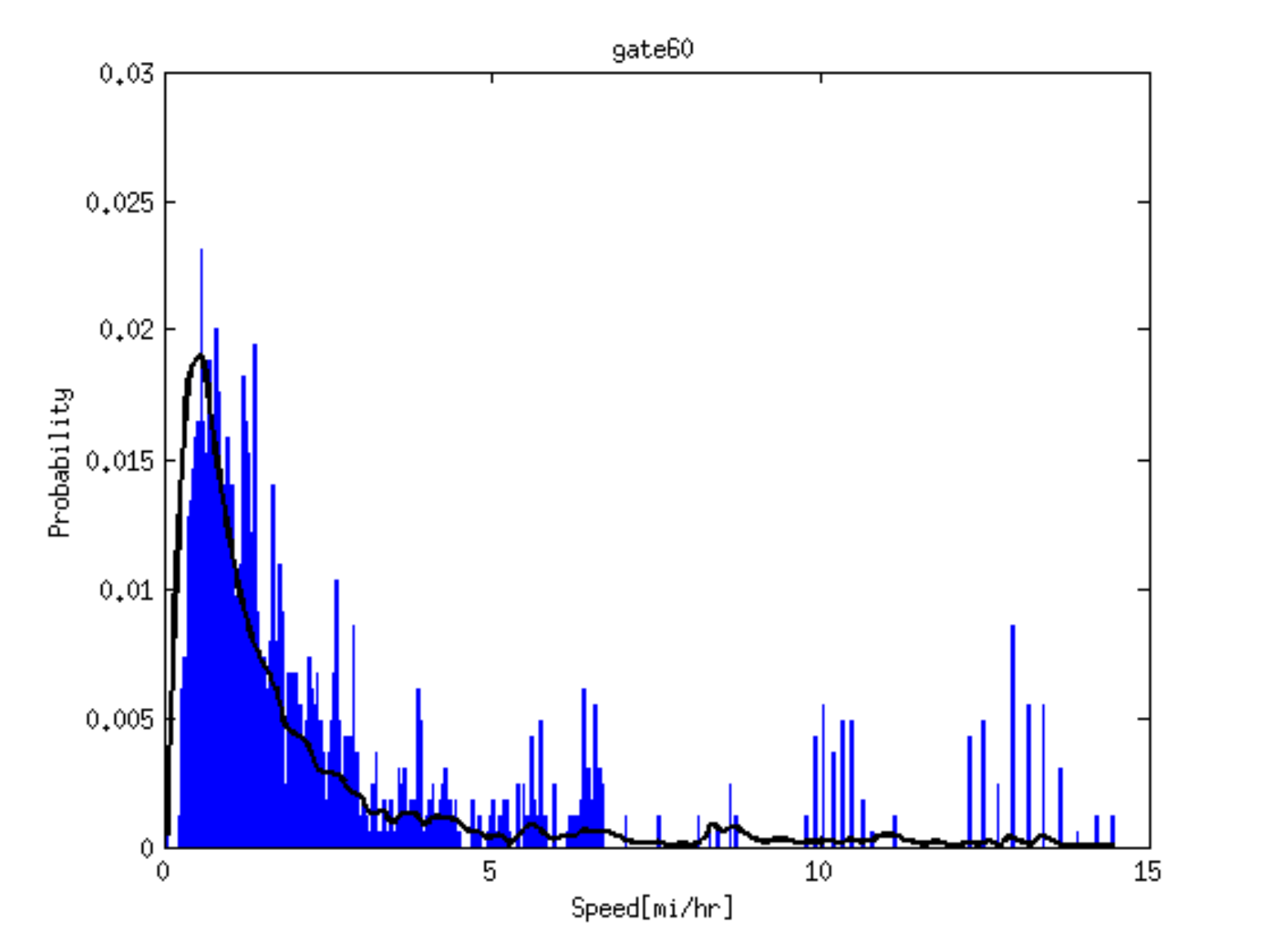
\caption{Walk Speed for gate 60}
\label{fig:walkSpeedGate60}
\end{figure}
When the walk speed data is aggregated, the secondary modes computed become less prevalent as shown on Figure \ref{fig:walkSpeed}. 
Since the first mode dominates all the others, we can ignore any mixtures, and treat passengers walk speeds as a single distribution. This may not be the most general solution, as the locations of the different distributions vary widely for different gates as observed in Table \ref{tab:walkSpeedsbyGate} \\

\begin{figure}[bH!]
\centering
\def\svgwidth{0.65\textwidth}
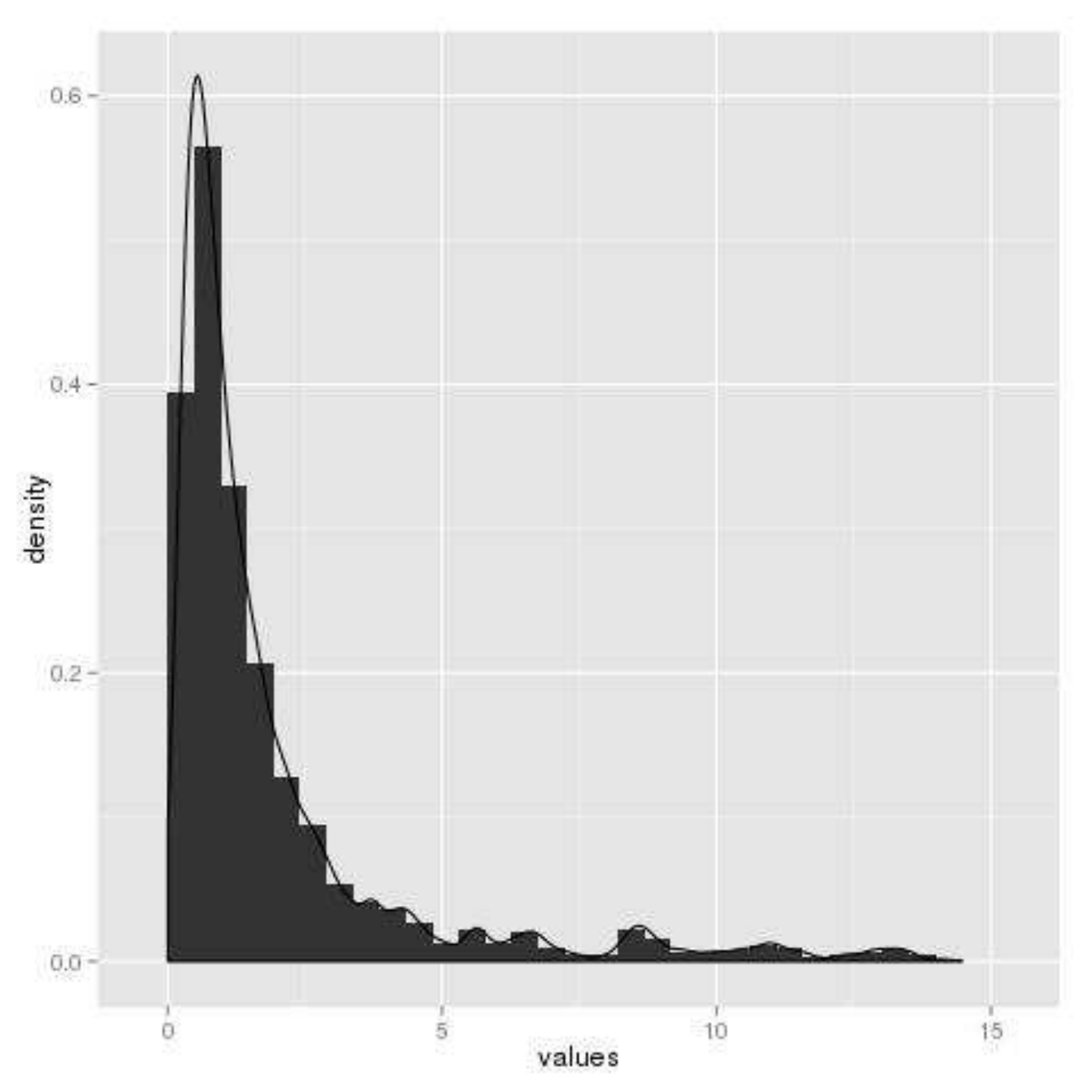
\caption{Walk Speed for all the gates}
\label{fig:fitAllGates}
\end{figure}

We instead opt for a mixture model to describe the walkspeed.
To build the model,  we use the two-step process described previously on the aggregated walk speed data for all gates..
The resulting distributions appear on figures \ref{fig:fitAllGates}. 

\begin{center}
\begin{table}[H]
\caption{Distribution of Walking Speeds per gate}
\centering
\begin{tabular}{|l|l|l|l|}
\hline
Gate & Mean & STD \\
\hline
All	& 	0.13765		&1.01290\\
Gate 53	&	-0.33383	& 1.03455\\
Gate 54 & 	0.00850		& 0.92939\\
Gate 55	&	0.03031		& 0.98285\\
Gate 56	& 	0.29650		& 0.96464\\
Gate 58	& 	0.10605		& 1.01170\\
Gate 59	& 	0.36459		& 0.96816\\
Gate 60	&	0.43726		& 0.92983\\
Gate 61 & 	0.29972		& 0.90701\\
\hline
\end{tabular}
\label{tab:walkSpeedsbyGate}
\end{table}
\end{center}

The error in the model is measured by looking at the mean and the standard deviation for the model, for the data between August and December excluding the month of September.\\
To calibrate our distribution, we compare the parameters of the walk speeds distributions in Table \ref{tab:walkSpeedsbyGate}. 

The service time at immigration is estimated per individual desk. 
It allows the use of the same service rate throughout the day independently of the number of active desks.
This enables the use of a control scheme with the staffing level as a control parameter.\\

We use the highest achieved service rates per desk as the service rate per desk. To compute it, 10 days between August and October 2012 are selected from the DWELL database, with the longest time spent at immigration. For those days, the throughput at immigration is calculated from DIMIA, and the service rate computed by dividing the throughput by the number of active desks per 15 minutes.

\subsection{Service Rate}
\begin{figure}[bH!]
\centering
\def\svgwidth{0.65\textwidth}
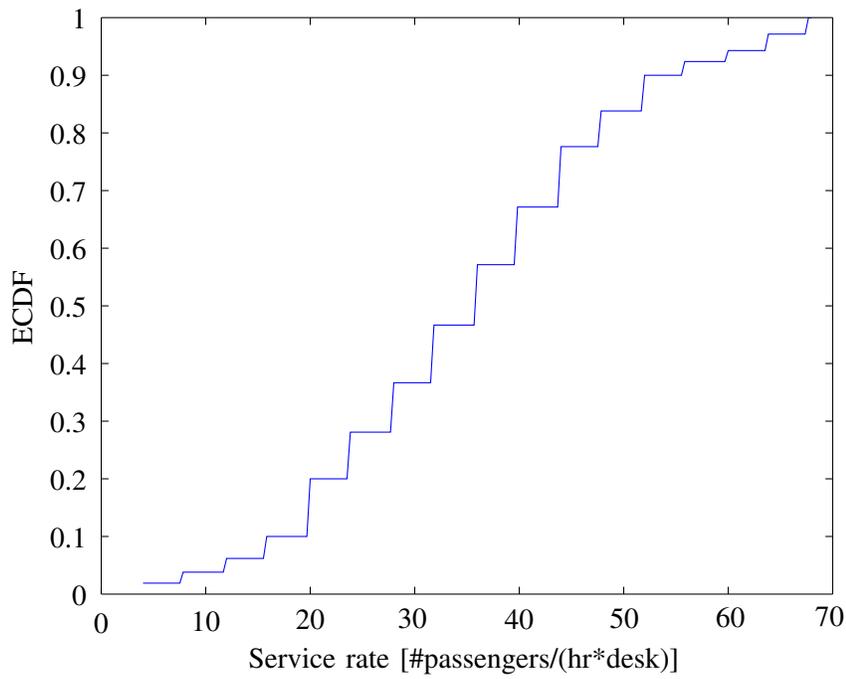
\caption{Service Rate distribution per desk}
\label{fig:deskSvceRate}
\end{figure}

In order to scale the service rate with the number of open desks, we have built the model of the service rate $\mu$for an individual server.
To obtain that service rate, the throughput at immigration for hours with an average wait time longer than 15 minutes are recorded. For these times,  the service rate is computed by dividing the throughput by the number of active desks. 
The result is the empirical distribution on Figure \ref{fig:deskSvceRate}.
The service rate $\mu(t)$ is then defined as the number of open desks at a given hour multiplied by a random number generated from this distribution.

\begin{figure}[bH!]
\centering
\def\svgwidth{0.65\textwidth}
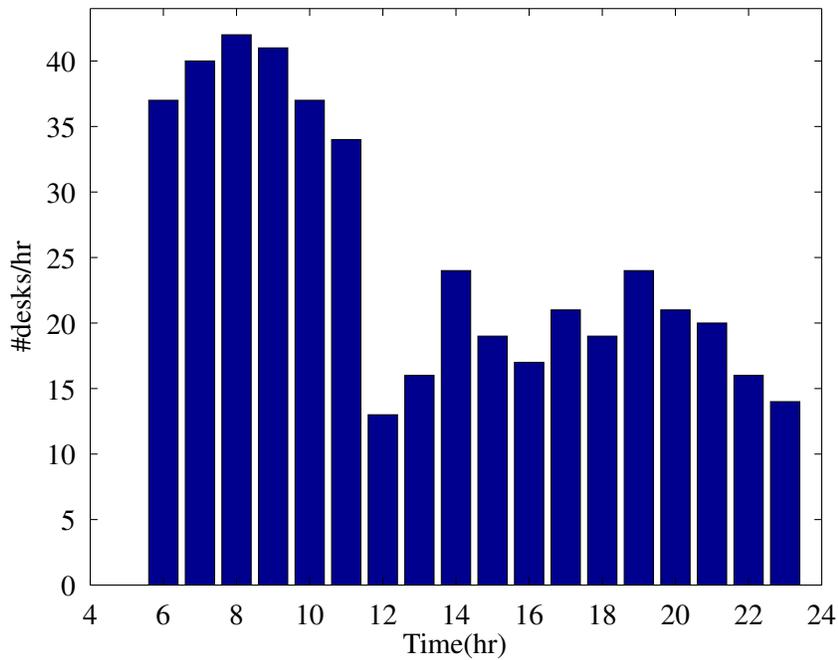
\caption{Number of Open Desks on November 26th 2012}
\label{fig:openDeskDist_1126}
\end{figure}

The actual service rate is consistent with the pattern of flight arrivals, as can be seen on figure \ref{fig:openDeskDist_1126}.
 There is a first peak in the staffing level around 8 am followed by a second peak around 7 pm.
It is to be noted that on several days, there is no record of any passenger crossing immigration around 1 pm. It is assumed that at that time, a minimum staffing level is maintained which would be the lowest staffing of the day.

\clearpage
\section{Results of the simulation}
In this section, the results from the simulation are analyzed and validated against estimated wait times and queue length data obtained from DWELL.
The wait time is defined as the time spent in the queue by a passenger from his or her arrival at the immigration zone to the beginning of service. 
The queue length is measured as the number of passengers left in the queue as a customer leaves the server.\\
12 days were simulated. Out of these days, 2 had an unstable queue, that grew unbounded as the arrival rate increased during the day.
\subsection{July 25th}
\begin{figure}[htp!]
\centering
\def\svgwidth{0.5\textwidth}
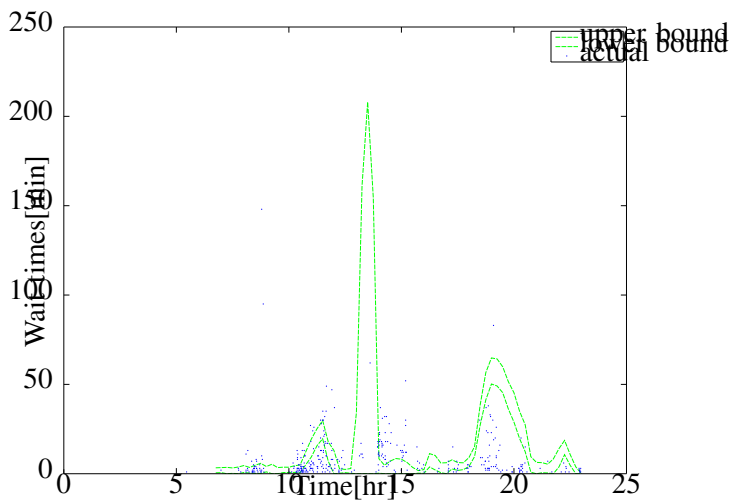
\caption{Average time spent in queue by a passenger on July 25th. }
\label{fig:waitTimes25Jul}
\end{figure}

As seen on figure \ref{fig:ActualVsSched25July} and figure \ref{fig:flightDelaysJuly25th}, most flights were on time on July 25th, except for a few morning flights who were late by almost 10 hours. Due to this delay, we expect service to be punctual in the morning as less passengers than expected present themselves at immigration, but slow in the afternoon.
This is mostly what we observe in the predicted and actual delays at immigration.
When comparing predicted delays to the delays information derived from DWELLon figure \ref{fig:waitTimes25Jul}, we can see that the simulation agrees with the actual wait times except for the large peak occuring before 3pm. This is due to a low number of open desks in our model.

\begin{figure}[htp!]
\centering
\def\svgwidth{0.5\textwidth}
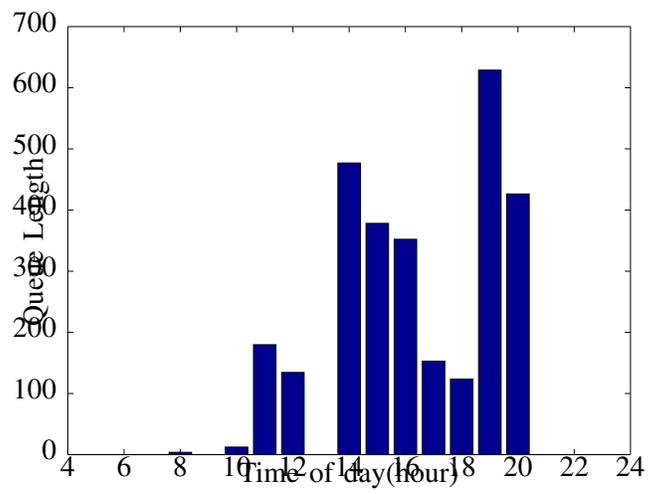
\caption{Average number of passengers at immigration on July 25th . }
\label{fig:queue25Jul}
\end{figure}

The length of the queue on figure \ref{fig:queue25Jul} is low in the morning due to the high number of open desks available in the morning. It increases in the afternoon due to a decrease in the number of open desks, and the late arrivals of passengers from delayed flights.
The change in queue length is not as dramatic as the rise in wait times in the afternoon. 
It indicates that despite having larger delays in the afternoon, those delays affect few passengers.
It is to be noted that whenever the service rate exceeds demand our model does not predict the formation of any queue.
\clearpage

\subsection{July 26th}
\begin{figure}[htp!]
\centering
\def\svgwidth{0.5\textwidth}
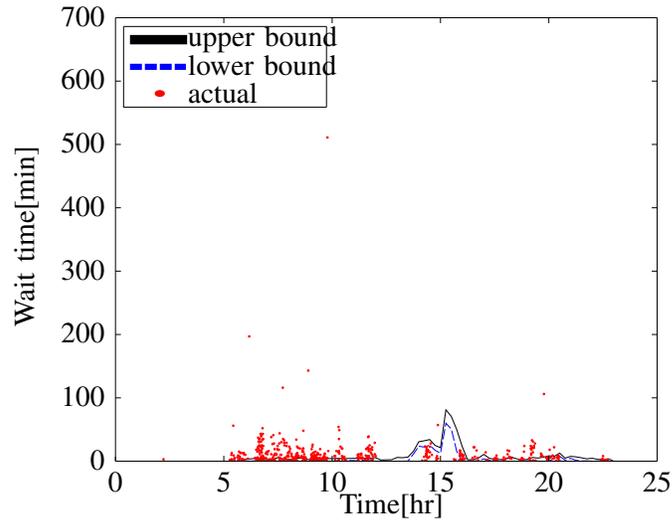
\caption{Average time spent in queue by a passenger on July 26th. }
\label{fig:waitTimes26Jul}
\end{figure}

\begin{figure}[htp!]
\centering
\def\svgwidth{0.5\textwidth}
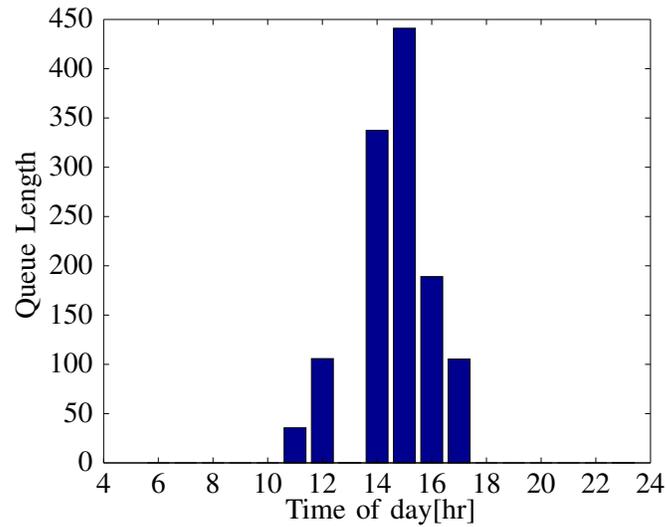
\caption{Simulated queue length July 26th. }
\label{fig:queueLength26Jul}
\end{figure}

On July 26th, predicted and actual delays remained low for most passengers as observed on figure \ref{fig:waitTimes26Jul}.
The model does not account at all for the wait times exceeding 100 minutes, and underestimate the waiting times at the begining of the day. 
As for July 25th, wait times occuring during the slow period of the day(2-3pm) are overestimated.
The predicted queue length on figure \ref{fig:queueLength26Jul} is also likely overestimated. 

\clearpage 
\subsection{December 11th}
\begin{figure}[htp!]
\centering
\def\svgwidth{0.5\textwidth}
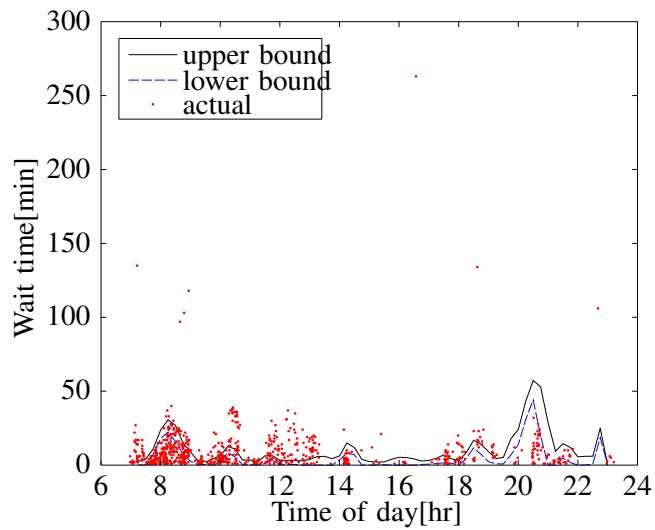
\caption{Average time spent in queue by a passenger on December 11th. }
\label{fig:waitTimes11Dec}
\end{figure}

\begin{figure}[htp!]
\centering
\def\svgwidth{0.5\textwidth}
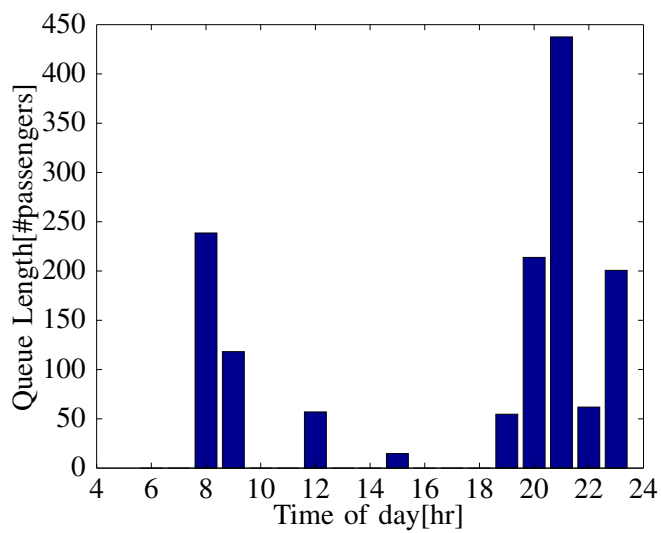
\caption{Simulated queue length December 11th. }
\label{fig:queueLength11Dec}
\end{figure}

The model tends to agree with actual wait times on December 11th, see figure \ref{fig:waitTimes11Dec}. 
It slightly underpredicts delay at immigration in the morning, and overestimates it in the afternoon. \\
Because staffing levels are lower in the afternoon, the last peak in demandoccuring at 8pm provokes longer queues as seen on \ref{fig:queueLength11Dec}

For all simulation results that were compared to the actual wait times, we observed that the simulated wait times were largely higher around 2pm than the actual results. It can be attributed by an error in the number of recorded desks at this time. For some of the days, there is no immigration records between 1:30pm and 2:00pm.
As explained in our analysis, only few of the data points can be used to obtain the time spent at immigration. This means that our actual wait times are probably a lower bound on the actual delays at immigration.
\newpage
\clearpage
\section{Conclusion}
In this paper, we have considered the problem of modeling the arrival process of passengers at the immigration services of an international airport. 
 
In our analysis, we have performed an investigation of the factors affecting passengers delays at immigration. We have generalized the notion of passener walk time to a model that is independent of the gate of origin, by using mixture models. Our model has been validated against a year of operational data.\\
Further research, would be on how to extend the model to other areas of the airport, and how to refine Wi-FI information to obtain finer passenger location estimates.
\newpage
\bibliographystyle{IEEEtran}
\bibliography{bibliography}
\end{document}

%% file: DIMIA_PAXCountImmiYear_arrivals.pdf_tex
\begingroup%
  \makeatletter%
  \providecommand\color[2][]{%
    \errmessage{(Inkscape) Color is used for the text in Inkscape, but the package 'color.sty' is not loaded}%
    \renewcommand\color[2][]{}%
  }%
  \providecommand\transparent[1]{%
    \errmessage{(Inkscape) Transparency is used (non-zero) for the text in Inkscape, but the package 'transparent.sty' is not loaded}%
    \renewcommand\transparent[1]{}%
  }%
  \providecommand\rotatebox[2]{#2}%
  \ifx\svgwidth\undefined%
    \setlength{\unitlength}{576bp}%
    \ifx\svgscale\undefined%
      \relax%
    \else%
      \setlength{\unitlength}{\unitlength * \real{\svgscale}}%
    \fi%
  \else%
    \setlength{\unitlength}{\svgwidth}%
  \fi%
  \global\let\svgwidth\undefined%
  \global\let\svgscale\undefined%
  \makeatother%
  \begin{picture}(1,0.78110567)%
    \put(0,0){\includegraphics[width=\unitlength]{DIMIA_PAXCountImmiYear_arrivals.pdf}}%
  \end{picture}%
\endgroup%

%% file: DWELL_PAXCountImmiYear_arrival.pdf_tex
\begingroup%
  \makeatletter%
  \providecommand\color[2][]{%
    \errmessage{(Inkscape) Color is used for the text in Inkscape, but the package 'color.sty' is not loaded}%
    \renewcommand\color[2][]{}%
  }%
  \providecommand\transparent[1]{%
    \errmessage{(Inkscape) Transparency is used (non-zero) for the text in Inkscape, but the package 'transparent.sty' is not loaded}%
    \renewcommand\transparent[1]{}%
  }%
  \providecommand\rotatebox[2]{#2}%
  \ifx\svgwidth\undefined%
    \setlength{\unitlength}{579.24603472bp}%
    \ifx\svgscale\undefined%
      \relax%
    \else%
      \setlength{\unitlength}{\unitlength * \real{\svgscale}}%
    \fi%
  \else%
    \setlength{\unitlength}{\svgwidth}%
  \fi%
  \global\let\svgwidth\undefined%
  \global\let\svgscale\undefined%
  \makeatother%
  \begin{picture}(1,0.76779016)%
    \put(0,0){\includegraphics[width=\unitlength]{DWELL_PAXCountImmiYear_arrival.pdf}}%
  \end{picture}%
\endgroup%

%% file: walkSpeedArrivals_clustersGates50.pdf_tex
\begingroup%
  \makeatletter%
  \providecommand\color[2][]{%
    \errmessage{(Inkscape) Color is used for the text in Inkscape, but the package 'color.sty' is not loaded}%
    \renewcommand\color[2][]{}%
  }%
  \providecommand\transparent[1]{%
    \errmessage{(Inkscape) Transparency is used (non-zero) for the text in Inkscape, but the package 'transparent.sty' is not loaded}%
    \renewcommand\transparent[1]{}%
  }%
  \providecommand\rotatebox[2]{#2}%
  \ifx\svgwidth\undefined%
    \setlength{\unitlength}{448bp}%
    \ifx\svgscale\undefined%
      \relax%
    \else%
      \setlength{\unitlength}{\unitlength * \real{\svgscale}}%
    \fi%
  \else%
    \setlength{\unitlength}{\svgwidth}%
  \fi%
  \global\let\svgwidth\undefined%
  \global\let\svgscale\undefined%
  \makeatother%
  \begin{picture}(1,0.75)%
    \put(0,0){\includegraphics[width=\unitlength]{walkSpeedArrivals_clustersGates50.pdf}}%
    \put(0.51660714,0.00681686){\color[rgb]{0,0,0}\makebox(0,0)[b]{\smash{Time (min)}}}%
    \put(0.03195218,0.38623913){\color[rgb]{0,0,0}\rotatebox{90}{\makebox(0,0)[b]{\smash{Frequency}}}}%
    \put(0.51660714,0.70804286){\color[rgb]{0,0,0}\makebox(0,0)[b]{\smash{Walk times from Pier A}}}%
    \put(0.13,0.04792543){\color[rgb]{0,0,0}\makebox(0,0)[b]{\smash{0}}}%
    \put(0.21531621,0.04633645){\color[rgb]{0,0,0}\makebox(0,0)[b]{\smash{5}}}%
    \put(0.29824896,0.04633645){\color[rgb]{0,0,0}\makebox(0,0)[b]{\smash{10}}}%
    \put(0.38833393,0.04633645){\color[rgb]{0,0,0}\makebox(0,0)[b]{\smash{15}}}%
    \put(0.47444464,0.04871991){\color[rgb]{0,0,0}\makebox(0,0)[b]{\smash{20}}}%
    \put(0.56214434,0.04871991){\color[rgb]{0,0,0}\makebox(0,0)[b]{\smash{25}}}%
    \put(0.64746057,0.04713093){\color[rgb]{0,0,0}\makebox(0,0)[b]{\smash{30}}}%
    \put(0.73357307,0.04951441){\color[rgb]{0,0,0}\makebox(0,0)[b]{\smash{35}}}%
    \put(0.82047827,0.05030889){\color[rgb]{0,0,0}\makebox(0,0)[b]{\smash{40}}}%
    \put(0.90182204,0.05189788){\color[rgb]{0,0,0}\makebox(0,0)[b]{\smash{45}}}%
    \put(0.1145,0.07125){\color[rgb]{0,0,0}\makebox(0,0)[rb]{\smash{0}}}%
    \put(0.1145,0.16528929){\color[rgb]{0,0,0}\makebox(0,0)[rb]{\smash{0.02}}}%
    \put(0.1145,0.25932679){\color[rgb]{0,0,0}\makebox(0,0)[rb]{\smash{0.04}}}%
    \put(0.1145,0.35336607){\color[rgb]{0,0,0}\makebox(0,0)[rb]{\smash{0.06}}}%
    \put(0.1145,0.44740357){\color[rgb]{0,0,0}\makebox(0,0)[rb]{\smash{0.08}}}%
    \put(0.1145,0.54144286){\color[rgb]{0,0,0}\makebox(0,0)[rb]{\smash{0.1}}}%
    \put(0.1145,0.63548036){\color[rgb]{0,0,0}\makebox(0,0)[rb]{\smash{0.12}}}%
    \put(0.64884553,0.64104495){\color[rgb]{0,0,0}\makebox(0,0)[lb]{\smash{gates 50-52}}}%
    \put(0.64884553,0.6084858){\color[rgb]{0,0,0}\makebox(0,0)[lb]{\smash{gates 53, 55, 57, 58}}}%
    \put(0.64884553,0.57592867){\color[rgb]{0,0,0}\makebox(0,0)[lb]{\smash{gates 54, 56, 59, 60, 61}}}%
  \end{picture}%
\endgroup%

%% file: walkSpeedDistribution.pdf_tex
\begingroup%
  \makeatletter%
  \providecommand\color[2][]{%
    \errmessage{(Inkscape) Color is used for the text in Inkscape, but the package 'color.sty' is not loaded}%
    \renewcommand\color[2][]{}%
  }%
  \providecommand\transparent[1]{%
    \errmessage{(Inkscape) Transparency is used (non-zero) for the text in Inkscape, but the package 'transparent.sty' is not loaded}%
    \renewcommand\transparent[1]{}%
  }%
  \providecommand\rotatebox[2]{#2}%
  \ifx\svgwidth\undefined%
    \setlength{\unitlength}{448bp}%
    \ifx\svgscale\undefined%
      \relax%
    \else%
      \setlength{\unitlength}{\unitlength * \real{\svgscale}}%
    \fi%
  \else%
    \setlength{\unitlength}{\svgwidth}%
  \fi%
  \global\let\svgwidth\undefined%
  \global\let\svgscale\undefined%
  \makeatother%
  \begin{picture}(1,0.75555891)%
    \put(0,0){\includegraphics[width=\unitlength]{walkSpeedDistribution.pdf}}%
    \put(0.5132364,0.00678125){\color[rgb]{0,0,0}\makebox(0,0)[b]{\smash{Speed in mi/hr}}}%
    \put(0.02731186,0.39212713){\color[rgb]{0,0,0}\rotatebox{90}{\makebox(0,0)[b]{\smash{PDF}}}}%
    \put(0.13,0.04633391){\color[rgb]{0,0,0}\makebox(0,0)[b]{\smash{0}}}%
    \put(0.38833393,0.04633391){\color[rgb]{0,0,0}\makebox(0,0)[b]{\smash{5}}}%
    \put(0.64666607,0.04633391){\color[rgb]{0,0,0}\makebox(0,0)[b]{\smash{10}}}%
    \put(0.905,0.04633391){\color[rgb]{0,0,0}\makebox(0,0)[b]{\smash{15}}}%
    \put(0.1145,0.07680891){\color[rgb]{0,0,0}\makebox(0,0)[rb]{\smash{0}}}%
    \put(0.1145,0.13237676){\color[rgb]{0,0,0}\makebox(0,0)[rb]{\smash{0.01}}}%
    \put(0.1145,0.18794462){\color[rgb]{0,0,0}\makebox(0,0)[rb]{\smash{0.02}}}%
    \put(0.1145,0.24351426){\color[rgb]{0,0,0}\makebox(0,0)[rb]{\smash{0.03}}}%
    \put(0.1145,0.29908212){\color[rgb]{0,0,0}\makebox(0,0)[rb]{\smash{0.04}}}%
    \put(0.1145,0.35464998){\color[rgb]{0,0,0}\makebox(0,0)[rb]{\smash{0.05}}}%
    \put(0.1145,0.41021784){\color[rgb]{0,0,0}\makebox(0,0)[rb]{\smash{0.06}}}%
    \put(0.1145,0.46578569){\color[rgb]{0,0,0}\makebox(0,0)[rb]{\smash{0.07}}}%
    \put(0.1145,0.52135355){\color[rgb]{0,0,0}\makebox(0,0)[rb]{\smash{0.08}}}%
    \put(0.1145,0.57692319){\color[rgb]{0,0,0}\makebox(0,0)[rb]{\smash{0.09}}}%
    \put(0.1145,0.63249105){\color[rgb]{0,0,0}\makebox(0,0)[rb]{\smash{0.1}}}%
    \put(0.70793945,0.6623924){\color[rgb]{0,0,0}\rotatebox{0.50480072}{\makebox(0,0)[lb]{\smash{Actual histogram}}}}%
    \put(0.70784805,0.63035331){\color[rgb]{0,0,0}\makebox(0,0)[lb]{\smash{Lognormal fit}}}%
  \end{picture}%
\endgroup%

%% file: arrival_process.pdf_tex
\begingroup%
  \makeatletter%
  \providecommand\color[2][]{%
    \errmessage{(Inkscape) Color is used for the text in Inkscape, but the package 'color.sty' is not loaded}%
    \renewcommand\color[2][]{}%
  }%
  \providecommand\transparent[1]{%
    \errmessage{(Inkscape) Transparency is used (non-zero) for the text in Inkscape, but the package 'transparent.sty' is not loaded}%
    \renewcommand\transparent[1]{}%
  }%
  \providecommand\rotatebox[2]{#2}%
  \ifx\svgwidth\undefined%
    \setlength{\unitlength}{536.925bp}%
    \ifx\svgscale\undefined%
      \relax%
    \else%
      \setlength{\unitlength}{\unitlength * \real{\svgscale}}%
    \fi%
  \else%
    \setlength{\unitlength}{\svgwidth}%
  \fi%
  \global\let\svgwidth\undefined%
  \global\let\svgscale\undefined%
  \makeatother%
  \begin{picture}(1,0.47493415)%
    \put(0,0){\includegraphics[width=\unitlength]{arrival_process.pdf}}%
    \put(0.35910177,0.43470501){\makebox(0,0)[b]{\smash{Gates}}}%
    \put(0.80613151,0.40765456){\makebox(0,0)[b]{\smash{Immigration}}}%
    \put(0.07600823,0.26335892){\makebox(0,0)[b]{\smash{Schedule}}}%
    \put(0.23990449,0.32295756){\makebox(0,0)[b]{\smash{Scheduled}}}%
    \put(0.2354346,0.24845926){\makebox(0,0)[b]{\smash{arrival}}}%
    \put(0.2354346,0.17694089){\makebox(0,0)[b]{\smash{times}}}%
    \put(0.56769702,0.24249939){\makebox(0,0)[b]{\smash{walk times}}}%
    \put(0.82103117,0.06083509){\makebox(0,0)[b]{\smash{Nb.}}}%
    \put(0.82103117,0.02805584){\makebox(0,0)[b]{\smash{Active desks}}}%
    \put(0.13708104,0.09719988){\makebox(0,0)[b]{\smash{DIMIA}}}%
    \put(0.13708104,0.06740056){\makebox(0,0)[b]{\smash{FIDS}}}%
    \put(0.13708104,0.03760124){\makebox(0,0)[b]{\smash{DWELL}}}%
    \put(0.92532879,0.25865796){\makebox(0,0)[b]{\smash{Departure}}}%
    \put(0.92532879,0.2392884){\makebox(0,0)[b]{\smash{times}}}%
  \end{picture}%
\endgroup%

%% file: Flights_SchedvsActual_12_8.pdf_tex
\begingroup%
  \makeatletter%
  \providecommand\color[2][]{%
    \errmessage{(Inkscape) Color is used for the text in Inkscape, but the package 'color.sty' is not loaded}%
    \renewcommand\color[2][]{}%
  }%
  \providecommand\transparent[1]{%
    \errmessage{(Inkscape) Transparency is used (non-zero) for the text in Inkscape, but the package 'transparent.sty' is not loaded}%
    \renewcommand\transparent[1]{}%
  }%
  \providecommand\rotatebox[2]{#2}%
  \ifx\svgwidth\undefined%
    \setlength{\unitlength}{1555.2bp}%
    \ifx\svgscale\undefined%
      \relax%
    \else%
      \setlength{\unitlength}{\unitlength * \real{\svgscale}}%
    \fi%
  \else%
    \setlength{\unitlength}{\svgwidth}%
  \fi%
  \global\let\svgwidth\undefined%
  \global\let\svgscale\undefined%
  \makeatother%
  \begin{picture}(1,0.63026363)%
    \put(0,0.030){\includegraphics[width=\unitlength]{Flights_SchedvsActual_12_8.pdf}}%
    \put(0.10839139,0.03){\color[rgb]{0.26666667,0.26666667,0.26666667}\makebox(0,0)[b]{\smash{6}}}%
    \put(0.2061846,0.03){\color[rgb]{0.26666667,0.26666667,0.26666667}\makebox(0,0)[b]{\smash{8}}}%
    \put(0.30397781,0.03){\color[rgb]{0.26666667,0.26666667,0.26666667}\makebox(0,0)[b]{\smash{10}}}%
    \put(0.40177102,0.03){\color[rgb]{0.26666667,0.26666667,0.26666667}\makebox(0,0)[b]{\smash{12}}}%
    \put(0.49956423,0.03){\color[rgb]{0.26666667,0.26666667,0.26666667}\makebox(0,0)[b]{\smash{14}}}%
    \put(0.59735744,0.03){\color[rgb]{0.26666667,0.26666667,0.26666667}\makebox(0,0)[b]{\smash{16}}}%
    \put(0.69515065,0.03){\color[rgb]{0.26666667,0.26666667,0.26666667}\makebox(0,0)[b]{\smash{18}}}%
    \put(0.79294386,0.03){\color[rgb]{0.26666667,0.26666667,0.26666667}\makebox(0,0)[b]{\smash{20}}}%
    \put(0.89073707,0.03){\color[rgb]{0.26666667,0.26666667,0.26666667}\makebox(0,0)[b]{\smash{22}}}%
    \put(0.10839139,0.11485746){\color[rgb]{0.26666667,0.26666667,0.26666667}\makebox(0,0)[rb]{\smash{10}}}%
    \put(0.10839139,0.1836738){\color[rgb]{0.26666667,0.26666667,0.26666667}\makebox(0,0)[rb]{\smash{20}}}%
    \put(0.10839139,0.2524953){\color[rgb]{0.26666667,0.26666667,0.26666667}\makebox(0,0)[rb]{\smash{30}}}%
    \put(0.10839139,0.32131164){\color[rgb]{0.26666667,0.26666667,0.26666667}\makebox(0,0)[rb]{\smash{40}}}%
    \put(0.10839139,0.39013313){\color[rgb]{0.26666667,0.26666667,0.26666667}\makebox(0,0)[rb]{\smash{50}}}%
    \put(0.10839139,0.45894947){\color[rgb]{0.26666667,0.26666667,0.26666667}\makebox(0,0)[rb]{\smash{60}}}%
    \put(0.10839139,0.52777097){\color[rgb]{0.26666667,0.26666667,0.26666667}\makebox(0,0)[rb]{\smash{70}}}%
    \put(0.52197972,0.01006552){\color[rgb]{0.26666667,0.26666667,0.26666667}\makebox(0,0)[b]{\smash{Time[hr]}}}%
    \put(0.03790712,0.31293393){\color[rgb]{0.26666667,0.26666667,0.26666667}\rotatebox{90}{\makebox(0,0)[b]{\smash{Frequency[\# flights]}}}}%
    \put(0.9,0.51){\color[rgb]{0.26666667,0.26666667,0.26666667}\makebox(0,0)[lb]{\smash{Scheduled}}}%
    \put(0.9,0.47){\color[rgb]{0.26666667,0.26666667,0.26666667}\makebox(0,0)[lb]{\smash{Actual}}}%
  \end{picture}%
\endgroup%

%% file: flightDelays_August12.pdf_tex
\begingroup%
  \makeatletter%
  \providecommand\color[2][]{%
    \errmessage{(Inkscape) Color is used for the text in Inkscape, but the package 'color.sty' is not loaded}%
    \renewcommand\color[2][]{}%
  }%
  \providecommand\transparent[1]{%
    \errmessage{(Inkscape) Transparency is used (non-zero) for the text in Inkscape, but the package 'transparent.sty' is not loaded}%
    \renewcommand\transparent[1]{}%
  }%
  \providecommand\rotatebox[2]{#2}%
  \ifx\svgwidth\undefined%
    \setlength{\unitlength}{1555.2bp}%
    \ifx\svgscale\undefined%
      \relax%
    \else%
      \setlength{\unitlength}{\unitlength * \real{\svgscale}}%
    \fi%
  \else%
    \setlength{\unitlength}{\svgwidth}%
  \fi%
  \global\let\svgwidth\undefined%
  \global\let\svgscale\undefined%
  \makeatother%
  \begin{picture}(1,0.60596708)%
    \put(0.05,0){\includegraphics[width=\unitlength]{flightDelays_August12.pdf}}%
    \put(0.09115226,0.04146357){\color[rgb]{0,0,0}\makebox(0,0)[b]{\smash{0}}}%
    \put(0.27469136,0.04146357){\color[rgb]{0,0,0}\makebox(0,0)[b]{\smash{5}}}%
    \put(0.45823045,0.04146357){\color[rgb]{0,0,0}\makebox(0,0)[b]{\smash{10}}}%
    \put(0.64176955,0.04146357){\color[rgb]{0,0,0}\makebox(0,0)[b]{\smash{15}}}%
    \put(0.82530864,0.04146357){\color[rgb]{0,0,0}\makebox(0,0)[b]{\smash{20}}}%
    \put(0.09012346,0.08615622){\color[rgb]{0,0,0}\makebox(0,0)[rb]{\smash{0}}}%
    \put(0.09012346,0.14406774){\color[rgb]{0,0,0}\makebox(0,0)[rb]{\smash{50}}}%
    \put(0.09012346,0.20198441){\color[rgb]{0,0,0}\makebox(0,0)[rb]{\smash{100}}}%
    \put(0.09012346,0.25989593){\color[rgb]{0,0,0}\makebox(0,0)[rb]{\smash{150}}}%
    \put(0.09012346,0.31780746){\color[rgb]{0,0,0}\makebox(0,0)[rb]{\smash{200}}}%
    \put(0.09012346,0.37572412){\color[rgb]{0,0,0}\makebox(0,0)[rb]{\smash{250}}}%
    \put(0.09012346,0.43363565){\color[rgb]{0,0,0}\makebox(0,0)[rb]{\smash{300}}}%
    \put(0.09012346,0.49155231){\color[rgb]{0,0,0}\makebox(0,0)[rb]{\smash{350}}}%
    \put(0.55,0.01100556){\color[rgb]{0,0,0}\makebox(0,0)[b]{\smash{Hour of scheduled arrival[hr]}}}%
    \put(0.01645794,0.29567812){\color[rgb]{0,0,0}\rotatebox{90}{\makebox(0,0)[b]{\smash{Delays[min]}}}}%
  \end{picture}%
\endgroup%

%% file: ImmiActualDepartures_12_8.pdf_tex
\begingroup%
  \makeatletter%
  \providecommand\color[2][]{%
    \errmessage{(Inkscape) Color is used for the text in Inkscape, but the package 'color.sty' is not loaded}%
    \renewcommand\color[2][]{}%
  }%
  \providecommand\transparent[1]{%
    \errmessage{(Inkscape) Transparency is used (non-zero) for the text in Inkscape, but the package 'transparent.sty' is not loaded}%
    \renewcommand\transparent[1]{}%
  }%
  \providecommand\rotatebox[2]{#2}%
  \ifx\svgwidth\undefined%
    \setlength{\unitlength}{855.6bp}%
    \ifx\svgscale\undefined%
      \relax%
    \else%
      \setlength{\unitlength}{\unitlength * \real{\svgscale}}%
    \fi%
  \else%
    \setlength{\unitlength}{\svgwidth}%
  \fi%
  \global\let\svgwidth\undefined%
  \global\let\svgscale\undefined%
  \makeatother%
  \begin{picture}(1,1.00724238)%
    \put(0,0.015){\includegraphics[width=\unitlength]{ImmiActualDepartures_12_8.pdf}}%
    \put(0.24276765,0.03725641){\color[rgb]{0.26666667,0.26666667,0.26666667}\makebox(0,0)[b]{\smash{5}}}%
    \put(0.45271622,0.03725641){\color[rgb]{0.26666667,0.26666667,0.26666667}\makebox(0,0)[b]{\smash{10}}}%
    \put(0.66267415,0.03725641){\color[rgb]{0.26666667,0.26666667,0.26666667}\makebox(0,0)[b]{\smash{15}}}%
    \put(0.87262272,0.03725641){\color[rgb]{0.26666667,0.26666667,0.26666667}\makebox(0,0)[b]{\smash{20}}}%
    \put(0.00293128,0.04473654){\color[rgb]{0.26666667,0.26666667,0.26666667}\makebox(0,0)[rb]{\smash{0}}}%
    \put(0.09293128,0.22438123){\color[rgb]{0.26666667,0.26666667,0.26666667}\makebox(0,0)[rb]{\smash{1000}}}%
    \put(0.09293128,0.40403528){\color[rgb]{0.26666667,0.26666667,0.26666667}\makebox(0,0)[rb]{\smash{2000}}}%
    \put(0.09293128,0.58367997){\color[rgb]{0.26666667,0.26666667,0.26666667}\makebox(0,0)[rb]{\smash{3000}}}%
    \put(0.09293128,0.76333401){\color[rgb]{0.26666667,0.26666667,0.26666667}\makebox(0,0)[rb]{\smash{4000}}}%
    \put(0.09293128,0.94297871){\color[rgb]{0.26666667,0.26666667,0.26666667}\makebox(0,0)[rb]{\smash{5000}}}%
    \put(0.53669939,0.00172579){\color[rgb]{0.26666667,0.26666667,0.26666667}\makebox(0,0)[b]{\smash{Time[hr]}}}%
    \put(0.000,0.5081877){\color[rgb]{0.26666667,0.26666667,0.26666667}\rotatebox{90}{\makebox(0,0)[b]{\smash{Frequency[\# passengers]}}}}%
  \end{picture}%
\endgroup%

%% file: StaffingLevel_August12.pdf_tex
\begingroup%
  \makeatletter%
  \providecommand\color[2][]{%
    \errmessage{(Inkscape) Color is used for the text in Inkscape, but the package 'color.sty' is not loaded}%
    \renewcommand\color[2][]{}%
  }%
  \providecommand\transparent[1]{%
    \errmessage{(Inkscape) Transparency is used (non-zero) for the text in Inkscape, but the package 'transparent.sty' is not loaded}%
    \renewcommand\transparent[1]{}%
  }%
  \providecommand\rotatebox[2]{#2}%
  \ifx\svgwidth\undefined%
    \setlength{\unitlength}{585.778125bp}%
    \ifx\svgscale\undefined%
      \relax%
    \else%
      \setlength{\unitlength}{\unitlength * \real{\svgscale}}%
    \fi%
  \else%
    \setlength{\unitlength}{\svgwidth}%
  \fi%
  \global\let\svgwidth\undefined%
  \global\let\svgscale\undefined%
  \makeatother%
  \begin{picture}(1,0.57873396)%
    \put(0,0){\includegraphics[width=\unitlength]{StaffingLevel_August12.pdf}}%
    \put(0.0664234,0.03587387){\color[rgb]{0.26666667,0.26666667,0.26666667}\makebox(0,0)[b]{\smash{0}}}%
    \put(0.25133983,0.03587387){\color[rgb]{0.26666667,0.26666667,0.26666667}\makebox(0,0)[b]{\smash{5}}}%
    \put(0.43625626,0.03587387){\color[rgb]{0.26666667,0.26666667,0.26666667}\makebox(0,0)[b]{\smash{10}}}%
    \put(0.62117269,0.03587387){\color[rgb]{0.26666667,0.26666667,0.26666667}\makebox(0,0)[b]{\smash{15}}}%
    \put(0.80608912,0.03587387){\color[rgb]{0.26666667,0.26666667,0.26666667}\makebox(0,0)[b]{\smash{20}}}%
    \put(0.99100555,0.03587387){\color[rgb]{0.26666667,0.26666667,0.26666667}\makebox(0,0)[b]{\smash{25}}}%
    \put(0.06369199,0.06679951){\color[rgb]{0.26666667,0.26666667,0.26666667}\makebox(0,0)[rb]{\smash{0}}}%
    \put(0.06369199,0.14329264){\color[rgb]{0.26666667,0.26666667,0.26666667}\makebox(0,0)[rb]{\smash{10}}}%
    \put(0.06369199,0.21978576){\color[rgb]{0.26666667,0.26666667,0.26666667}\makebox(0,0)[rb]{\smash{20}}}%
    \put(0.06369199,0.29627889){\color[rgb]{0.26666667,0.26666667,0.26666667}\makebox(0,0)[rb]{\smash{30}}}%
    \put(0.06369199,0.37277202){\color[rgb]{0.26666667,0.26666667,0.26666667}\makebox(0,0)[rb]{\smash{40}}}%
    \put(0.06369199,0.44926514){\color[rgb]{0.26666667,0.26666667,0.26666667}\makebox(0,0)[rb]{\smash{50}}}%
    \put(0.52871448,0.00397709){\color[rgb]{0.26666667,0.26666667,0.26666667}\makebox(0,0)[b]{\smash{Hour of the day[hr]}}}%
    \put(0.01452662,0.26824097){\color[rgb]{0.26666667,0.26666667,0.26666667}\rotatebox{90}{\makebox(0,0)[b]{\smash{\# Active Desks}}}}%
  \end{picture}%
\endgroup%

%% file: Flights_SchedvsActual_10_11.pdf_tex
\begingroup%
  \makeatletter%
  \providecommand\color[2][]{%
    \errmessage{(Inkscape) Color is used for the text in Inkscape, but the package 'color.sty' is not loaded}%
    \renewcommand\color[2][]{}%
  }%
  \providecommand\transparent[1]{%
    \errmessage{(Inkscape) Transparency is used (non-zero) for the text in Inkscape, but the package 'transparent.sty' is not loaded}%
    \renewcommand\transparent[1]{}%
  }%
  \providecommand\rotatebox[2]{#2}%
  \ifx\svgwidth\undefined%
    \setlength{\unitlength}{1717.1066296bp}%
    \ifx\svgscale\undefined%
      \relax%
    \else%
      \setlength{\unitlength}{\unitlength * \real{\svgscale}}%
    \fi%
  \else%
    \setlength{\unitlength}{\svgwidth}%
  \fi%
  \global\let\svgwidth\undefined%
  \global\let\svgscale\undefined%
  \makeatother%
  \begin{picture}(1,0.55269974)%
    \put(0,0.02){\includegraphics[width=\unitlength]{Flights_SchedvsActual_10_11.pdf}}%
    \put(0.13156238,0.0307494){\color[rgb]{0.26666667,0.26666667,0.26666667}\makebox(0,0)[b]{\smash{6}}}%
    \put(0.22534339,0.0307494){\color[rgb]{0.26666667,0.26666667,0.26666667}\makebox(0,0)[b]{\smash{8}}}%
    \put(0.31912906,0.0307494){\color[rgb]{0.26666667,0.26666667,0.26666667}\makebox(0,0)[b]{\smash{10}}}%
    \put(0.41291008,0.0307494){\color[rgb]{0.26666667,0.26666667,0.26666667}\makebox(0,0)[b]{\smash{12}}}%
    \put(0.50669575,0.0307494){\color[rgb]{0.26666667,0.26666667,0.26666667}\makebox(0,0)[b]{\smash{14}}}%
    \put(0.60047676,0.0307494){\color[rgb]{0.26666667,0.26666667,0.26666667}\makebox(0,0)[b]{\smash{16}}}%
    \put(0.69425778,0.0307494){\color[rgb]{0.26666667,0.26666667,0.26666667}\makebox(0,0)[b]{\smash{18}}}%
    \put(0.78804345,0.0307494){\color[rgb]{0.26666667,0.26666667,0.26666667}\makebox(0,0)[b]{\smash{20}}}%
    \put(0.88182446,0.0307494){\color[rgb]{0.26666667,0.26666667,0.26666667}\makebox(0,0)[b]{\smash{22}}}%
    \put(0.13063058,0.10525038){\color[rgb]{0.26666667,0.26666667,0.26666667}\makebox(0,0)[rb]{\smash{10}}}%
    \put(0.13063058,0.15601951){\color[rgb]{0.26666667,0.26666667,0.26666667}\makebox(0,0)[rb]{\smash{20}}}%
    \put(0.13063058,0.20679329){\color[rgb]{0.26666667,0.26666667,0.26666667}\makebox(0,0)[rb]{\smash{30}}}%
    \put(0.13063058,0.25756708){\color[rgb]{0.26666667,0.26666667,0.26666667}\makebox(0,0)[rb]{\smash{40}}}%
    \put(0.13063058,0.30834086){\color[rgb]{0.26666667,0.26666667,0.26666667}\makebox(0,0)[rb]{\smash{50}}}%
    \put(0.13063058,0.35910998){\color[rgb]{0.26666667,0.26666667,0.26666667}\makebox(0,0)[rb]{\smash{60}}}%
    \put(0.13063058,0.40988377){\color[rgb]{0.26666667,0.26666667,0.26666667}\makebox(0,0)[rb]{\smash{70}}}%
    \put(0.13063058,0.46065755){\color[rgb]{0.26666667,0.26666667,0.26666667}\makebox(0,0)[rb]{\smash{80}}}%
    \put(0.13063058,0.51142667){\color[rgb]{0.26666667,0.26666667,0.26666667}\makebox(0,0)[rb]{\smash{90}}}%
    \put(0.51513168,0.0){\color[rgb]{0.26666667,0.26666667,0.26666667}\makebox(0,0)[b]{\smash{Time[hr]}}}%
    \put(0.0000,0.27269249){\color[rgb]{0.26666667,0.26666667,0.26666667}\rotatebox{90}{\makebox(0,0)[b]{\smash{Frequency[\# flights]}}}}%
    \put(0.86922979,0.49950327){\color[rgb]{0.26666667,0.26666667,0.26666667}\makebox(0.1,0.1)[lb]{\smash{Scheduled}}}%
    \put(0.86922979,0.46901985){\color[rgb]{0.26666667,0.26666667,0.26666667}\makebox(0.1,0.1)[lb]{\smash{Actual}}}%
  \end{picture}%
\endgroup%

%% file: flightDelays_Nov10.pdf_tex
\begingroup%
  \makeatletter%
  \providecommand\color[2][]{%
    \errmessage{(Inkscape) Color is used for the text in Inkscape, but the package 'color.sty' is not loaded}%
    \renewcommand\color[2][]{}%
  }%
  \providecommand\transparent[1]{%
    \errmessage{(Inkscape) Transparency is used (non-zero) for the text in Inkscape, but the package 'transparent.sty' is not loaded}%
    \renewcommand\transparent[1]{}%
  }%
  \providecommand\rotatebox[2]{#2}%
  \ifx\svgwidth\undefined%
    \setlength{\unitlength}{1555.2bp}%
    \ifx\svgscale\undefined%
      \relax%
    \else%
      \setlength{\unitlength}{\unitlength * \real{\svgscale}}%
    \fi%
  \else%
    \setlength{\unitlength}{\svgwidth}%
  \fi%
  \global\let\svgwidth\undefined%
  \global\let\svgscale\undefined%
  \makeatother%
  \begin{picture}(1,0.60596708)%
    \put(0.02,0.02){\includegraphics[width=\unitlength]{flightDelays_Nov10.pdf}}%
    \put(0.05754996,0.07443187){\color[rgb]{0,0,0}\makebox(0,0)[b]{\smash{0}}}%
    \put(0.24108905,0.07443187){\color[rgb]{0,0,0}\makebox(0,0)[b]{\smash{5}}}%
    \put(0.42462815,0.07443187){\color[rgb]{0,0,0}\makebox(0,0)[b]{\smash{10}}}%
    \put(0.60816724,0.07443187){\color[rgb]{0,0,0}\makebox(0,0)[b]{\smash{15}}}%
    \put(0.79170634,0.07443187){\color[rgb]{0,0,0}\makebox(0,0)[b]{\smash{20}}}%
    \put(0.06012346,0.1714249){\color[rgb]{0,0,0}\makebox(0,0)[rb]{\smash{100}}}%
    \put(0.06012346,0.24873971){\color[rgb]{0,0,0}\makebox(0,0)[rb]{\smash{200}}}%
    \put(0.06012346,0.32605967){\color[rgb]{0,0,0}\makebox(0,0)[rb]{\smash{300}}}%
    \put(0.06012346,0.40337963){\color[rgb]{0,0,0}\makebox(0,0)[rb]{\smash{400}}}%
    \put(0.06012346,0.48069444){\color[rgb]{0,0,0}\makebox(0,0)[rb]{\smash{500}}}%
    \put(0.06012346,0.5580144){\color[rgb]{0,0,0}\makebox(0,0)[rb]{\smash{600}}}%
    \put(0.7,0.03748971){\color[rgb]{0,0,0}\makebox(0,0)[b]{\smash{Hour of the day[hr]}}}%
    \put(0.0,0.30000089){\color[rgb]{0,0,0}\rotatebox{90}{\makebox(0,0)[b]{\smash{Delays[min]}}}}%
  \end{picture}%
\endgroup%

%% file: ImmiActualDepartures_10_11.pdf_tex
\begingroup%
  \makeatletter%
  \providecommand\color[2][]{%
    \errmessage{(Inkscape) Color is used for the text in Inkscape, but the package 'color.sty' is not loaded}%
    \renewcommand\color[2][]{}%
  }%
  \providecommand\transparent[1]{%
    \errmessage{(Inkscape) Transparency is used (non-zero) for the text in Inkscape, but the package 'transparent.sty' is not loaded}%
    \renewcommand\transparent[1]{}%
  }%
  \providecommand\rotatebox[2]{#2}%
  \ifx\svgwidth\undefined%
    \setlength{\unitlength}{855.2bp}%
    \ifx\svgscale\undefined%
      \relax%
    \else%
      \setlength{\unitlength}{\unitlength * \real{\svgscale}}%
    \fi%
  \else%
    \setlength{\unitlength}{\svgwidth}%
  \fi%
  \global\let\svgwidth\undefined%
  \global\let\svgscale\undefined%
  \makeatother%
  \begin{picture}(1,1.00870465)%
    \put(0,0){\includegraphics[width=\unitlength]{ImmiActualDepartures_10_11.pdf}}%
    \put(0.12104771,0.01826499){\color[rgb]{0.26666667,0.26666667,0.26666667}\makebox(0,0)[b]{\smash{5}}}%
    \put(0.35210477,0.01826499){\color[rgb]{0.26666667,0.26666667,0.26666667}\makebox(0,0)[b]{\smash{10}}}%
    \put(0.58316183,0.01826499){\color[rgb]{0.26666667,0.26666667,0.26666667}\makebox(0,0)[b]{\smash{15}}}%
    \put(0.8142189,0.01826499){\color[rgb]{0.26666667,0.26666667,0.26666667}\makebox(0,0)[b]{\smash{20}}}%
    \put(0.07896539,0.04574862){\color[rgb]{0.26666667,0.26666667,0.26666667}\makebox(0,0)[rb]{\smash{0}}}%
    \put(0.07896539,0.22092168){\color[rgb]{0.26666667,0.26666667,0.26666667}\makebox(0,0)[rb]{\smash{1000}}}%
    \put(0.07896539,0.39610409){\color[rgb]{0.26666667,0.26666667,0.26666667}\makebox(0,0)[rb]{\smash{2000}}}%
    \put(0.07896539,0.57127715){\color[rgb]{0.26666667,0.26666667,0.26666667}\makebox(0,0)[rb]{\smash{3000}}}%
    \put(0.07896539,0.74645956){\color[rgb]{0.26666667,0.26666667,0.26666667}\makebox(0,0)[rb]{\smash{4000}}}%
    \put(0.07896539,0.92163262){\color[rgb]{0.26666667,0.26666667,0.26666667}\makebox(0,0)[rb]{\smash{5000}}}%
    \put(0.53695042,0.000){\color[rgb]{0.26666667,0.26666667,0.26666667}\makebox(0,0)[b]{\smash{Time[hr]}}}%
    \put(0.0,0.50599184){\color[rgb]{0.26666667,0.26666667,0.26666667}\rotatebox{90}{\makebox(0,0)[b]{\smash{Frequency[\# passengers]}}}}%
  \end{picture}%
\endgroup%

%% file: StaffingLevel_Nov10.pdf_tex
\begingroup%
  \makeatletter%
  \providecommand\color[2][]{%
    \errmessage{(Inkscape) Color is used for the text in Inkscape, but the package 'color.sty' is not loaded}%
    \renewcommand\color[2][]{}%
  }%
  \providecommand\transparent[1]{%
    \errmessage{(Inkscape) Transparency is used (non-zero) for the text in Inkscape, but the package 'transparent.sty' is not loaded}%
    \renewcommand\transparent[1]{}%
  }%
  \providecommand\rotatebox[2]{#2}%
  \ifx\svgwidth\undefined%
    \setlength{\unitlength}{829.7859375bp}%
    \ifx\svgscale\undefined%
      \relax%
    \else%
      \setlength{\unitlength}{\unitlength * \real{\svgscale}}%
    \fi%
  \else%
    \setlength{\unitlength}{\svgwidth}%
  \fi%
  \global\let\svgwidth\undefined%
  \global\let\svgscale\undefined%
  \makeatother%
  \begin{picture}(1,1.08651422)%
    \put(0,0){\includegraphics[width=\unitlength]{StaffingLevel_Nov10.pdf}}%
    \put(0.07718858,0.03943694){\color[rgb]{0.26666667,0.26666667,0.26666667}\makebox(0,0)[b]{\smash{0}}}%
    \put(0.26056118,0.03943694){\color[rgb]{0.26666667,0.26666667,0.26666667}\makebox(0,0)[b]{\smash{5}}}%
    \put(0.44393377,0.03943694){\color[rgb]{0.26666667,0.26666667,0.26666667}\makebox(0,0)[b]{\smash{10}}}%
    \put(0.62730636,0.03943694){\color[rgb]{0.26666667,0.26666667,0.26666667}\makebox(0,0)[b]{\smash{15}}}%
    \put(0.81067896,0.03943694){\color[rgb]{0.26666667,0.26666667,0.26666667}\makebox(0,0)[b]{\smash{20}}}%
    \put(0.99405155,0.03943694){\color[rgb]{0.26666667,0.26666667,0.26666667}\makebox(0,0)[b]{\smash{25}}}%
    \put(0.07526037,0.04714977){\color[rgb]{0.26666667,0.26666667,0.26666667}\makebox(0,0)[rb]{\smash{0}}}%
    \put(0.07526037,0.21835537){\color[rgb]{0.26666667,0.26666667,0.26666667}\makebox(0,0)[rb]{\smash{10}}}%
    \put(0.07526037,0.38955133){\color[rgb]{0.26666667,0.26666667,0.26666667}\makebox(0,0)[rb]{\smash{20}}}%
    \put(0.07526037,0.56075693){\color[rgb]{0.26666667,0.26666667,0.26666667}\makebox(0,0)[rb]{\smash{30}}}%
    \put(0.07526037,0.73196254){\color[rgb]{0.26666667,0.26666667,0.26666667}\makebox(0,0)[rb]{\smash{40}}}%
    \put(0.07526037,0.9031585){\color[rgb]{0.26666667,0.26666667,0.26666667}\makebox(0,0)[rb]{\smash{50}}}%
    \put(0.53562007,0.00280099){\color[rgb]{0.26666667,0.26666667,0.26666667}\makebox(0,0)[b]{\smash{Hour of the day[hr]}}}%
    \put(0.04055263,0.52148898){\color[rgb]{0.26666667,0.26666667,0.26666667}\rotatebox{90}{\makebox(0,0)[b]{\smash{\# Active Desks}}}}%
    \put(1.05093369,0.97155204){\color[rgb]{0.26666667,0.26666667,0.26666667}\makebox(0,0)[lb]{\smash{trace 0}}}%
  \end{picture}%
\endgroup%

%% file: Flights_SchedvsActual_25_7.pdf_tex
\begingroup%
  \makeatletter%
  \providecommand\color[2][]{%
    \errmessage{(Inkscape) Color is used for the text in Inkscape, but the package 'color.sty' is not loaded}%
    \renewcommand\color[2][]{}%
  }%
  \providecommand\transparent[1]{%
    \errmessage{(Inkscape) Transparency is used (non-zero) for the text in Inkscape, but the package 'transparent.sty' is not loaded}%
    \renewcommand\transparent[1]{}%
  }%
  \providecommand\rotatebox[2]{#2}%
  \ifx\svgwidth\undefined%
    \setlength{\unitlength}{1689.65532942bp}%
    \ifx\svgscale\undefined%
      \relax%
    \else%
      \setlength{\unitlength}{\unitlength * \real{\svgscale}}%
    \fi%
  \else%
    \setlength{\unitlength}{\svgwidth}%
  \fi%
  \global\let\svgwidth\undefined%
  \global\let\svgscale\undefined%
  \makeatother%
  \begin{picture}(1,0.56043036)%
    \put(0,0){\includegraphics[width=\unitlength]{Flights_SchedvsActual_25_7.pdf}}%
    \put(0.29266847,0.03766807){\color[rgb]{0.26666667,0.26666667,0.26666667}\makebox(0,0)[b]{\smash{5}}}%
    \put(0.45596331,0.03766807){\color[rgb]{0.26666667,0.26666667,0.26666667}\makebox(0,0)[b]{\smash{10}}}%
    \put(0.61926289,0.03766807){\color[rgb]{0.26666667,0.26666667,0.26666667}\makebox(0,0)[b]{\smash{15}}}%
    \put(0.78255773,0.03766807){\color[rgb]{0.26666667,0.26666667,0.26666667}\makebox(0,0)[b]{\smash{20}}}%
    \put(0.94585257,0.03766807){\color[rgb]{0.26666667,0.26666667,0.26666667}\makebox(0,0)[b]{\smash{25}}}%
    \put(0.15430461,0.0911779){\color[rgb]{0.26666667,0.26666667,0.26666667}\makebox(0,0)[rb]{\smash{0}}}%
    \put(0.15430461,0.18269953){\color[rgb]{0.26666667,0.26666667,0.26666667}\makebox(0,0)[rb]{\smash{20}}}%
    \put(0.15430461,0.27422588){\color[rgb]{0.26666667,0.26666667,0.26666667}\makebox(0,0)[rb]{\smash{40}}}%
    \put(0.15430461,0.36575224){\color[rgb]{0.26666667,0.26666667,0.26666667}\makebox(0,0)[rb]{\smash{60}}}%
    \put(0.15430461,0.4572786){\color[rgb]{0.26666667,0.26666667,0.26666667}\makebox(0,0)[rb]{\smash{80}}}%
    \put(0.55698892,0.0){\color[rgb]{0.26666667,0.26666667,0.26666667}\makebox(0,0)[b]{\smash{Time[hr]}}}%
    \put(0.02996529,0.28010692){\color[rgb]{0.26666667,0.26666667,0.26666667}\rotatebox{90}{\makebox(0,0)[b]{\smash{Frequency[\# flights]}}}}%
    \put(0.88302033,0.50363977){\color[rgb]{0.26666667,0.26666667,0.26666667}\makebox(0,0)[lb]{\smash{Scheduled}}}%
    \put(0.88368346,0.4713282){\color[rgb]{0.26666667,0.26666667,0.26666667}\makebox(0,0)[lb]{\smash{Actual}}}%
  \end{picture}%
\endgroup%

%% file: flightDelays_July25.pdf_tex
\begingroup%
  \makeatletter%
  \providecommand\color[2][]{%
    \errmessage{(Inkscape) Color is used for the text in Inkscape, but the package 'color.sty' is not loaded}%
    \renewcommand\color[2][]{}%
  }%
  \providecommand\transparent[1]{%
    \errmessage{(Inkscape) Transparency is used (non-zero) for the text in Inkscape, but the package 'transparent.sty' is not loaded}%
    \renewcommand\transparent[1]{}%
  }%
  \providecommand\rotatebox[2]{#2}%
  \ifx\svgwidth\undefined%
    \setlength{\unitlength}{1555.2bp}%
    \ifx\svgscale\undefined%
      \relax%
    \else%
      \setlength{\unitlength}{\unitlength * \real{\svgscale}}%
    \fi%
  \else%
    \setlength{\unitlength}{\svgwidth}%
  \fi%
  \global\let\svgwidth\undefined%
  \global\let\svgscale\undefined%
  \makeatother%
  \begin{picture}(1,0.60596708)%
    \put(0,0){\includegraphics[width=\unitlength]{flightDelays_July25.pdf}}%
    \put(0.0404318,0.05052122){\color[rgb]{0.26666667,0.26666667,0.26666667}\makebox(0,0)[b]{\smash{0}}}%
    \put(0.2239709,0.05052122){\color[rgb]{0.26666667,0.26666667,0.26666667}\makebox(0,0)[b]{\smash{5}}}%
    \put(0.40750999,0.05052122){\color[rgb]{0.26666667,0.26666667,0.26666667}\makebox(0,0)[b]{\smash{10}}}%
    \put(0.59104909,0.05052122){\color[rgb]{0.26666667,0.26666667,0.26666667}\makebox(0,0)[b]{\smash{15}}}%
    \put(0.77458818,0.05052122){\color[rgb]{0.26666667,0.26666667,0.26666667}\makebox(0,0)[b]{\smash{20}}}%
    \put(0.95812728,0.05052122){\color[rgb]{0.26666667,0.26666667,0.26666667}\makebox(0,0)[b]{\smash{25}}}%
    \put(0.04012346,0.07359053){\color[rgb]{0.26666667,0.26666667,0.26666667}\makebox(0,0)[rb]{\smash{0}}}%
    \put(0.04012346,0.15091049){\color[rgb]{0.26666667,0.26666667,0.26666667}\makebox(0,0)[rb]{\smash{100}}}%
    \put(0.04012346,0.22822531){\color[rgb]{0.26666667,0.26666667,0.26666667}\makebox(0,0)[rb]{\smash{200}}}%
    \put(0.04012346,0.30554527){\color[rgb]{0.26666667,0.26666667,0.26666667}\makebox(0,0)[rb]{\smash{300}}}%
    \put(0.04012346,0.38286523){\color[rgb]{0.26666667,0.26666667,0.26666667}\makebox(0,0)[rb]{\smash{400}}}%
    \put(0.04012346,0.46018004){\color[rgb]{0.26666667,0.26666667,0.26666667}\makebox(0,0)[rb]{\smash{500}}}%
    \put(0.04012346,0.5375){\color[rgb]{0.26666667,0.26666667,0.26666667}\makebox(0,0)[rb]{\smash{600}}}%
    \put(0.5,0.01440329){\color[rgb]{0.26666667,0.26666667,0.26666667}\makebox(0,0)[b]{\smash{Hour of the day[hr]}}}%
    \put(0.01295941,0.30072135){\color[rgb]{0.26666667,0.26666667,0.26666667}\rotatebox{90}{\makebox(0,0)[b]{\smash{Delays[min]}}}}%
  \end{picture}%
\endgroup%

%% file: ImmiActualDepartures_25_7.pdf_tex
\begingroup%
  \makeatletter%
  \providecommand\color[2][]{%
    \errmessage{(Inkscape) Color is used for the text in Inkscape, but the package 'color.sty' is not loaded}%
    \renewcommand\color[2][]{}%
  }%
  \providecommand\transparent[1]{%
    \errmessage{(Inkscape) Transparency is used (non-zero) for the text in Inkscape, but the package 'transparent.sty' is not loaded}%
    \renewcommand\transparent[1]{}%
  }%
  \providecommand\rotatebox[2]{#2}%
  \ifx\svgwidth\undefined%
    \setlength{\unitlength}{855.2bp}%
    \ifx\svgscale\undefined%
      \relax%
    \else%
      \setlength{\unitlength}{\unitlength * \real{\svgscale}}%
    \fi%
  \else%
    \setlength{\unitlength}{\svgwidth}%
  \fi%
  \global\let\svgwidth\undefined%
  \global\let\svgscale\undefined%
  \makeatother%
  \begin{picture}(1,1.00870465)%
    \put(0,0){\includegraphics[width=\unitlength]{ImmiActualDepartures_25_7.pdf}}%
    \put(0.12104771,0.03826499){\color[rgb]{0.26666667,0.26666667,0.26666667}\makebox(0,0)[lb]{\smash{5}}}%
    \put(0.35210477,0.03826499){\color[rgb]{0.26666667,0.26666667,0.26666667}\makebox(0,0)[lb]{\smash{10}}}%
    \put(0.58316183,0.03826499){\color[rgb]{0.26666667,0.26666667,0.26666667}\makebox(0,0)[lb]{\smash{15}}}%
    \put(0.8142189,0.03826499){\color[rgb]{0.26666667,0.26666667,0.26666667}\makebox(0,0)[lb]{\smash{20}}}%
    \put(0.07296539,0.04574862){\color[rgb]{0.26666667,0.26666667,0.26666667}\makebox(0,0)[lb]{\smash{0}}}%
    \put(0.07296539,0.1803417){\color[rgb]{0.26666667,0.26666667,0.26666667}\makebox(0,0)[lb]{\smash{500}}}%
    \put(0.07296539,0.31492542){\color[rgb]{0.26666667,0.26666667,0.26666667}\makebox(0,0)[lb]{\smash{1000}}}%
    \put(0.07296539,0.4495185){\color[rgb]{0.26666667,0.26666667,0.26666667}\makebox(0,0)[lb]{\smash{1500}}}%
    \put(0.07296539,0.58411157){\color[rgb]{0.26666667,0.26666667,0.26666667}\makebox(0,0)[lb]{\smash{2000}}}%
    \put(0.07296539,0.7186953){\color[rgb]{0.26666667,0.26666667,0.26666667}\makebox(0,0)[lb]{\smash{2500}}}%
    \put(0.07296539,0.85328838){\color[rgb]{0.26666667,0.26666667,0.26666667}\makebox(0,0)[lb]{\smash{3000}}}%
    \put(0.53695042,0.00271775){\color[rgb]{0.26666667,0.26666667,0.26666667}\makebox(0,0)[lb]{\smash{Time[hr]}}}%
    \put(0.03928906,0.50599184){\color[rgb]{0.26666667,0.26666667,0.26666667}\rotatebox{90}{\makebox(0,0)[lb]{\smash{Frequency[\# passengers]}}}}%
  \end{picture}%
\endgroup%

%% file: StaffingLevel_July25.pdf_tex
\begingroup%
  \makeatletter%
  \providecommand\color[2][]{%
    \errmessage{(Inkscape) Color is used for the text in Inkscape, but the package 'color.sty' is not loaded}%
    \renewcommand\color[2][]{}%
  }%
  \providecommand\transparent[1]{%
    \errmessage{(Inkscape) Transparency is used (non-zero) for the text in Inkscape, but the package 'transparent.sty' is not loaded}%
    \renewcommand\transparent[1]{}%
  }%
  \providecommand\rotatebox[2]{#2}%
  \ifx\svgwidth\undefined%
    \setlength{\unitlength}{835.1296875bp}%
    \ifx\svgscale\undefined%
      \relax%
    \else%
      \setlength{\unitlength}{\unitlength * \real{\svgscale}}%
    \fi%
  \else%
    \setlength{\unitlength}{\svgwidth}%
  \fi%
  \global\let\svgwidth\undefined%
  \global\let\svgscale\undefined%
  \makeatother%
  \begin{picture}(1,1.07956193)%
    \put(0,0){\includegraphics[width=\unitlength]{StaffingLevel_July25.pdf}}%
    \put(0.07669468,0.0391846){\color[rgb]{0.26666667,0.26666667,0.26666667}\makebox(0,0)[lb]{\smash{0}}}%
    \put(0.25889392,0.0391846){\color[rgb]{0.26666667,0.26666667,0.26666667}\makebox(0,0)[lb]{\smash{5}}}%
    \put(0.44109317,0.0391846){\color[rgb]{0.26666667,0.26666667,0.26666667}\makebox(0,0)[lb]{\smash{10}}}%
    \put(0.62329242,0.0391846){\color[rgb]{0.26666667,0.26666667,0.26666667}\makebox(0,0)[lb]{\smash{15}}}%
    \put(0.80549166,0.0391846){\color[rgb]{0.26666667,0.26666667,0.26666667}\makebox(0,0)[lb]{\smash{20}}}%
    \put(0.98769091,0.0391846){\color[rgb]{0.26666667,0.26666667,0.26666667}\makebox(0,0)[lb]{\smash{25}}}%
    \put(0.0747788,0.04684808){\color[rgb]{0.26666667,0.26666667,0.26666667}\makebox(0,0)[lb]{\smash{0}}}%
    \put(0.0747788,0.22375952){\color[rgb]{0.26666667,0.26666667,0.26666667}\makebox(0,0)[lb]{\smash{10}}}%
    \put(0.0747788,0.40067097){\color[rgb]{0.26666667,0.26666667,0.26666667}\makebox(0,0)[lb]{\smash{20}}}%
    \put(0.0747788,0.57758241){\color[rgb]{0.26666667,0.26666667,0.26666667}\makebox(0,0)[lb]{\smash{30}}}%
    \put(0.0747788,0.75449386){\color[rgb]{0.26666667,0.26666667,0.26666667}\makebox(0,0)[lb]{\smash{40}}}%
    \put(0.0747788,0.9314053){\color[rgb]{0.26666667,0.26666667,0.26666667}\makebox(0,0)[lb]{\smash{50}}}%
    \put(0.53219279,0.00278306){\color[rgb]{0.26666667,0.26666667,0.26666667}\makebox(0,0)[lb]{\smash{Hour of the day[hr]}}}%
    \put(0.04029314,0.51815212){\color[rgb]{0.26666667,0.26666667,0.26666667}\rotatebox{90}{\makebox(0,0)[lb]{\smash{\# Active Desks}}}}%
    \put(1.04420908,0.96533536){\color[rgb]{0.26666667,0.26666667,0.26666667}\makebox(0,0)[lb]{\smash{trace 0}}}%
  \end{picture}%
\endgroup%

%% file: TotalThroughput.pdf_tex
\begingroup%
  \makeatletter%
  \providecommand\color[2][]{%
    \errmessage{(Inkscape) Color is used for the text in Inkscape, but the package 'color.sty' is not loaded}%
    \renewcommand\color[2][]{}%
  }%
  \providecommand\transparent[1]{%
    \errmessage{(Inkscape) Transparency is used (non-zero) for the text in Inkscape, but the package 'transparent.sty' is not loaded}%
    \renewcommand\transparent[1]{}%
  }%
  \providecommand\rotatebox[2]{#2}%
  \ifx\svgwidth\undefined%
    \setlength{\unitlength}{1555.2bp}%
    \ifx\svgscale\undefined%
      \relax%
    \else%
      \setlength{\unitlength}{\unitlength * \real{\svgscale}}%
    \fi%
  \else%
    \setlength{\unitlength}{\svgwidth}%
  \fi%
  \global\let\svgwidth\undefined%
  \global\let\svgscale\undefined%
  \makeatother%
  \begin{picture}(1,0.60596708)%
    \put(0,0){\includegraphics[width=\unitlength]{TotalThroughput.pdf}}%
    \put(0.10078753,0.04691892){\color[rgb]{0.26666667,0.26666667,0.26666667}\makebox(0,0)[b]{\smash{0}}}%
    \put(0.2490437,0.04691892){\color[rgb]{0.26666667,0.26666667,0.26666667}\makebox(0,0)[b]{\smash{100}}}%
    \put(0.39730502,0.04691892){\color[rgb]{0.26666667,0.26666667,0.26666667}\makebox(0,0)[b]{\smash{200}}}%
    \put(0.54556633,0.04691892){\color[rgb]{0.26666667,0.26666667,0.26666667}\makebox(0,0)[b]{\smash{300}}}%
    \put(0.69382251,0.04691892){\color[rgb]{0.26666667,0.26666667,0.26666667}\makebox(0,0)[b]{\smash{400}}}%
    \put(0.84208382,0.04691892){\color[rgb]{0.26666667,0.26666667,0.26666667}\makebox(0,0)[b]{\smash{500}}}%
    \put(0.05309176,0.07278322){\color[rgb]{0.26666667,0.26666667,0.26666667}\makebox(0,0)[rb]{\smash{0}}}%
    \put(0.05309176,0.12961964){\color[rgb]{0.26666667,0.26666667,0.26666667}\makebox(0,0)[rb]{\smash{200}}}%
    \put(0.05309176,0.1864612){\color[rgb]{0.26666667,0.26666667,0.26666667}\makebox(0,0)[rb]{\smash{400}}}%
    \put(0.05309176,0.24330277){\color[rgb]{0.26666667,0.26666667,0.26666667}\makebox(0,0)[rb]{\smash{600}}}%
    \put(0.05309176,0.30014433){\color[rgb]{0.26666667,0.26666667,0.26666667}\makebox(0,0)[rb]{\smash{800}}}%
    \put(0.05309176,0.35698589){\color[rgb]{0.26666667,0.26666667,0.26666667}\makebox(0,0)[rb]{\smash{1000}}}%
    \put(0.05309176,0.41382746){\color[rgb]{0.26666667,0.26666667,0.26666667}\makebox(0,0)[rb]{\smash{1200}}}%
    \put(0.05309176,0.47066388){\color[rgb]{0.26666667,0.26666667,0.26666667}\makebox(0,0)[rb]{\smash{1400}}}%
    \put(0.05309176,0.52750544){\color[rgb]{0.26666667,0.26666667,0.26666667}\makebox(0,0)[rb]{\smash{1600}}}%
    \put(0.5,0.01440329){\color[rgb]{0.26666667,0.26666667,0.26666667}\makebox(0,0)[b]{\smash{Queue length[\# of pasengers]}}}%
    \put(0.02160494,0.29783951){\color[rgb]{0.26666667,0.26666667,0.26666667}\rotatebox{90}{\makebox(0,0)[b]{\smash{Throughput[\#of pasengers/hr]}}}}%
    \put(0.15329218,0.54048354){\color[rgb]{0.26666667,0.26666667,0.26666667}\makebox(0,0)[lb]{\smash{raw}}}%
    \put(0.15329218,0.52556584){\color[rgb]{0.26666667,0.26666667,0.26666667}\makebox(0,0)[lb]{\smash{mean}}}%
    \put(0.15329218,0.51064815){\color[rgb]{0.26666667,0.26666667,0.26666667}\makebox(0,0)[lb]{\smash{Uncertainty}}}%
  \end{picture}%
\endgroup%

%% file: walkTimesgate54.pdf_tex
\begingroup%
  \makeatletter%
  \providecommand\color[2][]{%
    \errmessage{(Inkscape) Color is used for the text in Inkscape, but the package 'color.sty' is not loaded}%
    \renewcommand\color[2][]{}%
  }%
  \providecommand\transparent[1]{%
    \errmessage{(Inkscape) Transparency is used (non-zero) for the text in Inkscape, but the package 'transparent.sty' is not loaded}%
    \renewcommand\transparent[1]{}%
  }%
  \providecommand\rotatebox[2]{#2}%
  \ifx\svgwidth\undefined%
    \setlength{\unitlength}{448bp}%
    \ifx\svgscale\undefined%
      \relax%
    \else%
      \setlength{\unitlength}{\unitlength * \real{\svgscale}}%
    \fi%
  \else%
    \setlength{\unitlength}{\svgwidth}%
  \fi%
  \global\let\svgwidth\undefined%
  \global\let\svgscale\undefined%
  \makeatother%
  \begin{picture}(1,0.75927056)%
    \put(0,0){\includegraphics[width=\unitlength]{walkTimesgate54.pdf}}%
    \put(0.51660714,0.004875){\color[rgb]{0,0,0}\makebox(0,0)[b]{\smash{Time[s]}}}%
    \put(0.51660714,0.71169551){\color[rgb]{0,0,0}\makebox(0,0)[b]{\smash{gate54}}}%
    \put(0.13112358,0.04105691){\color[rgb]{0,0,0}\makebox(0,0)[b]{\smash{-500}}}%
    \put(0.21723429,0.04105691){\color[rgb]{0,0,0}\makebox(0,0)[b]{\smash{0}}}%
    \put(0.30334501,0.04105691){\color[rgb]{0,0,0}\makebox(0,0)[b]{\smash{500}}}%
    \put(0.38945751,0.04105691){\color[rgb]{0,0,0}\makebox(0,0)[b]{\smash{1000}}}%
    \put(0.47556822,0.04105691){\color[rgb]{0,0,0}\makebox(0,0)[b]{\smash{1500}}}%
    \put(0.56167894,0.04105691){\color[rgb]{0,0,0}\makebox(0,0)[b]{\smash{2000}}}%
    \put(0.64778965,0.04105691){\color[rgb]{0,0,0}\makebox(0,0)[b]{\smash{2500}}}%
    \put(0.73390215,0.04105691){\color[rgb]{0,0,0}\makebox(0,0)[b]{\smash{3000}}}%
    \put(0.82001287,0.04105691){\color[rgb]{0,0,0}\makebox(0,0)[b]{\smash{3500}}}%
    \put(0.90612358,0.04105691){\color[rgb]{0,0,0}\makebox(0,0)[b]{\smash{4000}}}%
    \put(0.11674716,0.08052056){\color[rgb]{0,0,0}\makebox(0,0)[rb]{\smash{0}}}%
    \put(0.11562358,0.17340691){\color[rgb]{0,0,0}\makebox(0,0)[rb]{\smash{20}}}%
    \put(0.11562358,0.27528191){\color[rgb]{0,0,0}\makebox(0,0)[rb]{\smash{40}}}%
    \put(0.11562358,0.37715691){\color[rgb]{0,0,0}\makebox(0,0)[rb]{\smash{60}}}%
    \put(0.11562358,0.47903191){\color[rgb]{0,0,0}\makebox(0,0)[rb]{\smash{80}}}%
    \put(0.11562358,0.58090691){\color[rgb]{0,0,0}\makebox(0,0)[rb]{\smash{100}}}%
    \put(0.11562358,0.68278191){\color[rgb]{0,0,0}\makebox(0,0)[rb]{\smash{120}}}%
  \end{picture}%
\endgroup%

%% file: walkTimesgate60.pdf_tex
\begingroup%
  \makeatletter%
  \providecommand\color[2][]{%
    \errmessage{(Inkscape) Color is used for the text in Inkscape, but the package 'color.sty' is not loaded}%
    \renewcommand\color[2][]{}%
  }%
  \providecommand\transparent[1]{%
    \errmessage{(Inkscape) Transparency is used (non-zero) for the text in Inkscape, but the package 'transparent.sty' is not loaded}%
    \renewcommand\transparent[1]{}%
  }%
  \providecommand\rotatebox[2]{#2}%
  \ifx\svgwidth\undefined%
    \setlength{\unitlength}{448bp}%
    \ifx\svgscale\undefined%
      \relax%
    \else%
      \setlength{\unitlength}{\unitlength * \real{\svgscale}}%
    \fi%
  \else%
    \setlength{\unitlength}{\svgwidth}%
  \fi%
  \global\let\svgwidth\undefined%
  \global\let\svgscale\undefined%
  \makeatother%
  \begin{picture}(1,0.75)%
    \put(0,0){\includegraphics[width=\unitlength]{walkTimesgate60.pdf}}%
    \put(0.51660714,0.00602237){\color[rgb]{0,0,0}\makebox(0,0)[b]{\smash{Time[s]}}}%
    \put(0.51660714,0.70804286){\color[rgb]{0,0,0}\makebox(0,0)[b]{\smash{gate60}}}%
    \put(0.13,0.040775){\color[rgb]{0,0,0}\makebox(0,0)[b]{\smash{-5000}}}%
    \put(0.285,0.040775){\color[rgb]{0,0,0}\makebox(0,0)[b]{\smash{0}}}%
    \put(0.44,0.040775){\color[rgb]{0,0,0}\makebox(0,0)[b]{\smash{5000}}}%
    \put(0.595,0.040775){\color[rgb]{0,0,0}\makebox(0,0)[b]{\smash{10000}}}%
    \put(0.75,0.040775){\color[rgb]{0,0,0}\makebox(0,0)[b]{\smash{15000}}}%
    \put(0.905,0.040775){\color[rgb]{0,0,0}\makebox(0,0)[b]{\smash{20000}}}%
    \put(0.11608898,0.08078389){\color[rgb]{0,0,0}\makebox(0,0)[rb]{\smash{0}}}%
    \put(0.1145,0.15857143){\color[rgb]{0,0,0}\makebox(0,0)[rb]{\smash{20}}}%
    \put(0.1145,0.24589286){\color[rgb]{0,0,0}\makebox(0,0)[rb]{\smash{40}}}%
    \put(0.1145,0.33321429){\color[rgb]{0,0,0}\makebox(0,0)[rb]{\smash{60}}}%
    \put(0.1145,0.42053571){\color[rgb]{0,0,0}\makebox(0,0)[rb]{\smash{80}}}%
    \put(0.1145,0.50785714){\color[rgb]{0,0,0}\makebox(0,0)[rb]{\smash{100}}}%
    \put(0.1145,0.59517857){\color[rgb]{0,0,0}\makebox(0,0)[rb]{\smash{120}}}%
    \put(0.1145,0.6825){\color[rgb]{0,0,0}\makebox(0,0)[rb]{\smash{140}}}%
  \end{picture}%
\endgroup%

%% file: paramWalkSpeed_gate53.pdf_tex
\begingroup%
  \makeatletter%
  \providecommand\color[2][]{%
    \errmessage{(Inkscape) Color is used for the text in Inkscape, but the package 'color.sty' is not loaded}%
    \renewcommand\color[2][]{}%
  }%
  \providecommand\transparent[1]{%
    \errmessage{(Inkscape) Transparency is used (non-zero) for the text in Inkscape, but the package 'transparent.sty' is not loaded}%
    \renewcommand\transparent[1]{}%
  }%
  \providecommand\rotatebox[2]{#2}%
  \ifx\svgwidth\undefined%
    \setlength{\unitlength}{384bp}%
    \ifx\svgscale\undefined%
      \relax%
    \else%
      \setlength{\unitlength}{\unitlength * \real{\svgscale}}%
    \fi%
  \else%
    \setlength{\unitlength}{\svgwidth}%
  \fi%
  \global\let\svgwidth\undefined%
  \global\let\svgscale\undefined%
  \makeatother%
  \begin{picture}(1,1)%
    \put(0,0){\includegraphics[width=\unitlength]{paramWalkSpeed_gate53.pdf}}%
  \end{picture}%
\endgroup%

%% file: walkSpeedGate54.pdf_tex
\begingroup%
  \makeatletter%
  \providecommand\color[2][]{%
    \errmessage{(Inkscape) Color is used for the text in Inkscape, but the package 'color.sty' is not loaded}%
    \renewcommand\color[2][]{}%
  }%
  \providecommand\transparent[1]{%
    \errmessage{(Inkscape) Transparency is used (non-zero) for the text in Inkscape, but the package 'transparent.sty' is not loaded}%
    \renewcommand\transparent[1]{}%
  }%
  \providecommand\rotatebox[2]{#2}%
  \ifx\svgwidth\undefined%
    \setlength{\unitlength}{448bp}%
    \ifx\svgscale\undefined%
      \relax%
    \else%
      \setlength{\unitlength}{\unitlength * \real{\svgscale}}%
    \fi%
  \else%
    \setlength{\unitlength}{\svgwidth}%
  \fi%
  \global\let\svgwidth\undefined%
  \global\let\svgscale\undefined%
  \makeatother%
  \begin{picture}(1,0.75)%
    \put(0,0){\includegraphics[width=\unitlength]{walkSpeedGate54.pdf}}%
  \end{picture}%
\endgroup%

%% file: walkSpeedGate60.pdf_tex
\begingroup%
  \makeatletter%
  \providecommand\color[2][]{%
    \errmessage{(Inkscape) Color is used for the text in Inkscape, but the package 'color.sty' is not loaded}%
    \renewcommand\color[2][]{}%
  }%
  \providecommand\transparent[1]{%
    \errmessage{(Inkscape) Transparency is used (non-zero) for the text in Inkscape, but the package 'transparent.sty' is not loaded}%
    \renewcommand\transparent[1]{}%
  }%
  \providecommand\rotatebox[2]{#2}%
  \ifx\svgwidth\undefined%
    \setlength{\unitlength}{448.039bp}%
    \ifx\svgscale\undefined%
      \relax%
    \else%
      \setlength{\unitlength}{\unitlength * \real{\svgscale}}%
    \fi%
  \else%
    \setlength{\unitlength}{\svgwidth}%
  \fi%
  \global\let\svgwidth\undefined%
  \global\let\svgscale\undefined%
  \makeatother%
  \begin{picture}(1,0.74999944)%
    \put(0,0){\includegraphics[width=\unitlength]{walkSpeedGate60.pdf}}%
  \end{picture}%
\endgroup%

%% file: walkSpeed_allGates.pdf_tex
\begingroup%
  \makeatletter%
  \providecommand\color[2][]{%
    \errmessage{(Inkscape) Color is used for the text in Inkscape, but the package 'color.sty' is not loaded}%
    \renewcommand\color[2][]{}%
  }%
  \providecommand\transparent[1]{%
    \errmessage{(Inkscape) Transparency is used (non-zero) for the text in Inkscape, but the package 'transparent.sty' is not loaded}%
    \renewcommand\transparent[1]{}%
  }%
  \providecommand\rotatebox[2]{#2}%
  \ifx\svgwidth\undefined%
    \setlength{\unitlength}{479.991bp}%
    \ifx\svgscale\undefined%
      \relax%
    \else%
      \setlength{\unitlength}{\unitlength * \real{\svgscale}}%
    \fi%
  \else%
    \setlength{\unitlength}{\svgwidth}%
  \fi%
  \global\let\svgwidth\undefined%
  \global\let\svgscale\undefined%
  \makeatother%
  \begin{picture}(1,1)%
    \put(0,0){\includegraphics[width=\unitlength]{walkSpeed_allGates.pdf}}%
  \end{picture}%
\endgroup%

%% file: svceRateDistribution_perDesk.pdf_tex
\begingroup%
  \makeatletter%
  \providecommand\color[2][]{%
    \errmessage{(Inkscape) Color is used for the text in Inkscape, but the package 'color.sty' is not loaded}%
    \renewcommand\color[2][]{}%
  }%
  \providecommand\transparent[1]{%
    \errmessage{(Inkscape) Transparency is used (non-zero) for the text in Inkscape, but the package 'transparent.sty' is not loaded}%
    \renewcommand\transparent[1]{}%
  }%
  \providecommand\rotatebox[2]{#2}%
  \ifx\svgwidth\undefined%
    \setlength{\unitlength}{448bp}%
    \ifx\svgscale\undefined%
      \relax%
    \else%
      \setlength{\unitlength}{\unitlength * \real{\svgscale}}%
    \fi%
  \else%
    \setlength{\unitlength}{\svgwidth}%
  \fi%
  \global\let\svgwidth\undefined%
  \global\let\svgscale\undefined%
  \makeatother%
  \begin{picture}(1,0.75237912)%
    \put(0,0){\includegraphics[width=\unitlength]{svceRateDistribution_perDesk.pdf}}%
    \put(0.51422367,0.0068125){\color[rgb]{0,0,0}\makebox(0,0)[b]{\smash{Service rate [\#passengers/(hr*desk)]}}}%
    \put(0.05696731,0.38861825){\color[rgb]{0,0,0}\rotatebox{90}{\makebox(0,0)[b]{\smash{ECDF}}}}%
    \put(0.13,0.04394862){\color[rgb]{0,0,0}\makebox(0,0)[b]{\smash{0}}}%
    \put(0.24071429,0.0455376){\color[rgb]{0,0,0}\makebox(0,0)[b]{\smash{10}}}%
    \put(0.35142857,0.04792106){\color[rgb]{0,0,0}\makebox(0,0)[b]{\smash{20}}}%
    \put(0.45975938,0.04871556){\color[rgb]{0,0,0}\makebox(0,0)[b]{\smash{30}}}%
    \put(0.57285714,0.04792106){\color[rgb]{0,0,0}\makebox(0,0)[b]{\smash{40}}}%
    \put(0.68436593,0.04951004){\color[rgb]{0,0,0}\makebox(0,0)[b]{\smash{50}}}%
    \put(0.7966692,0.05030454){\color[rgb]{0,0,0}\makebox(0,0)[b]{\smash{60}}}%
    \put(0.90182204,0.05109903){\color[rgb]{0,0,0}\makebox(0,0)[b]{\smash{70}}}%
    \put(0.1145,0.07362912){\color[rgb]{0,0,0}\makebox(0,0)[rb]{\smash{0}}}%
    \put(0.1145,0.13475412){\color[rgb]{0,0,0}\makebox(0,0)[rb]{\smash{0.1}}}%
    \put(0.1145,0.19587912){\color[rgb]{0,0,0}\makebox(0,0)[rb]{\smash{0.2}}}%
    \put(0.1145,0.25700412){\color[rgb]{0,0,0}\makebox(0,0)[rb]{\smash{0.3}}}%
    \put(0.1145,0.31812912){\color[rgb]{0,0,0}\makebox(0,0)[rb]{\smash{0.4}}}%
    \put(0.1145,0.37925412){\color[rgb]{0,0,0}\makebox(0,0)[rb]{\smash{0.5}}}%
    \put(0.1145,0.44037912){\color[rgb]{0,0,0}\makebox(0,0)[rb]{\smash{0.6}}}%
    \put(0.1145,0.50150412){\color[rgb]{0,0,0}\makebox(0,0)[rb]{\smash{0.7}}}%
    \put(0.1145,0.56262912){\color[rgb]{0,0,0}\makebox(0,0)[rb]{\smash{0.8}}}%
    \put(0.1145,0.62375412){\color[rgb]{0,0,0}\makebox(0,0)[rb]{\smash{0.9}}}%
    \put(0.1145,0.68487912){\color[rgb]{0,0,0}\makebox(0,0)[rb]{\smash{1}}}%
  \end{picture}%
\endgroup%

%% file: deskOcc_26thNovember.pdf_tex
\begingroup%
  \makeatletter%
  \providecommand\color[2][]{%
    \errmessage{(Inkscape) Color is used for the text in Inkscape, but the package 'color.sty' is not loaded}%
    \renewcommand\color[2][]{}%
  }%
  \providecommand\transparent[1]{%
    \errmessage{(Inkscape) Transparency is used (non-zero) for the text in Inkscape, but the package 'transparent.sty' is not loaded}%
    \renewcommand\transparent[1]{}%
  }%
  \providecommand\rotatebox[2]{#2}%
  \ifx\svgwidth\undefined%
    \setlength{\unitlength}{448bp}%
    \ifx\svgscale\undefined%
      \relax%
    \else%
      \setlength{\unitlength}{\unitlength * \real{\svgscale}}%
    \fi%
  \else%
    \setlength{\unitlength}{\svgwidth}%
  \fi%
  \global\let\svgwidth\undefined%
  \global\let\svgscale\undefined%
  \makeatother%
  \begin{picture}(1,0.75)%
    \put(0,0){\includegraphics[width=\unitlength]{deskOcc_26thNovember.pdf}}%
    \put(0.51660714,0.00999482){\color[rgb]{0,0,0}\makebox(0,0)[b]{\smash{Time(hr)}}}%
    \put(0.06688711,0.38862261){\color[rgb]{0,0,0}\rotatebox{90}{\makebox(0,0)[b]{\smash{\#desks/hr}}}}%
    \put(0.13,0.040775){\color[rgb]{0,0,0}\makebox(0,0)[b]{\smash{4}}}%
    \put(0.2075,0.040775){\color[rgb]{0,0,0}\makebox(0,0)[b]{\smash{6}}}%
    \put(0.285,0.040775){\color[rgb]{0,0,0}\makebox(0,0)[b]{\smash{8}}}%
    \put(0.3625,0.040775){\color[rgb]{0,0,0}\makebox(0,0)[b]{\smash{10}}}%
    \put(0.44,0.040775){\color[rgb]{0,0,0}\makebox(0,0)[b]{\smash{12}}}%
    \put(0.5175,0.040775){\color[rgb]{0,0,0}\makebox(0,0)[b]{\smash{14}}}%
    \put(0.595,0.040775){\color[rgb]{0,0,0}\makebox(0,0)[b]{\smash{16}}}%
    \put(0.6725,0.040775){\color[rgb]{0,0,0}\makebox(0,0)[b]{\smash{18}}}%
    \put(0.75,0.040775){\color[rgb]{0,0,0}\makebox(0,0)[b]{\smash{20}}}%
    \put(0.8275,0.040775){\color[rgb]{0,0,0}\makebox(0,0)[b]{\smash{22}}}%
    \put(0.905,0.040775){\color[rgb]{0,0,0}\makebox(0,0)[b]{\smash{24}}}%
    \put(0.1145,0.07125){\color[rgb]{0,0,0}\makebox(0,0)[rb]{\smash{0}}}%
    \put(0.1145,0.14071071){\color[rgb]{0,0,0}\makebox(0,0)[rb]{\smash{5}}}%
    \put(0.1145,0.21016964){\color[rgb]{0,0,0}\makebox(0,0)[rb]{\smash{10}}}%
    \put(0.1145,0.27963036){\color[rgb]{0,0,0}\makebox(0,0)[rb]{\smash{15}}}%
    \put(0.1145,0.34909107){\color[rgb]{0,0,0}\makebox(0,0)[rb]{\smash{20}}}%
    \put(0.1145,0.41855179){\color[rgb]{0,0,0}\makebox(0,0)[rb]{\smash{25}}}%
    \put(0.1145,0.48801071){\color[rgb]{0,0,0}\makebox(0,0)[rb]{\smash{30}}}%
    \put(0.1145,0.55747143){\color[rgb]{0,0,0}\makebox(0,0)[rb]{\smash{35}}}%
    \put(0.1145,0.62693214){\color[rgb]{0,0,0}\makebox(0,0)[rb]{\smash{40}}}%
  \end{picture}%
\endgroup%

%% file: 25thJul_waitTimes.pdf_tex
\begingroup%
  \makeatletter%
  \providecommand\color[2][]{%
    \errmessage{(Inkscape) Color is used for the text in Inkscape, but the package 'color.sty' is not loaded}%
    \renewcommand\color[2][]{}%
  }%
  \providecommand\transparent[1]{%
    \errmessage{(Inkscape) Transparency is used (non-zero) for the text in Inkscape, but the package 'transparent.sty' is not loaded}%
    \renewcommand\transparent[1]{}%
  }%
  \providecommand\rotatebox[2]{#2}%
  \ifx\svgwidth\undefined%
    \setlength{\unitlength}{968.8bp}%
    \ifx\svgscale\undefined%
      \relax%
    \else%
      \setlength{\unitlength}{\unitlength * \real{\svgscale}}%
    \fi%
  \else%
    \setlength{\unitlength}{\svgwidth}%
  \fi%
  \global\let\svgwidth\undefined%
  \global\let\svgscale\undefined%
  \makeatother%
  \begin{picture}(1,0.75557391)%
    \put(0,0){\includegraphics[width=\unitlength]{25thJul_waitTimes.pdf}}%
    \put(0.51667465,0.05435921){\color[rgb]{0,0,0}\makebox(0,0)[b]{\smash{Time[hr]}}}%
    \put(0.10358712,0.39018332){\color[rgb]{0,0,0}\rotatebox{90}{\makebox(0,0)[b]{\smash{Wait times[min]}}}}%
    \put(0.13,0.05715524){\color[rgb]{0,0,0}\makebox(0,0)[b]{\smash{0}}}%
    \put(0.285,0.05715524){\color[rgb]{0,0,0}\makebox(0,0)[b]{\smash{5}}}%
    \put(0.44,0.05715524){\color[rgb]{0,0,0}\makebox(0,0)[b]{\smash{10}}}%
    \put(0.595,0.05715524){\color[rgb]{0,0,0}\makebox(0,0)[b]{\smash{15}}}%
    \put(0.75,0.05715524){\color[rgb]{0,0,0}\makebox(0,0)[b]{\smash{20}}}%
    \put(0.905,0.05715524){\color[rgb]{0,0,0}\makebox(0,0)[b]{\smash{25}}}%
    \put(0.11450041,0.07791082){\color[rgb]{0,0,0}\makebox(0,0)[rb]{\smash{0}}}%
    \put(0.11450041,0.20106936){\color[rgb]{0,0,0}\makebox(0,0)[rb]{\smash{50}}}%
    \put(0.11450041,0.32422791){\color[rgb]{0,0,0}\makebox(0,0)[rb]{\smash{100}}}%
    \put(0.11450041,0.44738646){\color[rgb]{0,0,0}\makebox(0,0)[rb]{\smash{150}}}%
    \put(0.11450041,0.570545){\color[rgb]{0,0,0}\makebox(0,0)[rb]{\smash{200}}}%
    \put(0.11450041,0.69370355){\color[rgb]{0,0,0}\makebox(0,0)[rb]{\smash{250}}}%
    \put(0.84095045,0.67827746){\color[rgb]{0,0,0}\makebox(0,0)[lb]{\smash{upper bound}}}%
    \put(0.84095045,0.665){\color[rgb]{0,0,0}\makebox(0,0)[lb]{\smash{lower bound}}}%
    \put(0.84095045,0.65172337){\color[rgb]{0,0,0}\makebox(0,0)[lb]{\smash{actual}}}%
  \end{picture}%
\endgroup%

%% file: queueOn25thJul.pdf_tex
\begingroup%
  \makeatletter%
  \providecommand\color[2][]{%
    \errmessage{(Inkscape) Color is used for the text in Inkscape, but the package 'color.sty' is not loaded}%
    \renewcommand\color[2][]{}%
  }%
  \providecommand\transparent[1]{%
    \errmessage{(Inkscape) Transparency is used (non-zero) for the text in Inkscape, but the package 'transparent.sty' is not loaded}%
    \renewcommand\transparent[1]{}%
  }%
  \providecommand\rotatebox[2]{#2}%
  \ifx\svgwidth\undefined%
    \setlength{\unitlength}{748bp}%
    \ifx\svgscale\undefined%
      \relax%
    \else%
      \setlength{\unitlength}{\unitlength * \real{\svgscale}}%
    \fi%
  \else%
    \setlength{\unitlength}{\svgwidth}%
  \fi%
  \global\let\svgwidth\undefined%
  \global\let\svgscale\undefined%
  \makeatother%
  \begin{picture}(1,0.72406417)%
    \put(0,0){\includegraphics[width=\unitlength]{queueOn25thJul.pdf}}%
    \put(0.51696471,0.0424385){\color[rgb]{0,0,0}\makebox(0,0)[b]{\smash{Time of day(hour)}}}%
    \put(0.09574866,0.37309947){\color[rgb]{0,0,0}\rotatebox{90}{\makebox(0,0)[b]{\smash{Queue Length}}}}%
    \put(0.13,0.05017647){\color[rgb]{0,0,0}\makebox(0,0)[b]{\smash{4}}}%
    \put(0.20750053,0.05017647){\color[rgb]{0,0,0}\makebox(0,0)[b]{\smash{6}}}%
    \put(0.285,0.05017647){\color[rgb]{0,0,0}\makebox(0,0)[b]{\smash{8}}}%
    \put(0.36250053,0.05017647){\color[rgb]{0,0,0}\makebox(0,0)[b]{\smash{10}}}%
    \put(0.44,0.05017647){\color[rgb]{0,0,0}\makebox(0,0)[b]{\smash{12}}}%
    \put(0.51750053,0.05017647){\color[rgb]{0,0,0}\makebox(0,0)[b]{\smash{14}}}%
    \put(0.595,0.05017647){\color[rgb]{0,0,0}\makebox(0,0)[b]{\smash{16}}}%
    \put(0.67250053,0.05017647){\color[rgb]{0,0,0}\makebox(0,0)[b]{\smash{18}}}%
    \put(0.75,0.05017647){\color[rgb]{0,0,0}\makebox(0,0)[b]{\smash{20}}}%
    \put(0.82750053,0.05017647){\color[rgb]{0,0,0}\makebox(0,0)[b]{\smash{22}}}%
    \put(0.905,0.05017647){\color[rgb]{0,0,0}\makebox(0,0)[b]{\smash{24}}}%
    \put(0.11450053,0.07290909){\color[rgb]{0,0,0}\makebox(0,0)[rb]{\smash{0}}}%
    \put(0.11450053,0.1572107){\color[rgb]{0,0,0}\makebox(0,0)[rb]{\smash{100}}}%
    \put(0.11450053,0.2415123){\color[rgb]{0,0,0}\makebox(0,0)[rb]{\smash{200}}}%
    \put(0.11450053,0.3258139){\color[rgb]{0,0,0}\makebox(0,0)[rb]{\smash{300}}}%
    \put(0.11450053,0.41011658){\color[rgb]{0,0,0}\makebox(0,0)[rb]{\smash{400}}}%
    \put(0.11450053,0.49441818){\color[rgb]{0,0,0}\makebox(0,0)[rb]{\smash{500}}}%
    \put(0.11450053,0.57871979){\color[rgb]{0,0,0}\makebox(0,0)[rb]{\smash{600}}}%
    \put(0.11450053,0.66302139){\color[rgb]{0,0,0}\makebox(0,0)[rb]{\smash{700}}}%
  \end{picture}%
\endgroup%

%% file: waitTimes_26_07.pdf_tex
\begingroup%
  \makeatletter%
  \providecommand\color[2][]{%
    \errmessage{(Inkscape) Color is used for the text in Inkscape, but the package 'color.sty' is not loaded}%
    \renewcommand\color[2][]{}%
  }%
  \providecommand\transparent[1]{%
    \errmessage{(Inkscape) Transparency is used (non-zero) for the text in Inkscape, but the package 'transparent.sty' is not loaded}%
    \renewcommand\transparent[1]{}%
  }%
  \providecommand\rotatebox[2]{#2}%
  \ifx\svgwidth\undefined%
    \setlength{\unitlength}{448bp}%
    \ifx\svgscale\undefined%
      \relax%
    \else%
      \setlength{\unitlength}{\unitlength * \real{\svgscale}}%
    \fi%
  \else%
    \setlength{\unitlength}{\svgwidth}%
  \fi%
  \global\let\svgwidth\undefined%
  \global\let\svgscale\undefined%
  \makeatother%
  \begin{picture}(1,0.75282507)%
    \put(0,0){\includegraphics[width=\unitlength]{waitTimes_26_07.pdf}}%
    \put(0.5304401,0.00590907){\color[rgb]{0,0,0}\makebox(0,0)[b]{\smash{Time[hr]}}}%
    \put(0.03841901,0.38831349){\color[rgb]{0,0,0}\rotatebox{90}{\makebox(0,0)[b]{\smash{Wait time[min]}}}}%
    \put(0.15018916,0.03644965){\color[rgb]{0,0,0}\makebox(0,0)[b]{\smash{0}}}%
    \put(0.29940345,0.03644965){\color[rgb]{0,0,0}\makebox(0,0)[b]{\smash{5}}}%
    \put(0.44861774,0.03644965){\color[rgb]{0,0,0}\makebox(0,0)[b]{\smash{10}}}%
    \put(0.59783202,0.03644965){\color[rgb]{0,0,0}\makebox(0,0)[b]{\smash{15}}}%
    \put(0.74704631,0.03644965){\color[rgb]{0,0,0}\makebox(0,0)[b]{\smash{20}}}%
    \put(0.89626059,0.03644965){\color[rgb]{0,0,0}\makebox(0,0)[b]{\smash{25}}}%
    \put(0.13526774,0.06692465){\color[rgb]{0,0,0}\makebox(0,0)[rb]{\smash{0}}}%
    \put(0.13526774,0.15424608){\color[rgb]{0,0,0}\makebox(0,0)[rb]{\smash{100}}}%
    \put(0.13526774,0.2415675){\color[rgb]{0,0,0}\makebox(0,0)[rb]{\smash{200}}}%
    \put(0.13526774,0.32888893){\color[rgb]{0,0,0}\makebox(0,0)[rb]{\smash{300}}}%
    \put(0.13526774,0.41621036){\color[rgb]{0,0,0}\makebox(0,0)[rb]{\smash{400}}}%
    \put(0.13526774,0.50353179){\color[rgb]{0,0,0}\makebox(0,0)[rb]{\smash{500}}}%
    \put(0.13526774,0.59085322){\color[rgb]{0,0,0}\makebox(0,0)[rb]{\smash{600}}}%
    \put(0.13526774,0.67817465){\color[rgb]{0,0,0}\makebox(0,0)[rb]{\smash{700}}}%
    \put(0.25042616,0.65061277){\color[rgb]{0,0,0}\makebox(0,0)[lb]{\smash{upper bound}}}%
    \put(0.25042616,0.61067495){\color[rgb]{0,0,0}\makebox(0,0)[lb]{\smash{lower bound}}}%
    \put(0.25042616,0.57273381){\color[rgb]{0,0,0}\makebox(0,0)[lb]{\smash{actual}}}%
  \end{picture}%
\endgroup%

%% file: queueLength_26_07.pdf_tex
\begingroup%
  \makeatletter%
  \providecommand\color[2][]{%
    \errmessage{(Inkscape) Color is used for the text in Inkscape, but the package 'color.sty' is not loaded}%
    \renewcommand\color[2][]{}%
  }%
  \providecommand\transparent[1]{%
    \errmessage{(Inkscape) Transparency is used (non-zero) for the text in Inkscape, but the package 'transparent.sty' is not loaded}%
    \renewcommand\transparent[1]{}%
  }%
  \providecommand\rotatebox[2]{#2}%
  \ifx\svgwidth\undefined%
    \setlength{\unitlength}{448bp}%
    \ifx\svgscale\undefined%
      \relax%
    \else%
      \setlength{\unitlength}{\unitlength * \real{\svgscale}}%
    \fi%
  \else%
    \setlength{\unitlength}{\svgwidth}%
  \fi%
  \global\let\svgwidth\undefined%
  \global\let\svgscale\undefined%
  \makeatother%
  \begin{picture}(1,0.7523791)%
    \put(0,0){\includegraphics[width=\unitlength]{queueLength_26_07.pdf}}%
    \put(0.53352696,0.00854257){\color[rgb]{0,0,0}\makebox(0,0)[b]{\smash{Time of day[hr]}}}%
    \put(0.05745461,0.39932468){\color[rgb]{0,0,0}\rotatebox{90}{\makebox(0,0)[b]{\smash{Queue Length}}}}%
    \put(0.15178571,0.0431541){\color[rgb]{0,0,0}\makebox(0,0)[b]{\smash{4}}}%
    \put(0.22710714,0.0431541){\color[rgb]{0,0,0}\makebox(0,0)[b]{\smash{6}}}%
    \put(0.30242857,0.0431541){\color[rgb]{0,0,0}\makebox(0,0)[b]{\smash{8}}}%
    \put(0.37775,0.0431541){\color[rgb]{0,0,0}\makebox(0,0)[b]{\smash{10}}}%
    \put(0.45307143,0.0431541){\color[rgb]{0,0,0}\makebox(0,0)[b]{\smash{12}}}%
    \put(0.52839286,0.0431541){\color[rgb]{0,0,0}\makebox(0,0)[b]{\smash{14}}}%
    \put(0.60371429,0.0431541){\color[rgb]{0,0,0}\makebox(0,0)[b]{\smash{16}}}%
    \put(0.67903571,0.0431541){\color[rgb]{0,0,0}\makebox(0,0)[b]{\smash{18}}}%
    \put(0.75435714,0.0431541){\color[rgb]{0,0,0}\makebox(0,0)[b]{\smash{20}}}%
    \put(0.82967857,0.0431541){\color[rgb]{0,0,0}\makebox(0,0)[b]{\smash{22}}}%
    \put(0.905,0.0431541){\color[rgb]{0,0,0}\makebox(0,0)[b]{\smash{24}}}%
    \put(0.13672143,0.0736291){\color[rgb]{0,0,0}\makebox(0,0)[rb]{\smash{0}}}%
    \put(0.13672143,0.14154518){\color[rgb]{0,0,0}\makebox(0,0)[rb]{\smash{50}}}%
    \put(0.13672143,0.20946303){\color[rgb]{0,0,0}\makebox(0,0)[rb]{\smash{100}}}%
    \put(0.13672143,0.2773791){\color[rgb]{0,0,0}\makebox(0,0)[rb]{\smash{150}}}%
    \put(0.13672143,0.34529518){\color[rgb]{0,0,0}\makebox(0,0)[rb]{\smash{200}}}%
    \put(0.13672143,0.41321303){\color[rgb]{0,0,0}\makebox(0,0)[rb]{\smash{250}}}%
    \put(0.13672143,0.4811291){\color[rgb]{0,0,0}\makebox(0,0)[rb]{\smash{300}}}%
    \put(0.13672143,0.54904518){\color[rgb]{0,0,0}\makebox(0,0)[rb]{\smash{350}}}%
    \put(0.13672143,0.61696303){\color[rgb]{0,0,0}\makebox(0,0)[rb]{\smash{400}}}%
    \put(0.13672143,0.6848791){\color[rgb]{0,0,0}\makebox(0,0)[rb]{\smash{450}}}%
  \end{picture}%
\endgroup%

%% file: waitTimes_11_12.pdf_tex
\begingroup%
  \makeatletter%
  \providecommand\color[2][]{%
    \errmessage{(Inkscape) Color is used for the text in Inkscape, but the package 'color.sty' is not loaded}%
    \renewcommand\color[2][]{}%
  }%
  \providecommand\transparent[1]{%
    \errmessage{(Inkscape) Transparency is used (non-zero) for the text in Inkscape, but the package 'transparent.sty' is not loaded}%
    \renewcommand\transparent[1]{}%
  }%
  \providecommand\rotatebox[2]{#2}%
  \ifx\svgwidth\undefined%
    \setlength{\unitlength}{448bp}%
    \ifx\svgscale\undefined%
      \relax%
    \else%
      \setlength{\unitlength}{\unitlength * \real{\svgscale}}%
    \fi%
  \else%
    \setlength{\unitlength}{\svgwidth}%
  \fi%
  \global\let\svgwidth\undefined%
  \global\let\svgscale\undefined%
  \makeatother%
  \begin{picture}(1,0.75158462)%
    \put(0,0){\includegraphics[width=\unitlength]{waitTimes_11_12.pdf}}%
    \put(0.53225938,0.0068125){\color[rgb]{0,0,0}\makebox(0,0)[b]{\smash{Time of day[hr]}}}%
    \put(0.07320145,0.3838513){\color[rgb]{0,0,0}\rotatebox{90}{\makebox(0,0)[b]{\smash{Wait time[min]}}}}%
    \put(0.16607143,0.04235962){\color[rgb]{0,0,0}\makebox(0,0)[b]{\smash{6}}}%
    \put(0.248175,0.04235962){\color[rgb]{0,0,0}\makebox(0,0)[b]{\smash{8}}}%
    \put(0.33027857,0.04235962){\color[rgb]{0,0,0}\makebox(0,0)[b]{\smash{10}}}%
    \put(0.41238036,0.04235962){\color[rgb]{0,0,0}\makebox(0,0)[b]{\smash{12}}}%
    \put(0.49448393,0.04235962){\color[rgb]{0,0,0}\makebox(0,0)[b]{\smash{14}}}%
    \put(0.5765875,0.04235962){\color[rgb]{0,0,0}\makebox(0,0)[b]{\smash{16}}}%
    \put(0.65869107,0.04235962){\color[rgb]{0,0,0}\makebox(0,0)[b]{\smash{18}}}%
    \put(0.74079286,0.04235962){\color[rgb]{0,0,0}\makebox(0,0)[b]{\smash{20}}}%
    \put(0.82289643,0.04235962){\color[rgb]{0,0,0}\makebox(0,0)[b]{\smash{22}}}%
    \put(0.905,0.04235962){\color[rgb]{0,0,0}\makebox(0,0)[b]{\smash{24}}}%
    \put(0.15129286,0.07283462){\color[rgb]{0,0,0}\makebox(0,0)[rb]{\smash{0}}}%
    \put(0.15129286,0.17470962){\color[rgb]{0,0,0}\makebox(0,0)[rb]{\smash{50}}}%
    \put(0.15129286,0.27658462){\color[rgb]{0,0,0}\makebox(0,0)[rb]{\smash{100}}}%
    \put(0.15129286,0.37845962){\color[rgb]{0,0,0}\makebox(0,0)[rb]{\smash{150}}}%
    \put(0.15129286,0.48033462){\color[rgb]{0,0,0}\makebox(0,0)[rb]{\smash{200}}}%
    \put(0.15129286,0.58220962){\color[rgb]{0,0,0}\makebox(0,0)[rb]{\smash{250}}}%
    \put(0.15129286,0.68408462){\color[rgb]{0,0,0}\makebox(0,0)[rb]{\smash{300}}}%
    \put(0.27884921,0.63708853){\color[rgb]{0,0,0}\makebox(0,0)[lb]{\smash{upper bound}}}%
    \put(0.27884921,0.60179924){\color[rgb]{0,0,0}\makebox(0,0)[lb]{\smash{lower bound}}}%
    \put(0.27884921,0.56827424){\color[rgb]{0,0,0}\makebox(0,0)[lb]{\smash{actual}}}%
  \end{picture}%
\endgroup%

%% file: queueLength_11_12.pdf_tex
\begingroup%
  \makeatletter%
  \providecommand\color[2][]{%
    \errmessage{(Inkscape) Color is used for the text in Inkscape, but the package 'color.sty' is not loaded}%
    \renewcommand\color[2][]{}%
  }%
  \providecommand\transparent[1]{%
    \errmessage{(Inkscape) Transparency is used (non-zero) for the text in Inkscape, but the package 'transparent.sty' is not loaded}%
    \renewcommand\transparent[1]{}%
  }%
  \providecommand\rotatebox[2]{#2}%
  \ifx\svgwidth\undefined%
    \setlength{\unitlength}{448bp}%
    \ifx\svgscale\undefined%
      \relax%
    \else%
      \setlength{\unitlength}{\unitlength * \real{\svgscale}}%
    \fi%
  \else%
    \setlength{\unitlength}{\svgwidth}%
  \fi%
  \global\let\svgwidth\undefined%
  \global\let\svgscale\undefined%
  \makeatother%
  \begin{picture}(1,0.75158462)%
    \put(0,0){\includegraphics[width=\unitlength]{queueLength_11_12.pdf}}%
    \put(0.52670551,0.0068125){\color[rgb]{0,0,0}\makebox(0,0)[b]{\smash{Time of day[hr]}}}%
    \put(0.0628882,0.38861825){\color[rgb]{0,0,0}\rotatebox{90}{\makebox(0,0)[b]{\smash{Queue Length[\#passengers]}}}}%
    \put(0.15178571,0.04235962){\color[rgb]{0,0,0}\makebox(0,0)[b]{\smash{4}}}%
    \put(0.22710714,0.04235962){\color[rgb]{0,0,0}\makebox(0,0)[b]{\smash{6}}}%
    \put(0.30242857,0.04235962){\color[rgb]{0,0,0}\makebox(0,0)[b]{\smash{8}}}%
    \put(0.37775,0.04235962){\color[rgb]{0,0,0}\makebox(0,0)[b]{\smash{10}}}%
    \put(0.45307143,0.04235962){\color[rgb]{0,0,0}\makebox(0,0)[b]{\smash{12}}}%
    \put(0.52839286,0.04235962){\color[rgb]{0,0,0}\makebox(0,0)[b]{\smash{14}}}%
    \put(0.60371429,0.04235962){\color[rgb]{0,0,0}\makebox(0,0)[b]{\smash{16}}}%
    \put(0.67903571,0.04235962){\color[rgb]{0,0,0}\makebox(0,0)[b]{\smash{18}}}%
    \put(0.75435714,0.04235962){\color[rgb]{0,0,0}\makebox(0,0)[b]{\smash{20}}}%
    \put(0.82967857,0.04235962){\color[rgb]{0,0,0}\makebox(0,0)[b]{\smash{22}}}%
    \put(0.905,0.04235962){\color[rgb]{0,0,0}\makebox(0,0)[b]{\smash{24}}}%
    \put(0.13672143,0.07283462){\color[rgb]{0,0,0}\makebox(0,0)[rb]{\smash{0}}}%
    \put(0.13672143,0.1407507){\color[rgb]{0,0,0}\makebox(0,0)[rb]{\smash{50}}}%
    \put(0.13672143,0.20866855){\color[rgb]{0,0,0}\makebox(0,0)[rb]{\smash{100}}}%
    \put(0.13672143,0.27658462){\color[rgb]{0,0,0}\makebox(0,0)[rb]{\smash{150}}}%
    \put(0.13672143,0.3445007){\color[rgb]{0,0,0}\makebox(0,0)[rb]{\smash{200}}}%
    \put(0.13672143,0.41241855){\color[rgb]{0,0,0}\makebox(0,0)[rb]{\smash{250}}}%
    \put(0.13672143,0.48033462){\color[rgb]{0,0,0}\makebox(0,0)[rb]{\smash{300}}}%
    \put(0.13672143,0.5482507){\color[rgb]{0,0,0}\makebox(0,0)[rb]{\smash{350}}}%
    \put(0.13672143,0.61616855){\color[rgb]{0,0,0}\makebox(0,0)[rb]{\smash{400}}}%
    \put(0.13672143,0.68408462){\color[rgb]{0,0,0}\makebox(0,0)[rb]{\smash{450}}}%
  \end{picture}%
\endgroup%